# The dual-process approach to human sociality: Meta-analytic evidence for a theory of internalized heuristics for self-preservation

Valerio Capraro

University of Milan-Bicocca

Correspondence:

valerio.capraro@unimib.it

Abstract

Which social decisions are influenced by intuitive processes? Which by deliberative processes? The dual-process approach to human sociality has emerged in the last decades as a vibrant and exciting area of research. Yet, a perspective that integrates empirical and theoretical work is lacking. This review and meta-analysis synthesizes the existing literature on the cognitive basis of cooperation, altruism, truth-telling, positive and negative reciprocity, and deontology, and develops a framework that organizes the experimental regularities. The meta-analytic results suggest that intuition favours a set of heuristics that are related to the instinct for self-preservation: people avoid being harmed, avoid harming others (especially when there is a risk of harm to themselves), and are averse to disadvantageous inequalities. Finally, this paper highlights some key research questions to further advance our understanding of the cognitive foundations of human sociality.

# The dual-process approach to human sociality: Meta-analytic evidence for a theory of internalized heuristics for self-preservation

Humans are unique in the animal kingdom for their capacity to live in large societies made of thousands, if not millions, of unrelated individuals. While other species such as bees, ants, and naked mole-rats live in large societies, these typically consist of members sharing a substantial degree of biological relatedness. Other non-human primates live in groups of unrelated individuals, but these groups tend to be relatively small. This uniquely-human capacity to form large societies has led numerous scholars to argue that this unparalleled ability has been crucial for our evolutionary success as an animal species (Boyd & Richerson, 2005; Fehr & Fischbacher, 2003; Kaplan, Hill, Lancaster & Hurtado, 2000; Tomasello, Carpenter, Call, Behne & Moll, 2005).

An obvious consequence of living in such vast societies is that most of our decisions do not affect only ourselves, but they also affect other people or society as a whole. Given that the well-being of our societies highly depends on these "social decisions", understanding how people make these types of decisions has been a topic of interest for social scientists for decades. In the last twenty years, in particular, there has been a proliferation of studies viewing human sociality through a dual-process lens. Dual-process theories contend that people's choices result from the interplay between two cognitive systems, one that is fast and intuitive, and one that is slow and deliberative. By applying this dual-process lens to human sociality, scholars have started investigating questions such as: Which social choices are influenced by intuitive processes? Which by deliberative processes? These questions turned out to be extremely fruitful and have shed light on a series of exciting empirical results, setting the stage for innovative theoretical frameworks. At the same time, however, a number of important problems remain unsolved, making this field of research among the most thrilling and fascinating across all the behavioural sciences. The time is then mature to present a review of the state of the art. The goal of this work is fourfold: (1) to introduce new researchers to a rapidly expanding field with many open questions; (2) to provide an exhaustive literature review and meta-analysis for scholars already in the field; (3) to present a theoretical framework that organizes the empirical regularities; and (4) to highlight key unsolved problems essential for furthering our understanding of the cognitive basis of human sociality.

## Preliminaries

### Social decisions

This review will focus on social decisions. Broadly speaking, social decisions are defined as decisions affecting people beyond the decision maker. However, without further specification, it is unlikely that one could draw general conclusions regarding the cognitive foundations of all social decisions, due to the vast differences among them. Indeed, social decisions can vary along multiple dimensions, including their consequences and the ability of those affected to influence the decision maker. To narrow the focus, this review primarily examines cooperation, altruism, truth-telling, positive and negative reciprocity, and act utilitarianism, as these have been the primary focus of previous research. Exploring the cognitive basis of other types of decisions is certainly an important direction for future work. For this reason, the final section will discuss the few studies that have investigated the cognitive basis of equity-efficiency, ingroup favouritism, and aggression. The following list includes only the main decisions under consideration, each accompanied by a brief definition. More details about their measurement will be discussed in the corresponding sections.

**Cooperation**

Two or more decision-makers have to simultaneously decide whether to incur a cost to increase their collective payoff.

**Altruism**

A decision-maker has to decide whether to pay a cost to increase the payoff of another person or group of people.

**Truth-telling**

A decision-maker has to decide whether to lie or to tell the truth. The decision-maker's and another person's payoffs depend on whether the decision-maker lies or tells the truth. The other person might be abstract (e.g., the experimenter) or concrete (e.g., a fellow participant).

**Reciprocity**

After observing (or being the recipient of) an action, a decision-maker has to decide whether to incur a cost to punish or reward the actor. Several types of reciprocity exist, depending on the context. "Direct reciprocity" usually refers to situations in which the decision-maker was the recipient of the action (Trivers, 1971); "indirect reciprocity" usually refers to situations in which the decision-maker observed the action, but was not directly affected by it (Nowak & Sigmund, 2005). Additionally, one can distinguish between "positive reciprocity" and "negative reciprocity", depending on whether the decision-maker has observed (or was the recipient of) a benevolent or a malevolent action.

**Deontology vs. act utilitarianism**

A person has to decide whether to act so as to maximize the greater good, or to adhere to certain rights or duties, even if these lead to suboptimal outcomes.

Table 1 provides a detailed categorization of these social decisions, outlining the operational methods, and including illustrative examples for each.

*Table 1. List of social decisions, methods by which they are operationalized, and illustrative examples.*

| Social decision | Task | Example |
|---|---|---|
| Cooperation | Prisoner's dilemma, Public Goods game | Two players each receive 1 dollar and must simultaneously decide whether to give it to the other player or not. The given amount will be doubled and earned by the other player. |
| Altruism | Dictator Game | The dictator receives 1 dollar and must decide how much, if any, to give to a recipient, who starts with nothing. The recipient is passive. |
| Truth-telling | Sender-Receiver Game, Die-rolling task, Various self-serving cheating tasks | A player secretly rolls a die and is paid according to the reported outcome: 1 dollar for reporting 1, 2 dollars for reporting 2, and so on. |
| Negative reciprocity | Ultimatum Game, Various games involving prosocial behaviour followed by a punishing phase | The proposer decides how to split 10 dollars between themselves and the responder. The responder decides whether to accept or reject the offer. If the offer is rejected, both players end up with nothing. |
| Positive reciprocity | Trust Game, Various games involving prosocial behaviour followed by a rewarding phase | The investor receives 10 dollars and must decide how much to invest. The invested amount is tripled and given to the investee. The investee decides how much of the received amount to return to the investor. |
| Deontology vs Act utilitarianism | Sacrificial dilemmas | A person must decide whether to push a large man off a bridge to stop a runaway |

| Social decision | Task | Example |
|---|---|---|
| | | trolley, which otherwise would kill five people. |

**The dual-process approach**

Dual process theories refer to a set of frameworks sharing the core idea that people's choices result from the interplay between two cognitive systems, one that is fast and intuitive, and one that is slow and deliberative.

This idea has been formalized in several ways spanning multiple decades of research (Chaiken, 1980; Chen & Chaiken, 1999; De Neys & Pennycook, 2019; Epstein, 1994; Epstein & Pacini, 1999; Evans, 1989, 2006; Evans & Over, 1996; Evans, 2008; Evans & Stanovich, 2013; Fodor, 1983, 2001; Glöckner & Betsch, 2008; Glöckner & Witteman, 2010; Hammond, 1996; Kahneman & Frederick, 2005; Kahneman, 2011; Lieberman, 2003; Nisbett et al., 2001; Reber, 1993; Scheider & Shiffrin, 1977; Sloman, 1996; Smith & DeCoster, 2000; Stanovich, 1999, 2004; Stanovich, 2011; Strack & Deutsch, 2004; Toates, 2006; Wilson, 2002), leading to a class of models that differ from each other in many details, not only in the names given to the two cognitive systems, but also in the specific functions that these systems are assumed to perform, to the point that these theories do not exactly map onto each other (Evans, 2008; Evans & Stanovich, 2013).

Central to the discourse is the debate surrounding the nature of the two systems. Specifically, a major question pertains to whether the systems are qualitatively distinct (Evans & Stanovich, 2013) or merely quantitatively different (Kruglanski & Gigerenzer, 2008; Kruglanski, 2011). Current data do not conclusively favour either interpretation (De Neys, 2021). It is important to highlight that the present review does not hinge on the resolution of this debate.

It is also worth noting that another ongoing discussion involves the temporal sequencing of typically intuitive and typically deliberative responses. Traditional dual process theories tend to assume that intuitive responses are generated immediately after the stimulus, whereas deliberative responses come at a later time. This assumption has been challenged by a recent line of research suggesting that people can process basic logical principles intuitively (Ball, Thomson & Stupple, 2017; Banks & Hope, 2014; De Neys, 2012; Handley et al., 2011; Pennycook, Fugelsang & Koehler, 2015). This observation is leading towards a revision of dual process models. In "dual process 2.0" models (see De Neys and Pennycook, 2019, for a review) both "heuristic", intuitive responses and "logical", intuitive responses are generated automatically. If they are not in conflict, deliberation will favour the one with greater strength. On the other hand, if there is a conflict between multiple intuitive responses, deliberation will intervene to select the final response. It is crucial to emphasize, however, that although deliberation may sometime rationalize the heuristic intuition, its primary role is to favour "logical" intuitions (De Neys, 2023; Pennycook et al. 2015; Pennycook, 2023). Therefore, dual process 2.0 models differ from their 1.0 antecedents regarding the moment in which typically intuitive responses are produced, but not regarding which choice is typically preferred by deliberation.

This article will not advocate for a particular dual-process theory or delineate the specific, defining, functions of the two types of processes; instead, it will adopt a pragmatic approach by employing the so-called "typical correlates". These are measures that can be manipulated experimentally which *typically* correlate with the two cognitive processes. Importantly, these measures do *not* define the cognitive processes. For example, to say that intuitive processes tend to be fast, whereas deliberative processes tend to be slow, does not imply that intuitive processes will always be faster than deliberative processes. It does not even imply that intuitive responses will be generated earlier than deliberative responses. It simply means that, by experimentally manipulating people's response time, for example by including a time pressure that forces people to provide their answer within a certain window of time, the experimenter would be more likely to observe an increase in the frequency of intuitive responses, compared to the control condition, with no time pressure; symmetrically, by including a time delay that forces people to provide their answer after a certain amount of time, the experimenter would be more likely to observe an increase in the frequency of deliberative responses, compared to the control condition, with no time delay. This assumption does not require that deliberative responses are generated later than intuitive responses. In fact, this assumption is satisfied also in dual process 2.0 models, where "heuristic" and "logical" intuitive responses are assumed to be generated at the same time; but then, at a later time, deliberation will be more likely to override the heuristic intuition in favour of the logical intuition, than the converse.

This review will concentrate on studies that have experimentally manipulated one of the measures typically correlated with intuitive or deliberative processes during a social decision. Adopting this pragmatic approach is important because it enables drawing conclusions about the impact of intuitive and deliberative processes on social decisions, without taking a stance on which underlying dual-process theory is correct.

Coming to the specific "typical correlates", this review will focus on those that have been used in most previous research applying the dual process approach to human sociality. In this line of research, intuitive responses are assumed to be typically fast, autonomous, and effortless, while deliberative responses are assumed to be typically slow, controlled, and effortful. All these correlates appear in the list of correlates reported by Evans and Stanovich (2013), Table 1. In addition to these correlates, several experiments have utilized another one due to the simplicity with which it can be manipulated in the laboratory: reliance on emotions vs reliance on reason.

A critical property of all these correlates is indeed their amenability to experimental manipulation. The most popular manipulation techniques are time constraints, cognitive load, conceptual primes, ego-depletion, and the two-response paradigm.[2]

**Time constraints**

Since intuitive processes tend to be fast and deliberative processes tend to be slow, a simple technique to promote the use of intuition over deliberation is to ask participants to respond within a given window of time ("time pressure" condition). Conversely, to promote the use of deliberation over intuition, one can ask participants to stop and think for some time over the decision problem before making a decision ("time delay" condition).

Although useful, time constraints can be criticized along several dimensions. The first critical issue originates from the observation that time constraints must be set in the decision screen and not in the instruction screen, because, if placed in the instruction screen, they might interfere with task comprehension. However, this methodological necessity implies that will-be-under-time-pressure participants can actually stay as long as they want in the instruction screen, where they are free to start using deliberation. Therefore, it is unclear whether the decision that they then make under time pressure can be considered their intuitive decision. The second limitation regards the length of the time pressure, which is typically relatively long (5-10 seconds). On the one hand, relatively long time windows are needed for practical reasons; i.e., it requires a few seconds to inform participants that they have to make a decision. But, on the other hand, this raises the issue of whether decisions observed after 10 seconds can be considered intuitive, especially in light of evidence that first decisions are made almost instantaneously, and sometimes even before people become aware of them (Soon, Brass, Heinze & Haynes, 2008). A potential response to this criticism is that it is not really important that all participants under time pressure make an intuitive decision: it is enough that a significant proportion of them make an intuitive decision, in order to detect, if it exists, a significant difference between intuitive and deliberative decisions. However, this creates another issue: in case one observes no difference between decisions under time pressure and decisions under time delay, then it is not possible to disentangle whether this null result is driven by the fact that the time pressure does not affect the decision at hand. It could also be due to the inability of time pressure to promote intuitive processes in a significant proportion of participants.

The third limitation regards the compliance rate in the time pressure condition. While researchers can force participants to respond after a certain amount of time (e.g., revealing the "submit" button only after 10 seconds), it is not possible to force participants to respond within a certain timeframe. This generates a methodological conundrum. If the experimenter does not allow participants to submit their choice after a certain amount of time, there will be a selection bias; if the experimenter does allow participants to submit at any time, then there will be a proportion of participants who will not respect the time constraint. This is particularly problematic because participants are typically heterogeneous in their response time. Therefore, participants who fail to respond within the time constraint may systematically differ from the remaining participants in some (potentially unobservable) variables. This raises the question of how these participants should be treated in the analysis. In the seminal work by Rand et al. (2012), these participants were eliminated from the analysis. In subsequent work, experimenters have included all participants in the analysis (Bouwmeester et al., 2017; Capraro, 2017; Rand, Newman & Wurzbacher, 2015; Tinghög et al., 2013; Tinghög et al., 2016). More recently, scholars have developed different, more radical, solutions. One consists in letting participants practice in a series of games, before the actual decision, so that they get used to respond quickly, which helps to minimize the proportion of participants who fail to comply with the time pressure (Everett et al., 2017). Another solution is to include monetary disincentives, penalizing participants for each second spent on the decision screen (Alós-Ferrer & Garagnani, 2020).

**Cognitive load**

Deliberation relies heavily on working memory (Evans & Stanovich, 2013), the cognitive system with a limited capacity that is responsible for storing short-term memory (Miyake & Shah, 1999). Therefore, one can inhibit the use of deliberation by making participants use their short-term memory to solve a parallel, unrelated task (Gilbert & Hixon, 1991), as, for example, memorizing a sequence of digits (Swann, Hixon, Stein-Seroussi & Gilbert, 1990), or hearing a series of letters while having to press a button whenever they hear a character that was the same as two letters before (Gevins & Cutillo, 1993). Since participants need to store temporary information in their working memory, they will be less likely to use their working memory when making the primary, parallel decision. For example, Gilbert & Hixon (1991) found that cognitively loaded participants are more likely to rely on stereotypes. Shiv & Fedorikhin (1999) found that cognitively loaded participants are less likely to exert self-control when choosing between cake and fruit salad. Hinson, Jameson, and Whitney (2003) found that cognitively loaded participants have greater discounting of future rewards.

Studies based on cognitive load also have their limitations. First, it is possible that the task being used to impair participants' access to working memory interacts with the primary decision problem. Second, similarly to studies with time constraints, a number of participants typically fail to correctly perform the task aimed at taxing their working memory; how should the data from these participants be treated in the analysis? Similar to time constraint studies, to avoid selection bias, experimenters typically include all participants in the analysis. Third, cognitive load manipulations can be used to *undermine* deliberative processes, but cannot be used to *promote* them.

**Conceptual primes**

A technique used to promote both intuition and deliberation is known as conceptual priming. The idea is simple: before making a choice, participants are *primed* to rely either on intuition or deliberation. The priming can be done in several different ways.

Some scholars explicitly ask participants to use their intuition or deliberation while making their choice (Liu & Hao, 2011), others solicit participants to write a piece of text in which they have to describe a time in their life in which following their gut reactions (or thinking carefully) led them to good decision making (Rand, Greene & Nowak, 2012), yet others nudge participants to make a decision by “relying on emotion” vs “relying on reason” (Levine, Barasch, Rand, Berman & Small, 2018).[3]

The main limitation of these explicit primes is their direct use of terms like “intuition” or “deliberation”. This may create demand effects such that participants make a decision according to what they believe the intuitive or the deliberative choice should be (Rand, 2016). A potential solution would be to use subtler primes, as, for example, those used by Small, Loewenstein and Slovic (2007).

Another implicit priming technique consists in making participants read the instructions and express their decision in a foreign language. Since native language is intuitive and automatic, while foreign languages tend to be processed more slowly, foreign language might affect the cognitive process being used to make a decision. According to this idea, foreign language has been shown to reduce heuristics and biases (Costa et al., 2014a; Keysar et al., 2012) and to reduce emotionally-driven responses in moral dilemmas (Cipolletti, McFarlane & Weissglass, 2016; Corey et al., 2017; Costa et al., 2014b; Geipel, Hadjichristidis & Surian, 2015a; Hayakawa, Tannenbaum, Costa, Corey & Keysar, 2017). This line of research thus suggests that information processed in a second language reduces intuitive thinking.

However, priming through a foreign language also has two limitations. One is that several works have shown that information processed in a second language does not increase analytic thinking. For example, it does not improve performance on cognitive reflection tests (Costa et al., 2014a; Milczarski, Borkowska, Paruzel-Czachura, & Białek, 2022), nor in tasks involving conjunction fallacy or base rate neglect (Vives et al., 2018). It also does not improve detection of semantic illusions (Geipel et al., 2015b; but see Hadjichristidis, Geipel, & Surian, 2017). Therefore, at best, foreign language can be useful to impair intuitive responses, but not to promote deliberative thinking. The second limitation is potentially more critical. Reasoning in a second language is typically cognitively tiring and this might contribute to depletion of cognitive resources that works in the opposite direction of the prime (Volk, Köhler & Pudelko, 2014). For example, Bialek et al. (2020) discovered that foreign language can impede the detection of belief-logic conflicts, which results in a failure to engage in deliberation to override incorrect intuitions.

**Ego depletion**

Ego depletion is grounded on the theoretical assumption that deliberation requires the exertion of self-control to override immediate, autonomous, intuitive reactions. Although there is an ongoing debate about whether self-control is a limited resource (Baumeister, Heatherton & Tice, 1994; Baumeister & Tierney, 2011) or not (Inzlicht & Schmeichel, 2012; Inzlicht,

Schmeichel & Macrae, 2014), several studies have provided evidence that exerting self-control at Time 1 reduces the ability to exert self-control at Time 2. Following this line of work, social scientists have investigated how participants make social decisions at Time 2 after having completed, at Time 1, a task meant to undermine their self-control.

Self-control can be impaired in several ways, for example using the Stroop task (Stroop, 1935), or the e-hunting task (Moller, Deci & Ryan, 2006). Other experimental techniques include sleep deprivation,[4] hunger,[5] alcohol intoxication,[6] and mortality salience.[7]

Studies using ego depletion also have their shortcomings. First, while ego-depletion tasks can be used to impair deliberative processing, there is no symmetric equivalent to impair intuitive processing. A potentially more critical limitation is related to the ongoing debate about whether the ego-depletion effect actually exists. On the one hand, some meta-analyses have shown that, if one corrects for publication bias, the ego-depletion effect may not be different from zero (Carter, Kofler, Forster & McCullough, 2015; Carter & McCullough, 2014; Hagger et al., 2016). On the other hand, these meta-analyses have also been criticized along several dimensions, one of which is that they include also studies without a manipulation check. Therefore, it is possible that the overall lack of an effect is ultimately driven by the inclusion of studies that were unable to manipulate self-control. Consequently, whether ego-depletion is a real or an illusory effect is still discussed (Friese, Loschelder, Gieseler, Frankenbach & Inzlicht, 2019).

**The two-response paradigm**

The two-response paradigm, introduced by Thompson, Turner and Pennycook (2011), requires participants to make two consecutive decisions, one under time pressure or cognitive load, and one under no constraints (or under time delay). By analysing if and how participants switch between choices, one can determine whether the choices are affected by deliberative or intuitive processes.

The primary conceptual limitation of this method is that it creates an experimental demand for consistency, which may make participants reluctant to change their initial, intuitive choice. Classic work indeed shows that people have strong preferences for maintaining their previous decision (Samuelson & Zeckhauser, 1988). One potential remedy is to introduce a control treatment wherein participants make only the deliberative choice (Bago, Bonnefon & De Neys, 2019). Another limitation is that this technique promotes intuitive decisions by either time pressure or cognitive load, thereby inheriting the same limitations associated with these cognitive manipulations.

## The cognitive basis of cooperation

One of the most important forms of social behaviour is cooperation. Many scholars argue that the uniquely-human capacity to cooperate with unrelated others has been the secret of the enormous evolutionary success of Homo sapiens as an animal species (Axelrod & Hamilton, 1981; Bowles & Gintis, 2011; Darwin, 1871; Fehr & Fischbacher, 2004; Fehr & Gächter, 2002; Gintis, Bowles, Boyd & Fehr, 2003; Milinski, Semman & Krambeck, 2002; Nowak, 2006; Ostrom, 2000; Perc et al., 2017; Rand & Nowak, 2013; Trivers, 1971). Along these lines, research in psychology has suggested that the psychological basis of cooperation, *shared intentionality*, is what makes humans uniquely human, as it is possessed by children, but not by great apes (Tomasello et al., 2005). Therefore, it seems reasonable to start the central section of this article by reviewing the literature on the cognitive basis of cooperation.

**Measures of cooperation**

Cooperation can be operationally defined using social dilemmas. A social dilemma is any situation in which there is a conflict between individual interest and group interest (Capraro, 2013; Dawes, 1980; Fehr & Fischbacher, 2004; Hardin, 1968; Kollock, 1988; Nowak, 2006; Olson, 1965; Perc et al., 2017; Rand & Nowak, 2013). The main social dilemmas are the prisoner's dilemma and the public goods game.

**The prisoner's dilemma**

In the prisoner's dilemma, two players can either cooperate or defect. If both players cooperate, they receive the *reward* for cooperation, $R$; if both players defect, they receive the *punishment*, $P$. If one player cooperates and the other defects, the cooperator receives the *sucker*'s payoff, $S$, and the defector receives the *temptation* payoff, $T$. Payoffs are assumed to satisfy the inequalities $T > R > P > S$. These inequalities guarantee that both players would be better off if they both cooperate than if they both defect. However, unilateral defection brings the highest individual payoff. This generates the conflict between individual payoff and group payoff that is at the core of the social dilemma.

**The public goods game**

In the public goods game, N players start with an initial endowment. Each player has to choose how much, if any, to contribute to a public pool. The total contribution gets multiplied by a constant (greater than 1 and smaller than N) and then equally distributed among the N players. Therefore, the final payoff of each player is the part of the initial endowment that they decided not to contribute plus the proportion of the public good they received. As in the prisoner's dilemma, players would be better off if they all cooperate, compared to when they all defect; however, each individual is better off when they defect, independently of the other players' strategies. Therefore, also in this case there is a conflict between individual payoff and group payoff.

### Other social dilemmas

The prisoner's dilemma and the public goods game are not the only games characterized by a conflict between individual and group payoffs. Other instances include the traveller's dilemma and the Bertrand competition. However, previous work on the cognitive underpinnings of cooperation has overlooked these social dilemmas[11], which, consequently, will not be considered in this review.

### Other definitions of cooperation

There are other ways to define cooperative behaviour. Of particular interest for the current article is the distinction between *pure cooperation* and *strategic cooperation* as proposed by Rand (2016).

*Pure cooperation* refers to any game in which: (i) a player incurs a cost to confer a benefit to one or more other players; (ii) the payoff-maximizing strategy in the iterated variant of the game differs from the payoff-maximizing strategy in its one-shot, anonymous variant. One can easily see that one-shot prisoner's dilemmas and public goods games (as well as traveller's dilemmas and Bertrand competitions) possess these two properties. However, there are also other games which exhibit these properties. Consider the trust game. Player 1 is given a certain amount of money and has to decide how much, if any, to invest. The invested amount is multiplied by a factor greater than 1 and given to Player 2. Then, Player 2 decides how much of the received amount to return to Player 1. In one-shot anonymous interactions, Player 2 has no incentive to return any money, and therefore Player 2's payoff-maximizing strategy is to keep the entire amount. However, in repeated interactions, it might be optimal for Player 2 to return part of the amount in order to avoid Player 1 not investing any money in the next round, leading to a long-term loss for Player 2. Therefore, the trust game (played in the role of Player 2) exemplifies pure cooperation in the sense defined by Rand (2016).

*Strategic cooperation* refers to those games in which a player incurs a cost to confer a benefit to another player; however, whether this choice maximizes the player's payoff depends on the decision of the other player. The two prototypical cases considered by Rand (2016) are the one-shot and anonymous trust game (played as Player 1), and the one-shot and anonymous ultimatum game (played as Player 1). Indeed, in the one-shot and anonymous trust game, whether investing money maximizes Player 1's payoff depends on the expected behaviour of Player 2: if Player 1 believes that Player 2 will return a sufficient amount of money, then it is optimal for Player 1 to invest their money; otherwise, it is not. In the ultimatum game, Player 1 has to decide how to split a certain sum of money and Player 2 has to decide whether to accept or reject the proposed split. In case Player 2 accepts the offer, then the two players are paid according to the agreed split; in case Player 2 rejects the offer, both players end up with nothing. Therefore, Player 1 can make a larger offer (i.e., incurring a cost to confer a benefit to Player 2), but this choice will be optimal only if Player 1 believes that Player 2 will accept the corresponding offer and will reject any smaller offer. Other cases of strategic cooperation are the repeated public goods game and the repeated prisoner's dilemma.

**Review of the experimental evidence**

Several dozen experimental papers have explored whether intuitive or deliberative processing are causally linked to cooperative behaviour in one-shot social dilemmas. The full list is reported in Table 2.

*Table 2. List of papers exploring the cognitive basis of cooperation in one-shot social dilemmas. PGG stands for "public goods game"; PD stands for "prisoner's dilemma"; CPR stands for "common pool resource" (economically equivalent to the PGG, but in a take frame). The table first reports studies showing an overall positive effect of promoting intuition (or a negative effect of promoting deliberation) on cooperation, then studies showing a null effect, then studies showing a negative effect of intuition (or a positive effect of deliberation), and finally studies showing an effect depending on a moderating variable.*

| Studies showing a positive effect of promoting intuition (or a negative effect of promoting deliberation) on cooperative behaviour | | | |
|---|---|---|---|
| Paper | Manipulation | Measure | Main result |
| Rand et al. (2012) | Time constraints | PGG | Pressure *increases* cooperation, compared to Delay |
| Cone & Rand (2014) | Time constraints | Competitive PGG | Pressure *increases* cooperation, compared to Delay |
| Rand et al. (2015) | Time constraints | PGG with outgroups | Pressure *increases* cooperation, compared to Delay |
| Everett et al. (2017) | Time constraints | PD | Pressure *increases* cooperation, compared to Delay, also towards competitive out-groups |
| Isler et al. (2018) | Time constraints | PGG | Pressure *increases* cooperation, compared to Delay |
| Gao et al. (2020) | Time constraints | PGG | Pressure *increases* cooperation, compared to Delay |
| Gao et al. (2020) | Cognitive load | PD | Load *increases* cooperation |
| Rand et al. (2012) | Conceptual primes | PGG | Intuition *increases* cooperation, compared to Deliberation |

| Studies showing a positive effect of promoting intuition (or a negative effect of promoting deliberation) on cooperative behaviour | | | |
|---|---|---|---|
| Lotz (2015) | Conceptual primes | Asymmetric PD | Intuition *increases* cooperation, compared to Deliberation |
| Urbig et al. (2016) | Conceptual primes | PGG | Second language *decreases* cooperation |
| Levine et al. (2018) | Conceptual primes | PD | Emotion *increases* cooperation, compared to Reason |
| Gärtner et al. (2022) | Conceptual primes | PD, PGG, and DG | Emotion *promotes* pro-sociality compared to Reason |
| Studies showing a null effect of promoting intuition or deliberation on cooperative behaviour | | | |
| Paper | Manipulation | Measure | Main result |
| Tinghög et al. (2013) | Time constraints | PD | *No effect* of Pressure vs Delay |
| Tinghög et al. (2013) | Time constraints | Binary PGG | *No effect* of Pressure vs Delay |
| Verkoeijen & Bouwmeester (2014) | Time constraints | PGG | *No effect* of Pressure vs Delay |
| Capraro & Cococcioni (2015) | Time constraints | PD | *No effect* of Pressure vs Delay |
| Strømland et al. (2016) | Time constraints | PGG | *No effect* of Pressure vs Delay |
| Bouwmeester et al. (2017) | Time constraints | 21 preregistered PGGs | *No effect* of Pressure vs Delay on the whole sample. Pressure *increases* cooperation, among participants who respect the time constraint |
| Isler et al. (2021b) | Time constraints | PD | *No effect* of Pressure vs Delay |
| Gargalianou et al. (2017) | Conceptual primes | PGG | *No effect* of second language on cooperation |
| Camerer et al. (2018) | Conceptual primes | PGG | *No effect* of Intuition vs Deliberation |

Studies showing a positive effect of promoting intuition (or a negative effect of promoting deliberation) on cooperative behaviour

| Rantapuska et al. (2017) | Ego depletion | PGG | *No effect* of hunger on cooperation |
|---|---|---|---|
| Häusser et al. (2019) | Ego depletion | PGG | *No effect* of hunger on cooperation |
| McEvoy et al. (2021) | Ego depletion | PD | *No effect* of sleep restriction on cooperation |
| Bago et al. (2021) | Two-response | PGG | Deliberation has *no effect* on cooperation |

Studies showing a negative effect of promoting intuition (or a positive effect of promoting deliberation) on cooperative behaviour

| Paper | Manipulation | Measure | Main result |
|---|---|---|---|
| Capraro & Cococcioni (2016) | Time constraints | PD | Pressure *decreases* cooperation, compared to Delay |
| Goeschl & Lohse (2018) | Time constraints | PGG | Pressure *decreases* cooperation, compared to Baseline |
| Bilancini et al. (2021) | Time constraints | PGG | Pressure *decreases* cooperation with respect to "Motivated Delay", where participants have to motivate their action |
| Bogliacino & Gómez (in preparation) | Time constraints | PD | Pressure *decreases* cooperation, compared to Motivated Delay |
| Capraro & Cococcioni (2016) | Ego depletion | PD | Ego depletion *decreases* cooperation |

Studies showing that the effect of promoting intuition or deliberation on cooperative behaviour depends on some moderating variable

| Paper | Manipulation | Measure | Main result |
|---|---|---|---|
| Rand & Kraft-Todd (2014) | Time constraints | PGG | Pressure *increases* cooperation, compared |

| Studies showing a positive effect of promoting intuition (or a negative effect of promoting deliberation) on cooperative behaviour | | | |
| --- | --- | --- | --- |
| | | | to Delay, but only among participants that are naïve and trusting |
| Rand et al. (2014) | Time constraints | PGG | Pressure *increases* cooperation, compared to Delay, but only among inexperienced participants |
| Lohse (2016) | Time constraints | PGG | Pressure, compared to Baseline, *decreases* cooperation, but only among subjects with high CRT |
| Kessler et al. (2017) | Time constraints | PD with varying b/c | Pressure, compared to Delay, *increases* cooperation for small b/c and *decreases* cooperation for large b/c |
| Bird et al. (2019) | Time constraints | PGG only males | Pressure, compared to Delay, *increases* cooperation among risk-averse males; Pressure *decreases* cooperation among risk-seeking males |
| Alós-Ferrer & Garagnani (2020) | Time constraints | PGG | Pressure, compared to Delay, *increases* cooperation among prosocial participants; Pressure has *no effect* on individualist participants; Pressure *decreases* cooperation among competitive participants |
| Isler et al. (2021a) | Time constraints | PGG/CPR | Pressure, compared to Delay, *increases* cooperation in the PGG; pressure *decreases* cooperation in the CPR |

| Studies showing a positive effect of promoting intuition (or a negative effect of promoting deliberation) on cooperative behaviour | | | |
|---|---|---|---|
| Nava et al. (2023) | Time constraints | PGG | Pressure, compared to Delay, *increases* cooperation among adults; Pressure *decreases* cooperation among adolescents |

Considering the volume of studies available, this review will concentrate on meta-analyses as well as select key references identified for their exploration of effect heterogeneity through the examination of theoretically grounded individual-level moderating factors. A discourse on additional factors found to moderate the effect is reserved for the *Outlook and open problems* subsection.

The first meta-analysis was published by Rand (2016), who analyzed 67 experiments in which participants' cognitive processing was manipulated while making a pure cooperation decision; he included in the meta-analysis also studies using the trust game (only player 2), and sequential prisoner's dilemmas (only second movers). He also included two field experiments in which subjects had to decide whether to return a lost glove under different time constraints (Artavia-Mora, Bedi & Rieger, 2017; Artavia-Mora, Bedi & Rieger, 2018) and one study in which cooperating with one player implied competing with a third party (De Dreu, Dussel & Ten Velden, 2015). In doing so, he reported an overall positive effect of intuition on cooperation (effect size = 6.1 percentage points, 95% CI = [3.4, 8.9]). Moreover, this positive effect was not present in strategic cooperation (effect size = − 0.4 percentage points, 95% CI = [−3.0, 2.1]). He used this result to support the Social Heuristics Hypothesis (SHH). The SHH (Rand et al., 2012; Rand et al., 2014) contends that people internalize strategies that are optimal in everyday interactions, which are typically repeated, and use them as default strategies in novel settings, when their capacity to deliberate is impaired. Then, after deliberation, people override these social heuristics and tailor their behaviour towards the one that is optimal in the given interaction. Therefore, the SHH predicts that intuition favours pure but not strategic cooperation, which is exactly what Rand found in his meta-analysis.

However, this overall conclusion was subsequently challenged by Kvarven et al. (2020), who conducted an updated meta-analysis (82 experiments). In their meta-analysis, Kvarven and colleagues included only social dilemmas, meaning that they did not include the trust game, the sequential prisoner's dilemma, and the two natural field experiments by Artavia-Mora and colleagues (2017, 2018). In doing so, they did find an overall effect of intuition on cooperation (effect size = 2.19 percentage points, CI not reported, $p = 0.005$), but, critically, this effect was entirely driven by conceptual primes explicitly asking participants to rely on their emotion (effect size = 14.88 percentage points, CI not reported, $p < 0.001$). The authors argued that these primes might have induced a demand effect. Moreover, a qualitatively similar result was obtained when adding, in the analysis, the trust game and the sequential prisoner's dilemmas, but not Artavia-Mora and colleagues' field experiments, because they interpreted these works as studying altruistic behaviour and not cooperative behaviour (the person who lost the glove could not reciprocate the action). In a response paper, Rand (2019) argued that there is no theoretical reason to exclude the trust game, the sequential prisoner's dilemma and the two field experiments, because they all meet the criteria for being classified as "pure cooperation". Adding these papers, Rand (2019) found a positive effect of intuition on pure cooperation (effect size = 3.1 percentage points, 95% CI = [1.5, 4.7]), which remains significant even when restricting the analysis to studies that do not use the explicit primes (effect size = 1.6 percentage points, 95% CI

= [0.4, 2.8]). Since 2020, there have been eleven more experimental studies testing the effect of manipulating cognitive processes on cooperation using social dilemmas: five of them reported a positive effect of promoting intuition on cooperation, four of them reported a null effect, and two of them reported a negative effect. However, the negative effect was obtained by comparing a "time pressure" manipulation with a "motivated delay" treatment, where participants were asked to motivate their action, before implementing it (Bilancini et al., 2022; Bogliacino & Gómez, in preparation). This intervention may activate other processes than pure deliberation. In conclusion, the available experimental evidence points towards a null or positive (but small and very heterogeneous) effect of promoting intuition on cooperation.

Some works have sought to explain part of the heterogeneity across studies by investigating the role of theoretically grounded individual-level moderators. Rand et al. (2014) found that the effect of intuition on cooperation was driven by participants with no previous experience in laboratory experiments. Rand and Kraft-Todd (2014) found that time pressure increases cooperation, but only among participants who are both inexperienced and have high trust in people with whom they interact in their everyday life. These results fit well with the SHH, because this framework predicts that inexperienced participants living in a trustworthy environment are more likely to bring cooperative heuristics to the lab than participants who are either experienced or who live in an untrustworthy environment; the former group may have learned that defecting is optimal in the given one-shot interaction, while the latter may have internalized defective heuristics.

Another work investigated the cognitive foundations of cooperation as a function of participants' social value orientation (SVO). Specifically, Alós-Ferrer and Garagnani (2020) implemented incentive compatible time constraints as follows: participants under time delay were paid €1 if they responded after 10 seconds; participants under time pressure lost 30 cents for every second spent on the decision screen. In doing so, they found no overall effect of time pressure on cooperation in a one-shot public goods game; however, they did find that time constraints interacted with SVO, such that time pressure made prosocial individuals more prosocial, and competitors more competitive; individualists were unaffected by the time constraints. They also replicated this result in a repeated public goods game with random re-matching and no feedback. They concluded that intuition enhances intrinsic predispositions.

**Outlook and open problems**

As described above, the evidence regarding a potential causal link between cognitive process and cooperation is somewhat mixed. Rand's (2016) meta-analysis shows that, overall, intuitive processes favour cooperative behaviour, whereas deliberative processes promote non-cooperative behaviour. However, this conclusion was challenged by a larger meta-analysis, with different inclusion criteria (Kvarven et al., 2020), which, subsequently, was challenged by an even larger meta-analysis (Rand, 2019). All these meta-analyses however agree on one point: explicit conceptual primes of emotion increase cooperative behaviour. They disagree on the effect of other cognitive manipulations.

The overall conclusion that one can draw from this debate is that there is a need for more and better designed studies. The vast majority of experiments used time constraints or explicit primes of intuition. However, both these methods have been criticized. The methodological issues of time constraints have been resolved in later studies (Alós-Ferrer & Garagnani, 2020; Everett et al., 2017; Isler et al., 2018); therefore, future work using time constraints can prioritize these methods. Regarding conceptual primes, there is a need for studies using more implicit conceptual primes. Indeed, there has only been one study using conceptual primes of intuitions that are not explicit (i.e., one study that employed second-language as a manipulation). To increase generalizability, it would be helpful also to conduct studies using different cognitive manipulations, as there is only one study using cognitive load and four using ego-depletion.

Another important question regards the role of potential moderators. Alós-Ferrer and Garagnani (2020) found that the effect of time pressure on cooperation depends on the social value orientation, such that prosocial individuals tend to become more cooperative when deciding under pressure, whereas individualists tend to be unaffected by the manipulation, and competitive subjects even tend to become more competitive.[11] To explain these results, they proposed that time pressure makes people more likely to behave according to their social value orientation, while time delay makes people readapt this initial response towards what they believe to be the norm in the given situation. Testing this hypothesis is an important direction for future research, because it is at odds with the SHH in terms of the predicted effect of deliberation on cooperation, but not in terms of the predicted effect of intuition. This is because the SHH makes predictions about the moderating role of the environment in which participants live. More precisely, the SHH contends that intuitive processes promote cooperation among participants who live in a cooperative setting, but not among those who live in a non-cooperative setting, who are predicted to remain unaffected by the cognitive manipulation. To the best of current knowledge, there are no studies correlating a participant's SVO with the cooperativeness of the environment in which this participant lives, but it seems plausible to expect a correlation between these two measures. Therefore, the predictions of the SHH are very similar to those of Alós-Ferrer and Garagnani's hypothesis, in terms of the effect of intuition. By contrast, while the SHH predicts that deliberation can never increase cooperation, Alós-Ferrer and Garagnani's hypothesis predicts a positive effect of deliberation on cooperation among pro-self individuals.

Another relevant, but poorly explored, line of research regards the existence of a potential inverted-U effect of cognitive effort on cooperation. Study 3 in Moore and Tenbrunsel (2014) found that cognitive complexity has an inverted-U effect on cooperation in a non-incentivized, social dilemma based on the Shark Harvesting and Resource Conservation exercise. Capraro and Cococcioni (2016) used this result to explain why extreme cognitive depletions (extreme time pressure and ego-depletion) seem to lead to lower rates of cooperation: it is possible that the effect of cognitive effort on cooperative behaviour is non-linear. The theoretical rationale for this assumption is that cooperation might be partly driven by an abstract desire to do the morally right thing (Capraro & Rand, 2018); and it has been long argued that the application of abstract moral rules is not immediate, but appears only at later stages of moral reasoning (Kohlberg, 1963, 1981). Therefore, exploring a potential inverted-U effect of cognitive reflection on cooperation is a fascinating and non-trivial question, with obvious theoretical and practical implications. One remark is that this question is inherently difficult because implementing extreme cognitive manipulations (e.g., extreme time pressure, or very hard cognitive load or ego depletion tasks) may interfere with participants understanding of the game or increase drop-out rates, leading to potential selection biases.

Another open problem regards whether the effect of promoting intuition vs deliberation on cooperation depends on the benefit-to-cost ratio. Kessler et al. (2017) found that time pressure increases cooperation for relatively small *b/c*, but the effect is reversed for bigger values. If confirmed, this effect would have important theoretical implications, because it would not be consistent with any of the frameworks previously mentioned, and therefore would require a new theorization. As a potential explanation, Kessler et al. (2017) proposed that deliberative processes put more weight on social efficiency.

Finally, an observation regarding asymmetric social dilemmas, that is, social dilemmas where the cost and the benefit of cooperation are player-dependent. These dilemmas are important in light of practical applications, because, in reality, there is often power imbalance among people, which can be modelled in terms of asymmetries between costs and benefits of cooperation. In spite of this practical importance, however, only one study explored the cognitive underpinnings of cooperation in asymmetric dilemmas: Lotz (2015) used conceptual primes and found that intuition promotes cooperation. More studies are needed also on this topic.

## The cognitive basis of altruism

We often pay costs to help others, even in situations where the recipient has no way to reciprocate our actions, such as purchasing food for a homeless individual situated at the entrance of a food shop. Altruism encapsulates the essence of this behavioural domain.

### Measures of altruism

The quintessential measure of altruism is the dictator game, wherein the *dictator* is allocated a certain amount of money and has to decide how much, if any, to donate to the *recipient*, who is initially given nothing. The recipient has no choice but to accept whatever the dictator decides to give. In the standard version of the dictator game, the recipient is typically an individual. However, one can consider variants where the recipient is not a singular person. Specifically, a variant that has garnered considerable attention is when the recipient is a charitable organization. Since there is no theoretical reason to distinguish between these two variants, this review will encompass both.

Another variation can be devised by changing the ratio between the monetary cost and the monetary benefit of the altruistic action. For instance, when the monetary cost of helping is less than the monetary benefit generated, the altruistic action not only aids the other individual but also enhances social efficiency. However, this introduces a confounding factor since equity and efficiency might be cognitively different. To circumvent this, the section will concentrate on studies employing a dictator game where the monetary cost of the altruistic action is equal to its monetary benefit. The concluding section of this article will review the literature on the cognitive foundation of altruism in dictator games with varying cost-to-benefit ratios.

**Review of the experimental evidence**

Table 3 reports the list of papers that explored the cognitive basis of altruism, along with a brief description of the main result for each paper.

*Table 3. List of papers exploring the cognitive basis of altruism. DG stands for the standard dictator game; DG charity stands for the DG when the recipient is a charitable organization.*

| Studies showing a positive effect of promoting intuition (or a negative effect of promoting deliberation) on altruistic behaviour | | | |
|---|---|---|---|
| Paper | Manipulation | Measure | Main result |
| Artavia-Mora et al. (2017) | Time constraints | Field experiment | Pressure *increases* helping compared to Delay |
| Artavia-Mora et al. (2018) | Time constraints | Field experiment | Pressure *increases* helping, compared to Delay |
| Chuan et al. (2018) | Time constraints | DG charity | Longer and longer delay (there is no baseline) *decreases* giving |
| Grolleau et al. (2018) | Time constraints | DG | Longer and longer delay (no baseline) *decreases* giving |
| Schulz et al. (2014) | Cognitive load | DG | Load *increases* giving |
| Zhong (2011) | Conceptual primes | DG charity | “Feel” *increases* giving, compared to “Decide” |

Studies showing a positive effect of promoting intuition (or a negative effect of promoting deliberation) on altruistic behaviour

| | | | |
|---|---|---|---|
| Wang et al. (2019) | Ego depletion | Helping in high-crisis scenarios | Depletion *increases* helping in high-crisis scenarios |

Studies showing a null effect of promoting intuition (or deliberation) on altruistic behaviour

| Paper | Manipulation | Measure | Main result |
|---|---|---|---|
| Carlson et al. (2016) | | DG charity | *No effect* of Pressure compared to Delay |
| Jarke-Neuert & Lohse (2016) | Time constraints | DG | *No effect* of Pressure compared to Delay |
| Tinghög et al. (2016) | Time constraints | DG charity | *No effect* of Pressure compared to Delay |
| Kessler et al. (2017) | Time constraints | DG | Pressure has *no effect* compared to Baseline |
| Andersen et al. (2018) | Time constraints | DG | *No effect* of Delay compared to Baseline |
| Strømland & Torsvik (2019) | Time constraints | DG | Pressure has *no effect* on giving, compared to Delay |
| Benjamin et al. (2013) | Cognitive load | DG | Load has *no effect* on giving |
| Kessler & Meier (2014) | Cognitive load | DG charity | Load has *no effect* on giving |
| Hauge et al. (2016) | Cognitive load | DG | Load has *no effect* on giving |
| Tinghög et al. (2016) | Cognitive load | DG charity | Load has *no effect* on giving |
| Grossman & Van der Weele (2017) | Cognitive load | DG charity | Load has *no effect* on giving |
| Kinnunen & Windmann (2013) | Conceptual primes | DG charity | Intuition has *no effect* on giving, compared to Deliberation |
| Anderson & Dickinson (2010) | Ego depletion | DG | Sleep deprivation has *no effect* on giving |
| Aarøe & Petersen (2013) | Ego depletion | DG | Drinking Sprite (containing carbohydrates) has *no effect* on giving, compared to drinking |

Studies showing a positive effect of promoting intuition (or a negative effect of promoting deliberation) on altruistic behaviour

| Paper | Manipulation | Measure | Main result |
|---|---|---|---|
| | | | Sprite zero (an artificial sweetener) |
| Rantapuska et al. (2017) | Ego depletion | DG charity | Hunger has *no effect* on giving |
| Häusser et al. (2019) | Ego depletion | DG and a volunteering task | Hunger has *no effect* on giving and volunteering |
| Bago et al. (2021) | Two-response | DG | Deliberation has *no effect* on altruism |

Studies showing a negative effect of promoting intuition (or a positive effect of promoting deliberation) on altruistic behaviour

| Paper | Manipulation | Measure | Main result |
|---|---|---|---|
| Mrkva (2017) | Time constraints | DG charity | Delay *increases* giving, compared to Baseline |
| Bilancini et al. (2021) | Time constraints | DG | Motivated Delay *increases* giving compared to Baseline and Delay |
| Roch et al. (2000) | Cognitive load | Taking from common pool | Load *increases* taking |
| Briers et al. (2006) | Ego depletion | DG | Hunger *decreases* giving |
| Halali et al. (2013) | Ego depletion | DG | Depletion *decreases* frequency of equal split |
| Xu et al. (2014) | Ego depletion | DG | Depletion *decreases* giving |
| Achtziger et al. (2015) | Ego depletion | DG | Depletion *decreases* giving |
| Dickinson & McElroy (2017) | Ego depletion | DG | Sleep restriction *decreases* giving |
| Holbein et al. (2019) | Ego depletion | DG charity | Sleep restriction *decreases* giving |
| Gärtner (2018) | Ego depletion | DG | Pressure *decreases* giving, compared to Baseline, but only when the dictator game is presented with no status-quo distribution |

Studies showing a positive effect of promoting intuition (or a negative effect of promoting deliberation) on altruistic behaviour

Studies showing that the effect of promoting intuition or deliberation on altruistic behaviour depends on some moderating variable

| Paper | Manipulation | Measure | Main result |
|---|---|---|---|
| Cornelissen et al. (2011) | Cognitive load | DG | Load *increases* giving among prosocials; no effect among individualists |
| Small et al. (2007) | Conceptual primes | DG charity | Intuition *increases* giving compared to Deliberation, but only when the target is identifiable and not when is statistical |
| Shi et al. (2020) | Conceptual primes | Helping situations | Deliberation, compared to Intuition, *decreases* helping, especially for subjects with low Honesty-Humility (Study 1) or proself subjects (Study 2) |
| Nockur & Pfattheicher (2021) | Conceptual primes | DG | Intuition, compared to Baseline, *decreases* giving among people low in Honesty-Humility, and has *no effect* among people high in Honesty-Humility |
| Ferrara et al. (2015) | Ego depletion | DG | Sleep restriction *decreases* giving, but only for females |
| Harel & Kogut (2015) | Ego depletion | DG charity | Hunger *decreases* giving to a charity that help people to purchase food |
| Banker et al. (2017) | Ego depletion | DG | Depletion has *no effect* on giving, it only makes participants |

| Studies showing a positive effect of promoting intuition (or a negative effect of promoting deliberation) on altruistic behaviour | | | |
|---|---|---|---|
| Itzchakov et al. (2018) | Ego depletion | DG charity | more likely to choose the status quo<br>Depletion *decreases* giving, but only when giving is accompanied by a persuasion message |

Given the multitude of studies, also in this behavioural domain this review will concentrate on meta-analyses and key references identified for their examination of the moderating effects of theoretically grounded individual-level variables. Fromell, Nosenzo and Owens (2020) analysed 60 experiments (total N = 12,574) and found that cognitive processing does not influence altruistic behaviour (meta-analytic standardized mean difference (Cohen's *d*) = -0.015, 95% CI = [-0.070, 0.041]). Since 2020, four additional studies have been conducted, one indicating a positive effect of deliberation on giving, two yielding null results, and one demonstrating a negative effect. Therefore, the available research suggests that, overall, cognitive processing does not have a significant impact on altruistic behaviour.

To understand the heterogeneity across studies, scholars have explored the role of potential individual-level moderators. One variable that has attracted considerable attention, for theoretical reasons, is gender. Previous work suggests that men and women often conform to stereotypes regarding their respective social roles, with women being more communal and unselfish, and men being more agentic and independent (Eagly, 1987). Congruent with this perspective, empirical research has consistently found that women tend to exhibit greater altruism (Brañas-Garza, Capraro & Rascón-Ramírez, 2018; Engel, 2011). Moreover, women are expected to demonstrate greater altruism than men in dictator games (Aguiar, Brañas-Garza, Cobo-Reyes, Jimenez & Miller, 2009; Brañas-Garza, Capraro & Rascón-Ramírez, 2018). Furthermore, women, when perceived as insufficiently altruistic, face greater disadvantages than men, being liked less, less likely to be helped, hired, promoted, paid fairly, and given status (Heilman & Okimoto, 2007). Hence, if intuitive reactions reflect internalized strategies that people quickly access when they have no opportunity to reason about the specific situation at hand, it is plausible to hypothesize that intuition promotes altruism for women but not for men. Corroborating this theory, Rand, Brescoll, Everett, Capraro and Barcelo (2016) conducted a meta-analysis of 22 dictator game experiments (total N = 4,366), exploring the moderating role of gender. Their findings revealed a significant interaction between cognitive manipulation and gender (effect size = 5.5 percentage points, 95% CI [2.6, 8.5]). Specifically, fostering intuition enhanced altruism for women (effect size = 3.8 percentage points, 95% CI [1.9, 5.7]) but not for men (effect size = -2.0 percentage points, 95% CI [-4.2, 0]). Moreover, consistent with this interpretation, a second study published in the same article indicated that the more women described themselves using masculine attributes relative to feminine attributes, the more deliberative processes reduced their altruistic behaviour. This suggests that the negative effect of deliberation on altruism among women was partly explained by identification with the other gender. However, this result was not replicated by Fromell et al. (2020), who did not find an interaction between gender and cognitive mode, but found that promoting intuition had no effect on women (meta-analytic standardized mean difference = -0.01, 95% CI = [-0.08,0.07]), whereas it had a statistically significant negative effect on men (meta-analytic standardized mean difference = -0.08, 95% CI = [-0.13, -0.03]). In conclusion, two meta-analyses examining the cognitive underpinnings of altruistic behaviour discern a potential differential effect on women and men. Yet, there is a divergence on the direction of the simple effects: Rand et al. (2016) contend that promoting intuition favours altruism for women but not for men; Fromell et al. (2020) argue that promoting intuition disfavours altruism for men but not for women.

Scholars have also investigated whether the effect of cognitive mode on altruistic behaviour is moderated by individual measures of prosociality, finding mixed results. Specifically, Cornelissen et al. (2011) demonstrated that cognitive load favours dictator game giving, but only among prosocial people; conversely, non-cognitively loaded prosocial people donated as much as non-cognitively loaded individualists. By contrast, Shi et al. (2020) reported that inducing gut reactions reduced altruistic behaviour of both prosocial and proself individuals, with a more pronounced reduction observed for proself individuals. The evidence becomes further nuanced when considering, as an individual measure of prosocial predispositions, Honesty-Humility (HH; a personality factor from the HEXACO model of personality that has been associated with various prosocial behaviours, including altruism; Thielmann et al., 2020). Shi et al. (2020) reported that people high in HH are not affected by intuitive versus deliberative primes. Conversely, people low in HH tend to decrease altruistic behaviour under deliberation. However, these results were not fully replicated in subsequent investigations. Nockur and Pfattheicher (2021) validated the absence of an effect among people high in HH, but found that prompting intuitive decisions, compared to control, decreased donations among people low in HH.

**Outlook and open problems**

To summarize, two meta-analyses concur that intuitive and deliberative processes do not exhibit a clear overall effect on altruistic behaviour. Nonetheless, existing research highlights the potential moderating roles of gender and individual measures of prosociality. However, the details of these moderations are still poorly understood.

Another avenue warranting further attention pertains to the impact of default options. In general, promoting intuition may increase the status-quo bias, because the default option may serve as an anchor that is readily accessible without reflection (Samuelson & Zeckhauser, 1988; Tversky & Kahneman, 1974). In the standard dictator game, giving nothing is the natural default option. This raises a critical issue: Can prompting intuition in the dictator game in some cases increase self-interest not because it increases self-interest per se, but because self-interest happens to coincide with the default option? Banker et al. (2017) probed this question and found that ego-depletion did not affect altruistic behaviour, but simply made people more likely to choose the status-quo, whichever that was. Similarly, Gärtner (2018) found that both prosocial and proself status-quos affected altruistic behaviour, but without interacting with time pressure. Notably, Gärtner's findings align with those of Banker and colleagues, suggesting that promoting intuition may just bolster the default option in the dictator game. Future work should strive to illuminate this aspect more comprehensively.

## The cognitive basis of truth-telling

The tension between lying and truth-telling underscores a myriad of social and economic interactions characterized by communication with asymmetric information. In such scenarios, people may be tempted to misreport their private information for their benefit. Both empirical research and anecdotal experience indicate that a substantial proportion of people do indeed take advantage of their private information to further their own interests. As a consequence, lying has a considerable toll on people, businesses, and society as a whole. For instance, tax evasion is estimated to cost about 100 billion dollars annually to the U.S. Treasury alone (Gravelle, 2009).

Hence, discerning whether intuitive processes favour or deter truth-telling is a matter of primary importance.

**Measures of lying**

Behavioural scientists have developed a plethora of methods to measure an individual's propensity to misreport the truth within controlled laboratory experiments. Drawing on Köbis et al. (2019), these measures can be broadly segmented into two categories: those where lying harms an "abstract" person (e.g., the experimenter), and those where lying harms a "concrete" person (e.g., another participant in the experiment).[13]

**Lying tasks (lying harms "abstract others")**

A simple approach to measure people's (dis-)honesty in situations in which lying harms an abstract person is to let participants complete a task and then pay them according to the self-reported performance in the task, rather than according to the actual performance. In this paradigm, the abstract person harmed by the lie is the experimenter.

A popular task employed for this purpose is the "matrix task" (Mazar, Amir & Ariely, 2008). In its original version, participants are exposed to a sequence of twenty 4x3 matrices, with entries comprising three-digit numbers between 0 and 10 (e.g., 1.34). Participants are given four minutes to identify two numbers per matrix that add up to 10. At the end of the experiment, participants are paid according to the self-reported performance in the task. Other tasks with analogous purpose include unsolvable logic puzzles (Muraven, Pogarsky & Shmueli, 2006), mathematical problems (Zhong, 2011), and self-reported performance on the Raven test (Pitesa, Thau, & Pillutla, 2013).

Another frequently utilized lying task is the dice-under-cup task (or dice-rolling task), wherein participants covertly roll a die (e.g., under a cup) and are remunerated according to the reported outcome (Fischbacher & Föllmi-Heusi, 2013).

**The sender-receiver game (lying harms "concrete others")**

The scenario in which lying harms concrete others is typically examined through the sender-receiver game (Gneezy, 2005). In this game, both the *sender* and the *receiver* are informed that there are two possible allocations of money. However, only the sender is informed about the exact monetary payoffs corresponding to these options. Then, the sender conveys a message to the receiver, chosen among two possible alternatives: Message A: "Option A gives you a better payment than Option B"; Message B: "Option B gives you a better payment than Option A". After receiving the message, the receiver decides which of the two options to implement as a payment.

This game, however, introduces a potential confound, termed "sophisticated deception" (Sutter, 2009). There might be senders telling the truth not out of honesty, but as a stratagem to deceive the receiver, premised on the assumption that the receiver will not believe them. To obviate this confound, researchers have developed variants of the sender-receiver game wherein the receiver is passive and makes no decision (Biziou van Pol, Haenen, Novaro, Occhipinti-Liberman & Capraro, 2015) or their decision has no bearing on the sender's payoff (Cappelen, Sørensen & Tungodden, 2013; Gneezy, Rockenbach & Serra-Garcia, 2013).

**Review of the empirical literature**

Table 4 provides a succinct overview of each paper investigating the cognitive basis of truth-telling.

*Table 4. List of experimental studies exploring the cognitive basis of truth-telling. SR stands for sender-receiver game; DR stands for die-rolling task; SSC stands for self-serving cheating task.*

Studies showing a positive effect of promoting intuition (or a negative effect of promoting deliberation) on truth-telling

| Paper | Manipulation | Measure | Main result |
|---|---|---|---|
| Foerster et al. (2013) | Time constraints | DR | Pressure *increases* truth-telling, compared to Baseline |
| Capraro (2017) | Time constraints | SR | Pressure *increases* truth-telling, compared to Delay |
| Lohse et al. (2018) | Time constraints | SSC | Pressure, compared to Delay, *increases* truth-telling, by making participants less aware of the opportunity to lie |
| Capraro et al. (2019a) | Time constraints | SR | Pressure *increases* truth-telling, compared to Delay |
| van't Veer et al. (2014) | Cognitive load | DR | Load *increases* truth-telling |
| Zhong (2011) | Conceptual primes | SR | Emotion *increases* truth-telling, compared to Deliberation |
| Bereby-Meyer et al. (2020) | Conceptual primes | DR | Second language *decreases* truth-telling |

Studies showing a null effect of promoting intuition or deliberation on truth-telling

| Paper | Manipulation | Measure | Main result |
|---|---|---|---|
| Andersen et al. (2018) | Time constraints | SSC | Delay has *no effect* on truth-telling, compared to Baseline |
| Van der Cruyssen et al. (2020) | Time constraints | DR | Pressure has *no effect* on truth-telling, compared to Baseline |
| Mitkidis et al. (2022) | Cognitive load | SSC | Load has *no effect* on truth-telling |

Studies showing a positive effect of promoting intuition (or a negative effect of promoting deliberation) on truth-telling

| Paper | Manipulation | Measure | Main result |
|---|---|---|---|
| Welsh et al. (2014) | Ego depletion | SR | Ego depletion has *no effect* on truth-telling |

Studies showing a negative effect of promoting intuition (or a positive effect of promoting deliberation) on truth-telling

| Paper | Manipulation | Measure | Main result |
|---|---|---|---|
| Gunia et al. (2012) | Time constraints | SR | Delay *increases* truth-telling, compared to Baseline |
| Shalvi et al. (2012) | Time constraints | DR | Pressure *decreases* truth-telling, compared to Baseline |
| Muraven et al. (2006) | Ego depletion | SSC | Ego depletion *decreases* truth-telling |
| Mead et al. (2009) | Ego depletion | SSC | Ego depletion *decreases* truth-telling |
| Gotlib & Converse (2010) | Ego depletion | SSC | Ego depletion *decreases* truth-telling |
| Gino et al. (2011) | Ego depletion | SSC | Ego depletion *decreases* truth-telling |
| Chiou et al. (2017) | Ego depletion | SSC only males | Ego depletion *decreases* truth-telling |
| Yam (2018) | Ego depletion | SSC | Ego depletion *decreases* honesty |
| Dickinson & Masclet (2021) | Ego depletion | DR and SSC | Sleep restriction *decreases* truth-telling |

Studies showing that the effect of promoting intuition or deliberation on truth-telling depends on some moderating variable

| Paper | Manipulation | Measure | Main result |
|---|---|---|---|
| Reis et al. (2022) | Cognitive load | DR | Load *increases* truth-telling in the gain dimension; load *decreases* truth-telling in the loss dimension |

| Studies showing a positive effect of promoting intuition (or a negative effect of promoting deliberation) on truth-telling | | | |
|---|---|---|---|
| Cappelen et al. (2013) | Conceptual primes | SR Pareto lies | Intuition *increases* truth-telling, compared to Deliberation, but only for males |
| Pitesa et al. (2013) | Ego depletion | SSC | Ego depletion *decreases* truth-telling, but only when consequences on others are not salient |
| Yam et al. (2014) | Ego depletion | SSC | Ego depletion *decreases* truth-telling, but only when there is low social agreement that the dishonest action is immoral |

Köbis, Verschuere, Bereby-Meyer, Rand and Shalvi (2019) conducted a meta-analysis of all the studies manipulating cognitive processes during a task involving lying. They examined variability both in the proportion of liars (73 studies, $N = 12{,}711$) and in the magnitude of lying (50 studies, $N = 6{,}473$). In doing so, they found that promoting intuitive processes increases the proportion of liars (log $OR = 0.38$, 95% credibility interval (CrI) = $[-0.86, 1.63]$, $p = 0.0004$) and the magnitude of lying ($g = 0.261$, 95% CrI = $[-0.280, 0.802]$, $p < .0001$) in tasks where lying harms abstract others, but not in tasks where lying harms concrete others (proportion of liars: log $OR = -0.04$, 95% CrI = $[-1.32, 1.24]$, $p = 0.861$; magnitude of lying: $g = -0.076$, 95% CrI = $[-0.696, 0.544]$, $p = 0.63$). In the latter instances, promoting intuitive over deliberative processes has no effect on dishonest behaviour. Köbis and colleagues interpreted their findings within the framework of the Social Heuristics Hypothesis: when lying harms abstract others, the negative consequences of lying are not salient leading participants towards a self-interested heuristic.[14] Conversely, when lying harms concrete others, its negative consequences become salient, generating a conflict between the self-interested heuristic and a social heuristic, encapsulated by the principle “do no harm”, that pushes in the opposite direction, and nullifies the effect of promoting intuition on dishonesty. Subsequent to this meta-analysis, two additional studies were published, corroborating the meta-analysis’s result that inducing intuition increases lying in scenarios where lying harms abstract others (Bereby-Meyer et al., 2020; Dickinson & Masclet, 2021).

**Outlook and open problems**

In summary, the available evidence suggests that promoting intuition increases dishonesty when lying harms abstract others, but not when lying harms concrete others, in which case intuition exerts no influence on dishonesty. This result evokes the “identifiable victim effect”, whereby people are far more likely to help identifiable others than abstract others (Jenny & Loewenstein, 1997); this effect also diminishes when people are primed to deliberate (Small et al., 2007).

However, it is crucial to acknowledge that, in the majority of these experiments, participants could calculate (or easily deduce) their payoff maximizing strategy before the cognitive manipulation took place. Taking the sender-receiver game as an illustrative example, the sender is informed about the payoffs associated with the available options prior to any cognitive manipulation. This carries two consequences. First, participants can in principle make a decision before the cognitive manipulation. Second, the logic of the Social Heuristics Hypothesis cannot be applied to this context. According to the SHH, people bring heuristics learned in their everyday life to the laboratory. They use these heuristics when they are in a situation where they are depleted of the cognitive resources needed to compute the payoff maximizing strategy. Therefore, if the payoff maximizing strategy is evident before the cognitive manipulation, there is no possibility of applying the SHH. Motivated by this observation, two studies implemented a variant of the sender-receiver game where participants were initially uninformed about the payoff-maximizing option, requiring them to infer it during the cognitive manipulation (Capraro, 2017; Capraro et al. 2019a). Both these studies found evidence that time pressure *increased* truth-telling. Analogously, Lohse et al. (2018) found that time pressure increased truth-telling when people were not initially aware of the possibility of lying. These results provide preliminary evidence that, when participants do not have access to the payoff maximizing option, intuitive processing promotes honesty. Interpreted through the lens of the social heuristics hypothesis, this trend may manifest because truth-telling maximizes long-term payoffs, prompting people to internalize it as a useful heuristic. Yet, further work is necessary to validate this theory employing appropriate methodologies, such as lying tasks where the cognitive process manipulation is crossed with a manipulation of the timing in which participants are informed about the payoff maximizing option.

Another important avenue for future research regards the consequences of lying. Virtually all studies have focused on the cognitive underpinnings of black lies, defined as lies that benefit the sender at a cost to the receiver. However, in reality, lies can have different consequences. For instance, spiteful lies harm both the sender and the receiver, altruistic white lies help the receiver at a cost to the sender, Pareto white lies help both the sender and the receiver. Notably, there is very little research on the cognitive basis of these types of lies. To the best of current knowledge, only one study explored the cognitive foundations of Pareto white lies (Cappelen et al., 2013), and another one investigated altruistic white lies (Cantarero & Van Tilburg, 2014). Exploring the cognitive underpinnings of lies beyond the realm of black lies is important for understanding the full range of cognitive processes involved in deception. Pareto white lies in particular are of great theoretical interest, representing the setting in which one can measure *intrinsic* lying aversion, isolated from any other confound (e.g. in the case of black lies, telling the truth coincides with being altruistic). Interestingly, the only study that manipulated the cognitive process in Pareto white lie situations found that promoting intuition increases lying aversion, but only among men (Cappelen et al., 2013).

## The cognitive basis of reciprocity

The previous sections reviewed the cognitive basis of human sociality in situations where there is only one decision-maker (dictator game, lying tasks) or multiple decision-makers who make decisions simultaneously (social dilemmas) or without knowledge of each other's payoff-maximizing strategy (sender-receiver games). However, in reality, many interactions involve an element of reciprocity, where the first mover chooses between a self-regarding and an other-regarding option, knowing that the second mover or a third-party may punish or reward their choice. This type of interaction will be the focus of this section.

### Measures of reciprocity

Depending on whether the second mover can punish or reward, one can classify four social decisions involving reciprocity: negative reciprocity, positive reciprocity, expectations of negative reciprocity, and expectations of positive reciprocity.

#### Measures of negative reciprocity

Negative reciprocity refers to situations in which the second mover has the choice to incur a cost to decrease the payoff of the first mover in response to a choice made by the first mover. There are two ways to measure negative reciprocity. One is through the second player in the ultimatum game (see below), the other consists of social dilemmas followed by a punishment phase. In this case, the punisher might or might not be the target of the initial action. Situations in which the punisher is the target of the initial action take the name of second-party punishment; situations in which the punisher is not the target of the initial action take the name of third-party punishment.

#### *Second-party punishment*

In the ultimatum game, Player 1, the *proposer*, has to propose how to divide a sum of money between themselves and Player 2, the *responder.* The responder can either accept or reject the offer. In case the responder accepts the offer, then the sum of money is split between the two players as agreed; in case the responder rejects the offer, neither the proposer nor the responder receives any money. Therefore, rejecting the proposer's offer corresponds to incurring a cost (equal to the offer received) to decrease the proposer's payoff.

Another way of measuring second-party punishment is by means of a social dilemma followed by a punishment phase, in which participants can pay a cost to decrease the payoff of other players. Various implementations are possible, depending, for instance, on the social dilemma or, in the punishment phase, on whether punishers can choose to punish all players indiscriminately, rather than targeting a specific player.

#### *Third-party punishment*

Third-party punishment is similar to second-party punishment. The key difference is that the punisher is not the target of the first action. Instead, it merely observes the behaviour of other participants, and then decides whether to punish those participants or not, based on their behaviour.

### Measures of positive reciprocity

Positive reciprocity occurs when the second mover incurs a cost to increase the payoff of the first mover in response to their choice. There are two types of positive reciprocity: one in which the rewarder is the target of the decision-maker's choice and one where they are not.

***Second-party rewarding***

In the trust game, there are two players: Player 1 (the investor or trustor) and Player 2 (the investee or trustee). Player 1 has to decide how much of a sum of money to transfer to Player 2. Any amount invested is multiplied by a factor greater than 1 (typically equal to 3) and given to Player 2. Player 2, in turn, has to decide how much, if any, to return to Player 1. This means that returning money incurs a cost (equal to the amount returned) to increase the payoff of Player 1.

Another way of measuring second-party rewarding is through a social dilemma followed by a rewarding stage. In this scenario, second movers can pay a cost to increase each other's payoff in response to the social dilemma interaction. Instead, iterated social dilemmas will not be considered as a measure of positive reciprocity, because decisions in the second and subsequent rounds may not only depend on others' past behaviour, but also on the expectations about others' future behaviour. Therefore, it is in general not possible to disentangle reciprocity from expected reciprocity in iterated social dilemmas.

***Third-party rewarding***

Third-party rewarding differs from second-party rewarding in that the rewarder is not the recipient of the action to be rewarded, but an observer.

**Measures of expectations of reciprocity**

The complementary notion of reciprocity is the expectation of reciprocity, where the first mover decides whether to incur a cost to confer a benefit to someone else with the knowledge that a second mover will respond to this decision, either by punishing or rewarding it. This concept can be divided into two parts: expectations of negative reciprocity and expectations of positive reciprocity. Expectations of negative reciprocity can be measured through the ultimatum game (player 1) and social dilemmas followed by a punishment or rewarding stage. Expectations of positive reciprocity can be measured through the trust game (player 1), social dilemmas followed by a rewarding phase, and the first round of a repeated social dilemma.

**Review of the empirical evidence on negative reciprocity**

Table 5 reports the list of papers exploring the cognitive basis of negative reciprocity.

*Table 5. List of experimental studies exploring the cognitive basis of negative reciprocity. UG stands for ultimatum game; IG for impunity game; IPG for iterated prisoner's dilemma; SG for steal game.*

| Studies showing a positive effect of promoting intuition (or a negative effect of promoting deliberation) on negative reciprocity | | | |
|---|---|---|---|
| Paper | Manipulation | Measure | Main result |
| Cappelletti et al. (2011) | Time constraints | UG responder | Pressure *increases* rejection of low offers, compared to Delay |
| Grimm & Mengel (2011) | Time constraints | UG responder | Delay *decreases* rejection of low offers, compared to Baseline |
| Wang et al. (2011) | Time constraints | TG with punishment | Pressure *increases* punishment compared to Delay |
| Wang et al. (2011) | Time constraints | UG responder | Delay *decreases* rejection of low offers, compared to Baseline |
| Neo et al. (2013) | Time constraints | UG responder | Delay *decreases* rejection of low offers, compared to Baseline |

Studies showing a positive effect of promoting intuition (or a negative effect of promoting deliberation) on negative reciprocity

| | | | |
|---|---|---|---|
| Kirk et al. (2011) | Conceptual primes | UG responder | Reducing Emotion *increases* acceptance rates, compared to Baseline |
| Bieleke et al. (2017) | Conceptual primes | UG responder | Intuition *increases* rejection rates compared to Deliberation |
| Aina et al. (2020) | Conceptual primes | UG responder | Direct response *increases* rejection rates compared to strategy method |
| Crockett et al. (2008) | Ego depletion | UG responder | Serotonin depletion *increases* rejection rates of unfair offers |
| Crockett et al. (2010a) | Ego depletion | UG responder | Serotonin depletion *increases* rejection rates of unfair offers |
| Crockett et al. (2010b) | Ego depletion | UG responder | Enhancing serotonin *decreases* rejection rates of unfair offers |
| Halali et al. (2014) | Ego depletion | UG responder | Ego depletion *increases* rejection of low offers |
| Liu et al. (2015) | Ego depletion | UG with punishment | Ego depletion *increases* punishment of unfair offers |
| Liu et al. (2015) | Ego depletion | UG responder | Ego depletion *increases* rejection of low offers |

Studies showing a null effect of promoting intuition (or deliberation) on negative reciprocity

| Paper | Manipulation | Measure | Main result |
|---|---|---|---|
| Artavia-Mora et al. (2017) | Time constraints | Field experiment | Pressure has *no effect* on punishment, compared to Delay |
| Balafoutas & Tarek (2018) | Time constraints | IG responder | Pressure has *no effect* rejection rates |

Studies showing a positive effect of promoting intuition (or a negative effect of promoting deliberation) on negative reciprocity

| | | | |
|---|---|---|---|
| Darling (2010) | Cognitive load | DG with punishment | Load has *no effect* on punishment |
| Cappelletti et al. (2011) | Cognitive load | UG responder | Load has *no effect* on rejection of low offers |
| Olschewski et al. (2018) | Cognitive load | UG responder | Load has *no effect* on rejection of low offers |
| Achtziger et al. (2018) | Ego depletion | UG responder | Ego depletion has *no effect* on rejection of low offers |
| Fraser & Nettle (2020) | Ego depletion | UG responder | Hunger has *no effect* on rejection of low offers |
| Martin (2020) | Ego depletion | UG responder | Ego depletion has *no effect* on rejection rates |
| Bago et al. (2021) | Two response | UG responder | Deliberation has *no effect* on rejection rates: most rejection choices under deliberation are already rejection choices under intuition |

Studies showing a negative effect of promoting intuition (or a positive effect of promoting deliberation) on negative reciprocity

| Paper | Manipulation | Measure | Main result |
|---|---|---|---|
| Hochman et al. (2015) | Time constraints | UG responder | Pressure *decreases* rejection rates, compared to Baseline |
| Anderson & Dickinson (2010) | Ego depletion | UG responder | Sleep restriction *increases* minimum acceptable offers |
| Achtziger et al. (2016) | Ego depletion | UG responder | Ego depletion *decreases* rejection of low offers |

Studies showing that the effect of promoting intuition or deliberation on negative reciprocity depends on some moderating variable

Studies showing a positive effect of promoting intuition (or a negative effect of promoting deliberation) on negative reciprocity

| Paper | Manipulation | Measure | Main result |
|---|---|---|---|
| Oechssler et al. (2015) | Time constraints | UG responder | Delay, compared to baseline, *decreases* rejections rates when offers are made using lottery tickets and not when using cash |
| Harris et al. (2020) | Cognitive load | UG responder | Load has *no effect* on rejection of low offers coming from low status proposers; Load *increases* acceptance of low offers coming from high status proposer |
| Yudkin et al. (2016) | Cognitive load | SG with punishment | Load makes people punish outgroup members more severely than ingroup members |

Based on existing literature, no meta-analysis appears to have been conducted on studies investigating the effect of promoting intuitive processing on negative reciprocity (see Hallsson, Siebner & Hulme (2018) for a review). To address this gap, a systematic review of these studies was carried out following the PRISMA (2020) statement (see Supplementary Information, Figure S1, for the flow diagram). This review yielded 28 articles for inclusion in the meta-analysis, for a total of 52 experimental treatments, 38 measuring negative reciprocity and 14 measuring expectations of negative reciprocity. The current subsection focuses on negative reciprocity. Standardized mean differences (Cohen's *d*) along with 95% confidence intervals were calculated at the treatment level, followed by a random-effects meta-analysis. The results indicate a significant overall effect, with an increase in negative reciprocity observed when promoting

intuitive processing ($d = 0.215$, 95% CI = [0.107, 0.324], 95% PI = [-0.33, 0.76], $z = 3.895$, $p < 0.001$). The forest plot of the meta-analysis is presented in Figure 1. Although standardized mean differences could not be obtained for three papers (Achtizer et al., 2006, 2008; Crockett et al., 2010a), each of these experiments reported a qualitatively similar effect, demonstrating that promoting intuitive processing increases negative reciprocity. Therefore, the inclusion of these studies does not affect the overall result.

To test for small-study effects, which indicate the possibility of publication bias, Egger's regression test was employed. This test regresses standardized coefficients over standard errors, and the results provide evidence of a small-study effect ($b = 1.85$, $z = 2.55$, $p = 0.011$). This indicates that better-powered studies tend to have smaller coefficients, suggesting an overreporting of small studies with significant results. The funnel plot, presented in Figure S2, provides visual evidence of this effect. To account for publication bias, the trim-and-fill method (Duval & Tweedie, 2000) was utilized. This method estimates the number of missing studies, imputes them, and then computes a new meta-analytic effect with the imputed studies. However, no missing studies were found by the algorithm, providing evidence that, if publication bias is present, it has little influence on the overall effect size.

Finally, it is worth noting that there is significant heterogeneity across studies ($Q = 121.68$, $df = 37$, $p < 0.001$). This suggests the presence of one or more moderators. One could be the cognitive manipulation. Grouping studies by cognitive manipulation indeed finds significant heterogeneity between groups ($Q = 10.41$, $df = 4$, $p = 0.034$). The overall effect is driven by studies using time constraints ($d = 0.328$, 95% CI = [0.132, 0.523], 95% PI = [-0.38, 1.02], $z = 3.286$, $p = 0.001$), ego depletion ($d = 0.300$, 95% CI = [0.123, 0.477], 95% PI = [0.09, 0.51], $z = 3.318$, $p = 0.001$) and conceptual primes ($d = 0.284$, 95% CI = [0.089, 0.478], 95% PI = [-0.03,

0.60], z = 2.861, p = 0.004), while it is not statistically significant for studies using cognitive load (p = 0.962) and the two-response paradigm (p = 0.577). However, care should be taken when interpreting this analysis, as several categories are based on a low number of studies. Previous work appears to have afforded limited attention to other potential moderators, with one notable exception. Yudkin et al. (2016) found that promoting intuition increases punishment towards outgroup members, but has the opposite effect on punishment of ingroup members. This suggests that punishing ingroup members for breaking norms requires more deliberate thought than punishing outgroup members, who are viewed more automatically as outsiders. This may be linked to the fact that people may have an inherent preference for their own group, which is supported by emotions (Joseph & Haidt, 2004).

**Transparency and openness**

All data, analysis code and research materials are available at: https://osf.io/2v9uj/?view_only=ce8390f627104efeb737db61f32d61d3. Data were analysed using Stata, version 18.0. This study's design and its analysis were not pre-registered.

*Figure 1. Random-effects meta-analysis of the studies testing the effect of cognitive manipulations on negative reciprocity.*

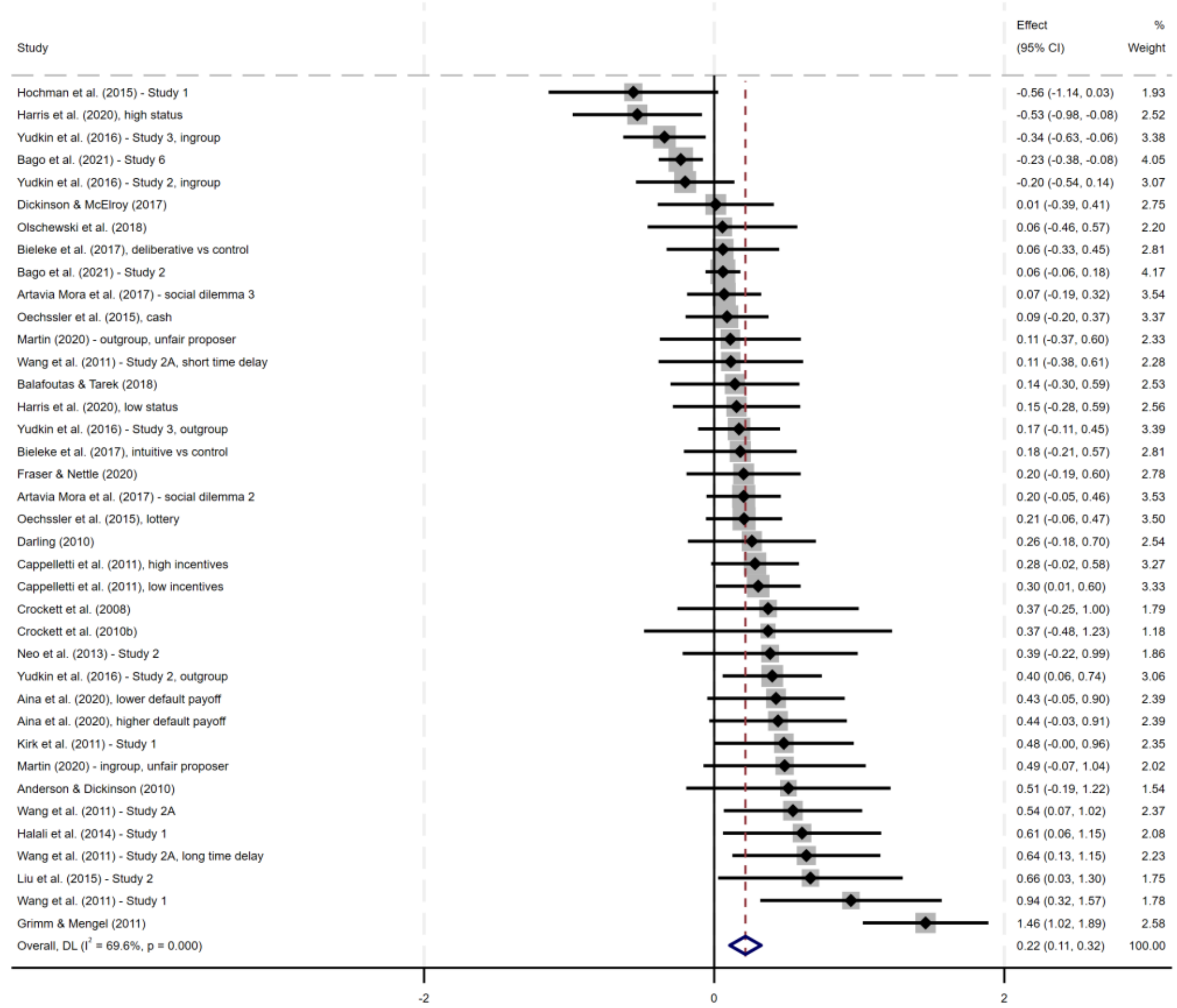

**Review of the empirical evidence on expectations of negative reciprocity**

Table 6 reports the list of papers investigating the cognitive underpinnings of expectations of negative reciprocity, including a short description for each of them.

*Table 6. List of experimental studies exploring the cognitive basis of expectations of negative reciprocity. UG stands for ultimatum game; PGG for public goods game; DG for dictator game.*

| Studies showing a positive effect of promoting intuition (or a negative effect of promoting deliberation) on expectations of negative reciprocity | | | |
|---|---|---|---|
| Paper | Manipulation | Measure | Main result |
| Cappelletti et al. (2011) | Time constraints | UG proposer | High pressure (15s) *increases* offers, compared to Low pressure (180s) |
| Halali et al. (2013) | Ego depletion | UG proposer | Depletion *increases* proportion of fair offers |
| Achtziger et al. (2016) | Ego depletion | UG proposer | Depletion *increases* offers |
| Frey et al. (2022) | Conceptual primes | UG proposer | Empathic focus on others *increase* offers |

| Studies showing a null effect of promoting intuition (or deliberation) on expectations of negative reciprocity | | | |
|---|---|---|---|
| Paper | Manipulation | Measure | Main result |
| Grimm & Mengel (2011) | Time constraints | UG proposer | Delay, compared to Baseline, has *no effect* on offers |
| Neo et al. (2013) | Time constraints | UG proposer | Delay, compared to Baseline, has *no effect* on offers |
| Frey et al. (2022) | Conceptual primes | UG proposer | Deliberation has *no effect* on offers |

| Studies showing a positive effect of promoting intuition (or a negative effect of promoting deliberation) on expectations of negative reciprocity | | | |
|---|---|---|---|
| Fraser & Nettle (2020) | Ego depletion | UG proposer | Hunger has *no effect* on offers |

| Studies showing a negative effect of promoting intuition (or a positive effect of promoting deliberation) on expectations of negative reciprocity | | | |
|---|---|---|---|
| Paper | Manipulation | Measure | Main result |
| Achtziger et al. (2018) | Ego depletion | UG proposer | Depletion *decreases* offers |

To shed light on the overall effect, a random-effects meta-analysis of standardized mean differences and 95% confidence intervals of the 14 experimental treatments identified through the previous systematic review was performed. The results reveal that promoting intuitive processing significantly increases expectations of negative reciprocity (d = 0.195, 95% CI = [0.019, 0.371], 95% PI = [-0.39, 0.78], z = 2.168, p = 0.030). See Figure 2 for the forest plot. The Egger's test finds no evidence of publication bias (b = 0.83, z = 0.67, p = 0.503) and the trim-and-fill method does not find any missing studies. See Figure S3 for the funnel plot. Furthermore, there is evidence of significant heterogeneity across studies (Q = 33.64, df = 13, p = 0.001), suggesting the existence of one or more moderators. The heterogeneity due to cognitive manipulations is this time not statistically significant (p = 0.263). However, this result should be taken with caution, due to the limited number of studies. Exploring this and other potential moderators is an interesting direction for future research.

*Figure 2. Random-effects meta-analysis of the studies testing the effect of cognitive manipulations on expectations of negative reciprocity.*

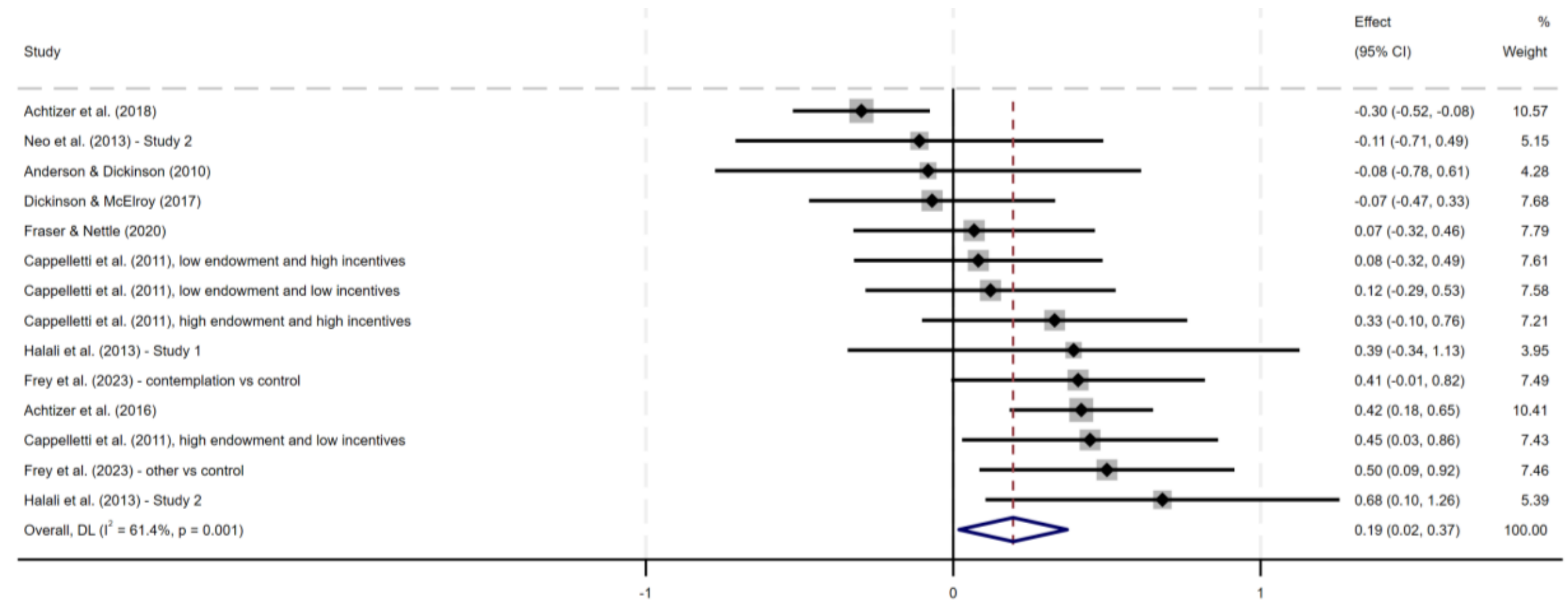

**Review of the empirical evidence on positive reciprocity**

Table 7 reports the list of papers investigating the effect of cognitive manipulation on positive reciprocity, including a short description for each of them.

*Table 7. List of experimental papers exploring the cognitive basis of positive reciprocity in one-shot games. TG stands for trust game; PGG for public goods game.*

| Studies showing a positive effect of promoting intuition (or a negative effect of promoting deliberation) on positive reciprocity | | | |
|---|---|---|---|
| Paper | Manipulation | Measure | Main result |
| Halali et al. (2014) | Ego depletion | TG investee | Ego depletion *increases* return rates |
| Rantapuska et al. (2017) | Ego depletion | TG investee | Hunger *increases* return rates |
| Katzir et al. (2021) | Ego depletion | TG investee | Ego depletion *increases* return rates |
| **Studies showing a null effect of promoting intuition (or deliberation) on positive reciprocity** | | | |
| Paper | Manipulation | Measure | Main result |
| Neo et al. (2013) | Time constraints | TG investee | Delay has *no effect* on return rates, compared to Baseline |
| Cabrales et al. (2022) | Time constraints | TG investee | Delay has *no effect* on return rates, compared to Pressure |
| Bogliacino & Gómez (in prep) | Cognitive load | TG investee | Load has *no effect* on return rates |
| **Studies showing a negative effect of promoting intuition (or a positive effect of promoting deliberation) on positive reciprocity** | | | |
| Paper | Manipulation | Measure | Main result |

| Studies showing a positive effect of promoting intuition (or a negative effect of promoting deliberation) on positive reciprocity | | | |
|---|---|---|---|
| Katzir et al. (2021) | Time constraints | TG investee | Pressure *decreases* return rates, compared to Delay |
| Urbig et al. (2016) | Conceptual primes | Sequential PGG | Second language *decreases* reciprocation of high contributions |
| Dickinson & McElroy (2017) | Ego depletion | TG investee | Sleep restriction *decreases* return rates |
| Studies showing the effect of promoting intuition or deliberation on positive reciprocity depends on some moderating variable | | | |
| Paper | Manipulation | Measure | Main result |
| Martin (2020) | Ego depletion | TG investee | Ego depletion *decreases* return rates to generous ingroup investors, and has *no effect* on return rates to generous outgroup investors |

To the best of current understanding, no meta-analysis has been conducted examining the effect of cognitive manipulations in the domain of positive reciprocity (see Hallsson et al. (2018) for a review). To bridge this gap, a systematic review of these studies was undertaken (see Supplementary Information, Figure S4, for the flow diagram). This review yielded 14 articles for inclusion in the meta-analysis, for a total of 27 experimental treatments, 14 measuring positive reciprocity and 13 measuring expectations of positive reciprocity. Standardized mean differences and 95% confidence intervals at the treatment level were calculated. A meta-analysis of the 14 treatments measuring positive reciprocity reveals that, overall, the effect of promoting intuitive processing on positive reciprocity is not statistically different from zero (d = -0.049, 95% CI = [- 0.220, 0.121], 95% PI = [-0.68, 0.58], $z = -0.567$, $p = 0.571$). See Figure 3 for the forest plot. Furthermore, the Egger's test finds no evidence of publication bias ($b = 0.07$, $z = 0.04$, $p = 0.967$). The trim-and-fill method does find two missing studies, but imputing them does not change the significance of the meta-analytic effect ($d = 0.039$, 95% CI = [-0.154, 0.231]). See Figure S5 for the funnel plot. Therefore, promoting intuitive processing does not have a statistically significant effect on positive reciprocity. However, there is significant heterogeneity across studies ($Q = 55.95$, $df = 13$, $p < 0.001$), suggesting the existence of one or more moderators. Cognitive manipulation, in this context, does not emerge as a significant moderator ($p = 0.159$). However, given the small number of studies per cognitive manipulation, this result should be taken with caution.

*Figure 3. Forest plot of the random-effects meta-analysis of the studies testing the effect of cognitive manipulations on positive reciprocity.*

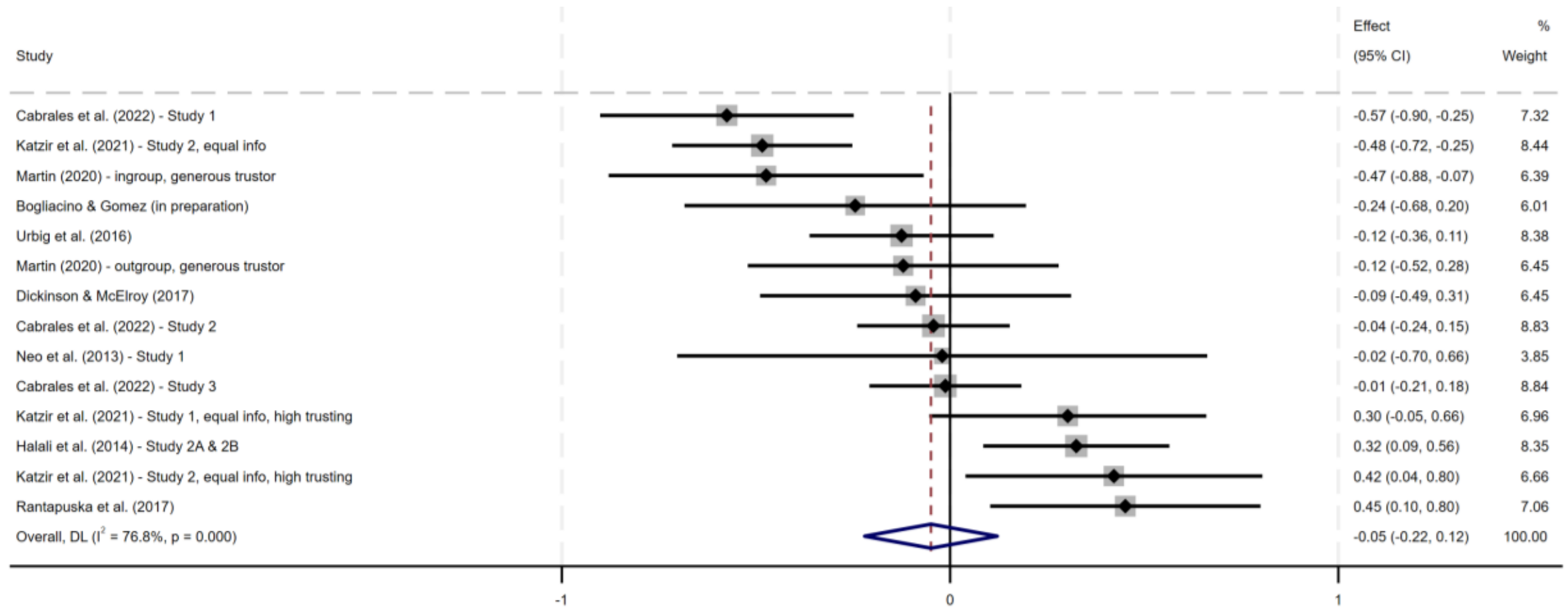

**Review of the empirical evidence on the expectations of positive reciprocity**

Table 8 includes a comprehensive list of papers that feature an experiment manipulating intuitive or deliberative processes in participants who are required to make decisions with the knowledge that their choices may be rewarded by another participant.

*Table 8. List of experimental studies exploring the cognitive basis of expectations of positive reciprocity. TG stands for trust game; PGG stands for public goods game.*

| Studies showing a positive effect of promoting intuition (or a negative effect of promoting deliberation) on expected positive reciprocity | | | |
|---|---|---|---|
| Paper | Manipulation | Measure | Main result |
| Jaeger et al. (2019) | Time constraints | TG investor | Pressure *increases* trust compared to Delay |
| **Studies showing a null effect of promoting intuition (or deliberation) on expected positive reciprocity** | | | |
| Paper | Manipulation | Measure | Main result |
| Neo et al. (2013) | Time constraints | TG investor | Delay, compared to Baseline, has *no effect* on trust |
| Rand et al. (2014) | Time constraints | TG investor | Delay, compared to Pressure, has *no effect* on trust |
| Duffy & Smith (2014) | Cognitive load | Iterated PD | Load has no effect on cooperation in the first rounds |
| Fabio et al. (2023) | Cognitive load | Iterated PD | Load has no effect on cooperation in the first rounds |
| Bogliacino & Gómez (in preparation) | Cognitive load | TG investor | Load has *no effect* on trust |
| Anderson & Dickinson (2010) | Ego depletion | TG investor | Sleep deprivation has *no effect* on trust |

Studies showing a positive effect of promoting intuition (or a negative effect of promoting deliberation) on expected positive reciprocity

| Evans et al. (2011) | Ego depletion | TG investor | Ego depletion *has no* effect on trust, but it makes people more likely to follow the default |
|---|---|---|---|
| Dickinson & McElroy (2017) | Ego depletion | TG investor | Sleep deprivation has *no effect* on trust |
| Rantapuska et al. (2017) | Ego depletion | TG investor | Hunger has *no effect* on trust |

Studies showing a negative effect of promoting intuition (or a positive effect of promoting deliberation) on expected positive reciprocity

| Paper | Manipulation | Measure | Main result |
|---|---|---|---|
| Ainsworth et al. (2014) | Ego depletion | TG investor | Ego depletion *decreases* trust |

To estimate the overall effect, a random-effects meta-analysis of the standardized mean differences and 95% confidence intervals of the 13 treatments identified in the previous systematic review was conducted. Data from the iterated prisoner's dilemma (two treatments) will be incorporated as robustness checks, given the unavailability of data for the initial round; only data from the first five rounds are accessible. It appears that, overall, intuitive processing does not exert a significant influence on expectations of positive reciprocity ($d = -0.108$, 95% CI = [-0.352, 0.135], 95% PI = [-1.00, 0.78], $z = -0.871$, $p = 0.383$). See Figure 4. Egger's test reveals a significant small-study effect ($b = -2.79$, $z = -2.73$, $p = 0.006$). The trim-and-fill method estimates four missing studies. Calibrating for these studies, the null effect of cognitive manipulation on expectations of positive reciprocity is confirmed ($d = 0.075$, 95% CI = [-0.179, 0.329]). See Figure S6 for the funnel plot before and after the inclusion of imputed studies. This conclusion remains consistent even when incorporating the two studies that utilized the iterated Prisoner's dilemma (d = -0.077, 95% CI = [-0.292, 0.137], 95% PI = [-0.89, 0.73], z = -0.705, p = 0.481). Although there is significant heterogeneity across studies ($Q = 79.44$, $df = 12$, $p < 0.001$), cognitive manipulation does not appear to moderate the effect of cognitive processing on expectations of positive reciprocity ($p = 0.249$). As in the previous cases, however, this result should be interpreted with care, given the limited number of studies per cognitive manipulation.

In summary, the available research does not furnish evidence that cognitive processing influences expectations of positive reciprocity.

*Figure 4. Random-effects meta-analysis of the studies on the cognitive basis of expectations of positive reciprocity.*

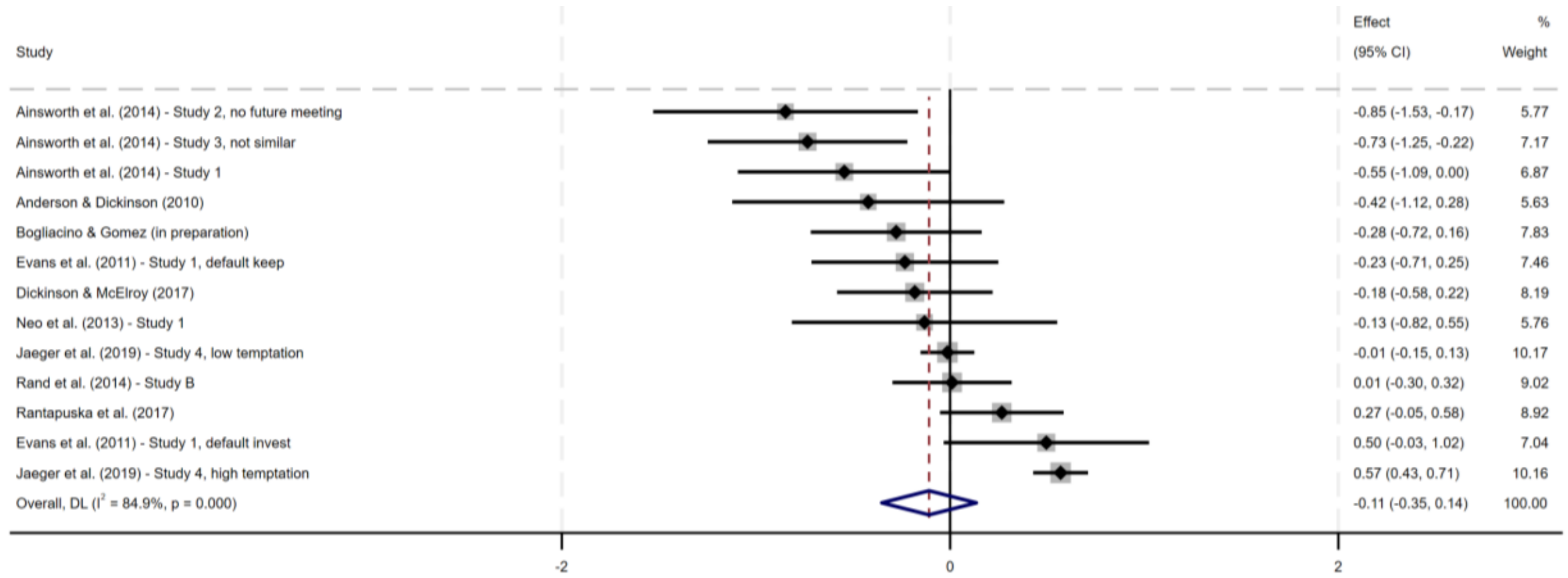

**Outlook and open problems**

To summarize, the evidence regarding the effect of cognitive process manipulations on negative reciprocity is clear because most studies report a positive effect of promoting intuition on both second- and third-party punishment. This effect is also reflected on the behaviour of first-players, as intuitive processing increases expectations of negative reciprocity as well. At the same time, however, little is known about the role of potential moderators. Both for negative reciprocity and expectations of negative reciprocity, the meta-analyses revealed significant heterogeneity across studies. Although part of this heterogeneity may be attributed to the specific experimental manipulation used to affect cognitive processing (especially in the case of negative reciprocity), there are likely more factors to uncover. For example, Yudkin et al. (2016) suggested that whether the person facing punishment is an ingroup member or an outgroup member may affect the intuitiveness of negative reciprocity. Further work could focus on these and other moderators.

It is important to note that the literature reviewed on negative reciprocity pertains to prosocial punishment, which refers to the punishment of antisocial behaviours. Yet, research evidence indicates occasional punishment of prosocial actions (Herrmann, Thoni, & Gachter, 2008). To the best of current knowledge, only one study has explored the cognitive basis of antisocial punishment. Pfattheicher, Keller, and Knezevic (2017) found that, compared to a baseline condition, conceptual priming of intuition increased antisocial punishment in a series of one-shot public goods games followed by a punishment stage. In a subsequent study within the same paper, Pfattheicher and colleagues also found that inhibiting intuition decreased antisocial punishment, but activating or inhibiting deliberation did not have the same effect. These results provide initial evidence suggesting that intuitive processing drives antisocial punishment.

Regarding positive reciprocity, the meta-analyses suggested that, overall, cognitive processing has little effect on it or its expectations. Nevertheless, in both behavioural domains, there are instances where intuitive processing either supports or undermines positive reciprocity and its expectations. Accordingly, the meta-analyses revealed significant heterogeneity across studies in both domains – despite the limited number of studies. It is therefore apparent that there are significant moderating factors at play. These moderators warrant further investigation, in order to better discern the contexts in which intuitive processing either promotes or hinders positive reciprocity and its expectations.

## The cognitive basis of deontology

Many choices involve a moral component. Philosophers and psychologists have been interested in how people make decisions in such situations for centuries. This has led to two main traditions: consequentialism and deontology. Consequentialist moral theories hold that an action's rightness or wrongness depends solely on its consequences. One popular consequentialist theory is act utilitarianism, which states that "actions are right in proportion as they tend to promote happiness, wrong as they tend to produce the reverse of happiness" (Bentham 1789/1983; Mill, 1863). Consequentialism can be logically objected in many ways. The primary non-consequentialist theory discussed in contemporary research is deontology, which holds that an action's morality depends on whether it respects certain rights or fulfils certain duties, independently of its consequences (Kant, 1797/2002). Specifically, research on the dual-process approach to moral judgments has focused on the "do no harm" proscription.

**Measures of the conflict between deontology and act utilitarianism**

**Personal moral dilemmas**

The conflict between the deontological proscription of "do no harm" and the consequentialist maxim of "save more lives" has been traditionally assessed through personal moral dilemmas. These are hypothetical scenarios where participants must decide whether it is morally acceptable to cause direct harm to one person to save a greater number of people. One of these dilemmas, considered among others in Greene et al.'s (2001) pioneering work, is the "footbridge dilemma". In this scenario, a runaway trolley is heading towards five people, and the only way to save them is by pushing a large stranger off a footbridge, causing his death but stopping the trolley. When faced with such dilemmas, most people experience a conflict between cost-benefit reasoning ("better to save more lives") and a negative emotional response to harming an innocent person ("don't push!"). Thus, these dilemmas offer a *within-subject* measure of the tension between deontology and consequentialism. Even if everyone made the same choice, most people would still experience internal tension. Therefore, personal moral dilemmas are particularly useful for dissociating cognitive processes within individuals.

**The process-dissociation approach**

In recent years, moral psychologists have adapted personal moral dilemmas, which were originally designed to dissociate cognitive processes within individuals, to measure individual differences between people. Indeed, personal moral dilemmas have limitations as a between-subject measure, as they cannot distinguish between individuals who may appear utilitarian due to a lack of emotional response, such as psychopaths, from those who genuinely seek to maximize total welfare (Patil et al. 2021).

To overcome this limitation, Conway and Gawronski (2013) proposed an approach based on Jacoby's (1991) process dissociation procedure. The starting observation of Conway and Gawronski was that classical personal moral dilemmas treat deontology and act utilitarianism as inversely related. For example, in the footbridge dilemma, choosing the deontological option also means rejecting the utilitarian option.

To address this, Conway and Gawronski compared responses to *incongruent dilemmas*, in which the deontological and utilitarian responses are misaligned, to responses to *congruent dilemmas*, in which they are aligned. An example of a congruent dilemma is torturing a person to prevent a paint bomb from exploding, while an example of an incongruent dilemma is torturing a person to prevent a pipe bomb from exploding. By comparing responses to these two types of dilemmas, the relative influence of deontological and utilitarian inclinations can be determined. This method allows one to estimate separately, within a single person, their deontological and utilitarian inclinations. Importantly, these inclinations are distinct and correlated with different factors. Deontological inclinations are associated with empathic concern, perspective-taking, and religiosity, while utilitarian inclinations are associated with a need for cognition (Conway & Gawronski, 2013). In summary, this process dissociation procedure turns personal moral dilemmas into useful measures of individual differences.

**A two-dimensional model of act utilitarianism**

Kahane et al. (2018) have proposed another approach for measuring individual differences in moral decision-making. Their goal is to sort people according to strictly philosophical criteria, by conceptually defining utilitarianism and then measuring the extent to which people behave consistently with this definition. Utilitarianism, according to Kahane and colleagues, is defined through two dimensions. One dimension is inspired by the work of utilitarian philosopher Peter Singer (1979) and requires people to value the well-being of sentient beings impartially, or "impartial beneficence". The other dimension is inspired by personal moral dilemmas and requires individuals to sacrifice lesser goods for greater goods, or "instrumental harm". Kahane and colleagues designed a scale, called the Oxford Utilitarianism Scale (OUS), to measure these two factors and showed that they are distinct in the general population. It is important to note that these factors do not exactly map onto the utilitarian and deontological inclinations as defined by Conway and Gawronski (2013), as the correlates are different.[15]

**Review of the empirical evidence on the cognitive basis of act utilitarianism**

Table 9 lists the experimental papers containing at least one study applying a cognitive manipulation to one of the measures introduced in the previous section, along with a short description of the main result of each paper.

*Table 9. List of experimental studies exploring the cognitive basis of act utilitarianism. PMS stands for personal moral dilemmas; OUS for Oxford Utilitarianism Scale.*

| Studies showing a positive effect of promoting intuition on deontological judgments or a negative effect of promoting deliberation on utilitarian judgments | | | |
|---|---|---|---|
| Paper | Manipulation | Measure | Main result |
| Suter & Hertwig (2011) | Time constraints | PMD | Pressure *increases* deontological choices, compared to Baseline |
| Cummins & Cummins (2012) | Time constraints | PMD | Pressure *increases* deontological choices, compared to Baseline |
| Conway and Gawronski (2013) | Cognitive load | Process dissociation | Load *reduces* utilitarian judgments, while leaving deontological judgments unaffected |

| Studies showing a positive effect of promoting intuition on deontological judgments or a negative effect of promoting deliberation on utilitarian judgments | | | |
|---|---|---|---|
| Białek & De Neys (2017) | Cognitive load | PMD | Load *reduces* utilitarian judgments |
| Li et al. (2018) | Cognitive load | Process dissociation | Load *reduces* utilitarian judgments, but has *no effect* on deontological judgments |
| Valdesolo & DeSteno (2006) | Conceptual primes | PMD | Affect induction *increases* deontological judgments, compared to Baseline |
| Suter & Hertwig (2011) | Conceptual primes | PMD | Fast induction *increases* deontological judgments, compared to Slow induction |
| Cummins & Cummins (2012) | Conceptual primes | PMD | Emotion induction *increases* deontological judgments compared to Baseline |
| Paxton et al. (2012) | Conceptual primes | PMD | Deliberation induction *increases* utilitarian judgments, compared to Baseline |
| Conway and Gawronski (2013) | Conceptual primes | Process dissociation | Emotion induction, compared to Baseline, *increases* deontological judgments, while leaving utilitarian judgments unaffected |
| Kvaran et al. (2013) | Conceptual primes | PMD | Emotional prime *increases* deontological judgments; analytical prime *increases* utilitarian judgments |
| Costa et al. (2014b) | Conceptual primes | PMD | Second language *increases* utilitarian judgments |

Studies showing a positive effect of promoting intuition on deontological judgments or a negative effect of promoting deliberation on utilitarian judgments

| | | | |
|---|---|---|---|
| Paxton et al. (2014) | Conceptual primes | PMD | Inducing Reflection *increases* utilitarian judgments, compared to Baseline |
| Geipel et al. (2015a) | Conceptual primes | PMD | Second language *increases* utilitarian judgments |
| Lee & Gino (2015) | Conceptual primes | PMD | Suppressing Emotion *increases* utilitarian judgments compared to Baseline |
| Cipolletti et al. (2016) | Conceptual primes | PMD | Second language *increases* utilitarian judgments |
| Corey et al. (2017) | Conceptual primes | PMD | Second language *increases* utilitarian judgments |
| Hayakawa et al. (2017) | Conceptual primes | Process dissociation | Second language *decreases* deontological judgments |
| Lyrintzis (2017) | Conceptual primes | PMD | Inducing Reflection *increases* utilitarian judgments, compared to Baseline |
| Muda et al. (2018) | Conceptual primes | Process dissociation | Second language *decreases* deontological judgments |
| Capraro et al. (2019b) | Conceptual primes | OUS | Intuition, compared to Deliberation, *increases* non-utilitarian judgments on the instrumental harm dimension; intuition has *no effect* on the impartial beneficence dimension |
| Driver (2020) | Conceptual primes | PMD | Second language *decreases* deontological judgments |

| Studies showing a positive effect of promoting intuition on deontological judgments or a negative effect of promoting deliberation on utilitarian judgments | | | |
|---|---|---|---|
| Dylman & Champoux-Larsson (2020) | Conceptual primes | PMD | Second language *decreases* deontological judgments only when it is dissimilar from first language |
| Brouwer (2021) | Conceptual primes | PMD | Second language *decreases* deontological judgments |
| Liu & Liao (2021) | Conceptual primes | PMD | Inducing Deliberation increases utilitarian judgments, compared to Baseline |
| Starcke et al. (2012) | Ego depletion | PMD | Inducing stress *increases* deontological decisions |
| Trémolière et al. (2012) | Ego depletion | PMD | Ego depletion *increases* deontological judgments |
| Youssef et al. (2012) | Ego depletion | PMD | Inducing stress *increases* deontological decisions |
| Cellini et al. (2017) | Ego depletion | PMD | Lunch nap *increases* utilitarian judgments |
| Timmons & Byrne (2019) | Ego depletion | PMD | Ego depletion *increases* deontological judgments |
| Li et al. (2021) | Ego depletion | PMD | Inducing stress *increases* deontological judgments |
| Cummins & Cummins (2012) | Two response | PMD | Second response *increases* utilitarian judgments |
| Studies showing a null effect of promoting intuition or deliberation on deontological judgments | | | |

Studies showing a positive effect of promoting intuition on deontological judgments or a negative effect of promoting deliberation on utilitarian judgments

| Paper | Manipulation | Measure | Main result |
| --- | --- | --- | --- |
| Ham & van den Bos (2010) | Time constraints | PMD | Delay, compared to Pressure, has *no effect* on deontological judgments |
| Houston (2010) | Time constraints | PMD | Pressure has *no effect* on deontological decisions, compared to Baseline |
| Tinghög et al. (2016) | Time constraints | PMD | Pressure, compared to Delay, has *no effect* on deontological choices |
| Kroneisen & Steghaus (2021) | Time constraints | PMD | Pressure, compared to Baseline, has *no effect* on deontological choices |
| Greene et al. (2008) | Cognitive load | PMD | Load has *no effect* on deontological choices |
| Ramsden (2015) | Cognitive load | PMD | Load has *no effect* on utilitarian judgments |
| Tinghög et al. (2016) | Cognitive load | PMD | Load has *no effect* on deontological choices |
| Rosas et al. (2019) | Cognitive load | PMD | Load has *no effect* on deontological judgments |
| Cummins & Cummins (2012) | Conceptual primes | PMD | Cognition induction has *no effect* compared to Baseline |
| Gürcay & Baron (2017) | Conceptual primes | PMD | Inducing Deliberation, compared to Intuition, has *no effect* on deontological decisions |
| Spears et al. (2018) | Conceptual primes | PMD | Attention prime has *no effect* on utilitarian judgments, compared to Baseline |
| Brouwer (2019) | Conceptual primes | PMD | Second language has *no effect* on |

| Studies showing a positive effect of promoting intuition on deontological judgments or a negative effect of promoting deliberation on utilitarian judgments | | | |
|---|---|---|---|
| | | | deontological judgments in high proficient speakers |
| Scherlippens (2019) | Conceptual primes | PMD | Second language has *no effect* on deontological judgments |
| Del Maschio et al. (2022) | Conceptual primes | PMD | Second language has *no effect* on deontological judgments |
| Weippert et al. (2018) | Ego depletion | PMD | Physical fatigue has *no effect* on deontological judgments |
| Francis et al. (2019) | Ego depletion | PMD | Alcohol has *no effect* on deontological judgments |
| Paruzel-Czachura et al. (2023) | Ego depletion | PMD/OUS | Alcohol consumption has *no effect* on deontological judgments |
| Bago & De Neys (2019) | Two response | PMD | Most utilitarian deliberative judgments were already utilitarian under intuition |

Studies showing a negative effect of promoting intuition on deontological judgments or a positive effect of promoting deliberation on utilitarian judgments

| Paper | Manipulation | Measure | Main result |
|---|---|---|---|
| Rosas & Aguilar-Pardo (2020) | Time constraints | PMD | Pressure, compared to Delay, *decreases* deontological choices |
| Liu & Liao (2021) | Cognitive load | PMD | Load decreases deontological judgments |
| Schwitzgebel & Cushman (2015) | Conceptual primes | PMD | Inducing Reflection along with a 15s time delay *increases* deontological |

Studies showing a positive effect of promoting intuition on deontological judgments or a negative effect of promoting deliberation on utilitarian judgments

| | | | |
|---|---|---|---|
| | | | judgments, compared to Baseline |
| Hayakawa et al. (2017) | Conceptual primes | Process dissociation | Second language *decreases* utilitarian judgments |
| Shin & Kim (2017) | Conceptual primes | PMD | Second language *decreases* utilitarian judgments |
| Muda et al. (2018) | Conceptual primes | Process dissociation | Second language *decreases* utilitarian judgments |

Studies showing that the effect of promoting intuition or deliberation on deontological or utilitarian judgments depends on some moderating variable

| Paper | Manipulation | Measure | Main result |
|---|---|---|---|
| Körner & Volk (2014) | Time constraints | PMD | Pressure, compared to Delay, *increases* deontological decisions, but only when induced to process information concretely |
| Trémolière & Bonnefon (2014) | Time constraints | PMD varying save/kill | Pressure, compared to Baseline, *increases* deontological choices for save/kill = 5 but not save/kill = 500 |
| Trémolière & Bonnefon (2014) | Time constraints | PMD varying save/kill | Load has *no effect* on deontological choices for save/kill = 5 but not save/kill = 500 |
| Killgore et al. (2007) | Ego depletion | PMD | Sleep restriction *decreases* deontological judgments among people with average emotional intelligence, but has *no effect* among people with |

| Studies showing a positive effect of promoting intuition on deontological judgments or a negative effect of promoting deliberation on utilitarian judgments | |
|---|---|
| | high emotional intelligence |

To the best of current knowledge, no meta-analysis has examined the overarching effect of cognitive manipulations on moral judgments in personal moral dilemmas. The meta-analyses by Klenk (2022) and Rehren (2022) did not differentiate between personal moral dilemmas, where harm is direct, from “impersonal” moral dilemmas, where harm is a side-effect (further discussion on impersonal moral dilemmas will follow). To address this gap, a systematic review of relevant studies was undertaken, leading to the identification of 52 articles suitable for inclusion in the meta-analysis, for a total of 111 experimental treatments (see Supplementary Information, Figure S7, for the flow diagram). Standardized mean differences and their 95% confidence intervals were calculated at the treatment level, followed by a random-effects meta-analysis. For studies using the process dissociation approach, the “traditional analysis” typically reported in such literature (e.g., Conway & Gawronski, 2013) was considered, focusing only on the incongruent dilemmas. Regarding the two studies using the OUS, only the instrumental harm dimension was included. The findings indicate that promoting intuitive processing reduces utilitarian choices overall ($d = -0.234$, 95% CI = [-0.300, -0.168], 95% PI = [-0.83, 0.37], $z = -6.949$, $p < 0.001$). The Egger’s test finds no statistical evidence of publication bias ($p = 0.106$). The trim-and-fill method does estimate 22 out of 132 missing studies, but even imputing these studies the effect remains abundantly significant ($d = -0.122$, 95% CI = [-0.195, -0.049]). See Figure 5 for the forest plot and Figure S8 for the funnel plot with and without imputed studies. There is also significant heterogeneity across studies ($Q = 577.47$, $df = 109$, $p < 0.001$), which appears to be partly due to the cognitive manipulations. Grouping by cognitive manipulation reveals significant heterogeneity across cognitive manipulations ($p = 0.006$). The meta-analytic effect is driven by studies using conceptual primes ($d = -0.307$, 95% CI = [-0.395, -0.219], 95% PI = [-0.92, 0.31], $z = -6.824$, $p < 0.001$) and ego depletion ($d = -0.271$, 95% CI = [-0.462, -0.080], 95% PI = [-1.01, 0.47], $z = -2.782$, $p = 0.005$). However, these results should be interpreted prudently, given the limited number of studies per experimental manipulation.

*Figure 5. Random-effects meta-analysis of the studies on the cognitive basis of utilitarianism using sacrificial dilemmas.*

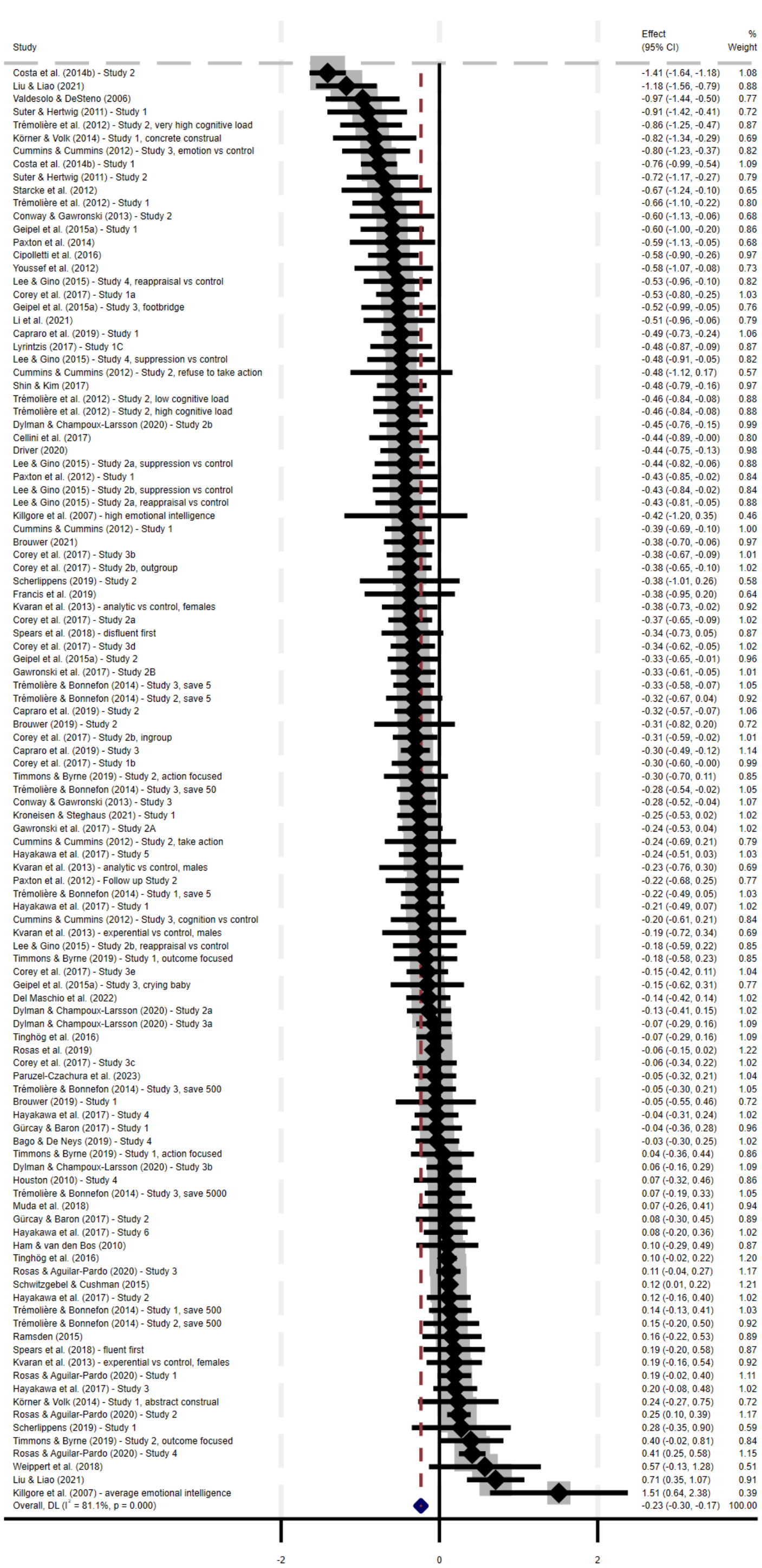

**Outlook and open problems**

To summarize, the overall evidence suggests that intuitive processing promotes deontological judgments in personal moral dilemmas, although there is significant heterogeneity across studies. Several moderators are worth exploring in future work. For example, Körner & Volk (2014) found that time pressure increases deontological decisions only when participants are induced to process information concretely, rather than abstractly. This may be because the emotional response that drives intuitive deontology is favoured by a concrete representation of the dilemma at hand. Trémolière & Bonnefon (2014) reported that increasing the saved-to-killed ratio not only makes people more utilitarian but also makes them more intuitively utilitarian. This may be because a large saved-to-killed ratio makes the utilitarian option more readily accessible. Understanding the effect of these moderators may be important from a practical perspective because real-life moral dilemmas may appear with very different saved-to-killed ratios or at different levels of concreteness.

## Theoretical framework

The objective of this section is to organize the results into a coherent theoretical framework. One possible approach is to generalize the Social Heuristics Hypothesis (SHH), which was originally introduced by Rand and colleagues (2014) in the domain of cooperative behaviour, to other social decisions. Using the authors' words, the SHH contends that "cooperation is typically advantageous in everyday life, leading to the formation of generalized cooperative intuitions. Deliberation, by contrast, adjusts behaviour towards the optimum for a given situation. Thus, in one-shot anonymous interactions where selfishness is optimal, intuitive responses tend to be more cooperative than deliberative responses".

However, a straightforward extension of the SHH does not explain all the experimental regularities discussed in the preceding sections. This is because any such extension would predict that deliberative processing increases self-regarding behaviour in one-shot and anonymous decisions. However, the evidence indicates that deliberative processing increases honesty (and therefore decreases self-interest) when dishonesty harms abstract others.

Nonetheless, the SHH may not be entirely incorrect. Although this counterexample violates the SHH's predictions regarding the role of deliberation, it does not violate those regarding the role of intuition. Therefore, the next question is whether a straightforward generalization of the other "half" of the SHH makes predictions that align with the experimental evidence. This paper proposes the following generalization of the first "half" of the SHH, termed GSHH1:

*Intuitive processing promotes heuristics that have been internalized in everyday social interactions that resemble the decision under consideration.*

The aim of the following subsection is to derive some social heuristics that many people are likely to internalize. It is worth noting that these heuristics were deduced post hoc and were not preregistered as a priori hypotheses. They were developed to capture the variability observed in the results, appealing to existing theories (the SHH) and the fundamental instinct for self-preservation.

**The instinct for self-preservation and its consequences**

The premise lies in the observation that self-preservation is one of the most fundamental animal instincts. All species, and humans in particular, have developed and internalized a set of automatic behaviours that preserve them from danger. This section argues that such behaviours generate, among humans, the evolution of two heuristics.

First, humans not only internalize heuristics that protect them from being harmed, but they also internalize heuristics that avoid harming others *directly*. This happens because, in quotidian interactions, causing direct harm to others usually comes with a substantial retaliatory cost. The argument is straightforward: if one were to physically harm another individual, it is plausible that the aggrieved party, propelled by their own instinct for self-preservation, would retaliate, endangering the instigator. Consequently, the instinct for self-preservation not only preserves one from receiving harm, but also from initiating harm.

Second, the instinct for self-preservation tends to make people averse to disadvantageous inequalities, and it does so for at least two reasons. At a very basic level, getting less resources than others may threaten the possibility of surviving or reproducing. This might explain why many human and even non-human interactions seem to be guided by expectations about how resources should be divided, to the point that inequity aversion is probably universal among humans (although with some variability; Henrich et al., 2001) and has been documented also among non-human primates (Brosnan & De Wall, 2003). At a slightly less basic level, there is another, probably uniquely-human, reason why inequity aversion may descend from the instinct for self-preservation. When we receive less than others without a justified cause, we feel that we have been left behind, that we hold little value, and this threatens our self-esteem, one of the most fundamental psychological mechanisms for self-preservation, so fundamental that some psychologists believe that it descends directly from the uniquely-human awareness of their own mortality (Becker, 1973; Greenberg, Pyszczynski & Solomon, 1986).

In summary, there are at least three heuristics that we may expect many people have internalized: (i) don't get harmed; (ii) don't harm others, especially when they can harm you back; (iii) don't get fewer resources than others. The next sections combine these heuristics with the GSHH1. The aim is to derive some hypotheses about the effect of intuitive processing on social decisions, and check whether they are reflected in the experimental evidence.

**Hypotheses 1a-1b: The Ultimatum Game**

Among UG responders, we may expect that intuitive processes promote aversion to disadvantageous inequalities and therefore increases rejections of low offers. Among proposers, we may expect that intuitive processes favour concerns about harming people that have the potential to harm them back, and therefore increases offers. In sum, the GSHH1 predicts that intuitive processes are positively related to (H1a) rejection rates and to (H1b) offers. The experimental evidence reviewed before provides support for both these predictions (H1a: d = 0.215; H1b: d = 0.195).

**Hypothesis 2: Comparing the Dictator Game to the Ultimatum Game**

The main difference between the DG and the UG, in the context of this analysis, is that the DG recipient has no means to retaliate against the dictator, which means that the dictator can make a small donation without fear of harm. As a result, we might expect the positive effect of intuitive processing on UG offers to be less pronounced in the DG. In other words, the GSHH1 predicts that (H2) the positive effect of intuitive processes on UG offers is stronger than its effect on DG giving. The experimental results align with this prediction, with meta-analytic findings indicating that intuitive processing favours UG offers (d = 0.195, 95% CI = [0.019, 0.371]), whereas there is no statistical evidence for an overall effect of intuitive processing on DG donations, as demonstrated in the meta-analysis by Fromell et al. (2020) (d = -0.015, 95% CI = [-0.070, 0.041]). It is important to note, however, that the confidence intervals are slightly overlapping. This may be partly driven by the fact that there are far more studies on the DG. This discrepancy manifests in the broader confidence interval associated with the UG. In light of this, we interpret this finding as preliminary evidence in line with H2. Future studies on the cognitive basis of UG offers may help sharpen the precision of the estimated effect size.

Another important remark is that the GSHH1 does not make any prediction about the effect of intuitive processing on the DG. It might be zero, or it may not be zero, particularly for participants who are punished in their everyday lives for acting altruistically or egoistically. The GSHH1 does not contradict the existence of specific participant groups for which intuitive processing is either favourable or unfavourable towards DG giving, as shown in studies exploring gender (Fromell et al., 2020; Rand et al., 2016) or personality factors (Nockur & Pfattheicher, 2021; Shi et al., 2020). It only predicts that, in general, the effect is likely to be weak and certainly weaker than in the UG.

**Hypothesis 3: Truth-telling**

Since intuitive processing makes people less likely to directly harm others, the GSHH1, when applied to the context of truth-telling, predicts that (H3) the relationship between intuitive processing and honesty when dishonesty harms concrete others is greater than the relationship between intuitive processing and honesty when dishonesty harms abstract others. This prediction aligns with the meta-analytic findings reported by Köbis and colleagues (2019). Their research indicated that the distinction as to whether dishonesty harms concrete vs abstract others moderates the effect of intuitive processing on the probability ($z = -1.94$) and the proportion of lying ($z = -1.99$) in the predicted direction. The link between intuitive processing and dishonesty is positive when dishonesty harms abstract others (probability of lying: log OR = 0.38, magnitude of lying: $g = 0.261$) and not significantly different from zero when dishonesty harms concrete others (probability of lying: log OR = -0.04, magnitude of lying: $g = -0.076$).

Similar to the DG case mentioned above, it is important to emphasize that the GSHH1 does not make any specific prediction about the overall effect of promoting intuitive processing on honesty. Promoting intuitive processing may increase or decrease honesty for different individuals or in different situations (Gai & Puntoni, 2021; Speer et al., 2020; Speer et al., 2022). The only specific prediction of the GSHH1 regards the comparison of the effect of intuitive processing when dishonesty harms concrete vs abstract others.

**Hypothesis 4: Direct harm in personal moral dilemmas**

Directly harming someone to save a greater number of people in hypothetical personal moral dilemmas such as the footbridge dilemma comes with no actual cost for the decision maker: the payoff of the decision maker is exactly the same whether they push the large man off the bridge or not. However, in reality directly harming another person usually does come with a cost – the person being harmed might resist or fight back. Therefore, since the GSHH1 is based on the idea that intuitive processing does not promote choices that are optimal in the given situation, but choices that are optimal in everyday interactions that resemble the given situation, it makes the prediction that (H4) intuitive processing increases rejection of direct harm in personal moral dilemmas. This hypothesis aligns with the meta-analytic results presented before ($d = 0.234$).

**Hypothesis 5: Harm as a side-effect in trolley problems**

Consider the classic trolley problem: a person must decide whether to divert a trolley onto another track to save five people, knowing that one person will be killed as a side-effect. In this dilemma, the harm is a foreseeable side-effect rather than direct. Therefore, the GSHH1 predicts that the association between intuitive processing and the deontological choice is stronger in sacrificial dilemmas involving direct harm than in those where harm is a side-effect (H5).

This prediction can be tested through a meta-analysis of available studies, using standardized mean differences and 95% confidence intervals. A systematic review led to the inclusion of 46 treatments in the meta-analysis. The overall effect was statistically significant ($d = -0.076$, 95% CI = [-0.147, -0.005], 95% PI = [-0.44, 0.29], $z = -2.088$, $p = 0.037$). However, importantly, in line with H5 the effect was smaller than the effect for sacrificial dilemmas: For moral dilemmas involving direct harm, the effect was -0.234 (95% CI = [-0.300, -0.168]); note that the two confidence intervals are non-overlapping. Further evidence supporting H5 comes from a meta-regression of the effect sizes, including a dummy variable representing the type of moral dilemma (1 = sacrificial, 0 = side-effect), which demonstrates that the effect sizes associated with sacrificial moral dilemmas tend to be more negative than those associated with moral dilemmas where harm is a side-effect ($d = -0.141$, 95% CI = [-0.255, -0.028], $t = -2.47$, $p = 0.014$). A similar result is obtained taking into consideration potential publication bias. Egger's test shows a significant small-study effect for side-effect dilemmas ($b = -2.91$, $z = -4.95$, $p < 0.001$). The trim-and-fill method estimates that there are thirteen missing studies, and imputing these studies results in a meta-analytic effect that is not statistically different from zero and only slightly overlaps with the interval for personal dilemmas ($d = 0.017$, 95% CI = [-0.054, 0.088] vs $d = -0.122$, 95% CI = [-0.195, -0.049]). See Figure 6 for the forest plot and Figure S9 for the funnel plot with and without imputed studies.

Figure 6. Forest plot of the studies exploring the cognitive basis of utilitarianism in side-effect dilemmas.

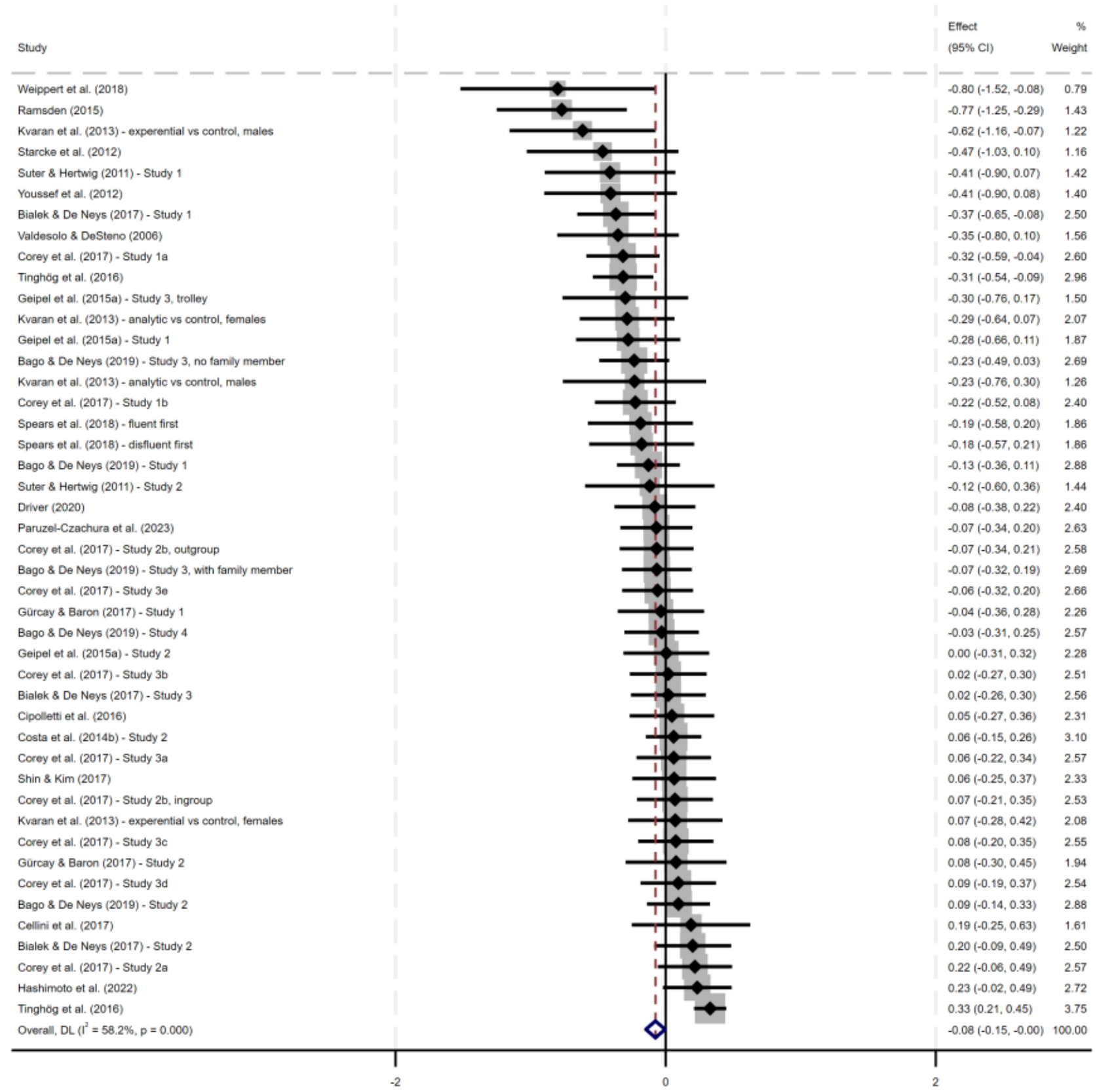

### Cooperation and positive reciprocity

The GSHH1 does not make any general predictions about the association between intuitive processing and cooperation or positive reciprocity. This is because these behaviours do not involve harming others in a context where they can harm you back. However, this does not mean that the GSHH1 predicts that intuitive processing never promotes cooperation or positive reciprocity. It may be the case that certain groups of people have internalized these behaviours as a default strategy for specific reasons, but this is not a general expectation. The results of the earlier meta-analyses support this conclusion. The debate about the effect of intuitive processing on cooperation has suggested that it promotes cooperation weakly and only for particular groups of people, such as inexperienced subjects living in cooperative societies (Rand et al., 2014), which is consistent with the GSHH1. For instance, people living in cooperative societies may have internalized cooperation to a greater extent than people in non-cooperative societies because non-cooperating in a cooperative society comes with a greater reputational cost. Regarding positive reciprocity, the available studies do not provide evidence of an effect of cognitive processing, and, to the best of current knowledge, there are no studies looking at potential moderators of the effect of cognitive processing on positive reciprocity.

## General open problems

The previous sections mentioned several open problems whose solution would further advance our understanding of the cognitive basis of specific social decisions. This concluding section discusses broader unresolved problems, that transcend specific social decisions and pertain to the overarching field of research.

### The effect of deliberative processing

A significant unresolved issue pertains to the impact of deliberative processing on social decision-making. According to the original Social Heuristics Hypothesis, thoughtful consideration of options should lead people to select the optimal choice in a given situation. However, the available evidence indicates that the picture may be more complex. For instance, in lying tasks where dishonesty harms abstract others, deliberation appears to increase honest behaviour, which decreases self-interest. On the other hand, in situations like the ultimatum game, deliberation may increase self-interest. Thus, it is challenging to categorize the effects of deliberation in simple terms that depend solely on the outcomes of available actions. It is plausible that deliberation makes people more emotionally detached from the situation, encouraging them to reason about it in abstract terms. This may lead to different consequences depending on the social decision. In the ultimatum game, deliberation may help people overcome the emotional reaction that prompts them to reject low offers. On the other hand, in lying tasks where dishonesty harms abstract others, deliberation may make people more likely to acknowledge the harm caused by dishonesty, leading to an increase in honest behaviour. In personal moral dilemmas, deliberation may lead one to represent the dilemma in abstract terms and therefore to endorse a utilitarian perspective, as suggested by Körner and Volk (2014). However, this remains speculative, and future research should shed more light on the effect of deliberative processing on social decisions.

**Understanding how people associate social contexts**

The SHH and the GSHH1 share two fundamental pillars: (i) People intuitively associate social contexts with shared characteristics; (ii) People transfer to experimental settings the strategies that are typically effective in those associated contexts they experience in quotidian life. These observations pave the way for several important theoretical questions: What criteria do people use to associate different social contexts? What features make certain social contexts appear analogous, and what differentiates them? Do participants genuinely apply everyday heuristics within experimental environments?

Existing literature in this domain is scarce, largely due to the challenge of measuring how people form associations between diverse contexts. Rand and colleagues argued that, given the iterative nature of many real-life situations, people might associate one-shot games with their iterated variant (Bear & Rand, 2016; Rand et al., 2014). This approach is sensible, particularly for symmetric games like the prisoner's dilemma. However, asymmetric games present complications. In iterated versions of such games, roles might interchange, necessitating a consideration of the frequency of each role undertaken by players. Moreover, in daily life people typically have the possibility to choose their interaction partner, creating dynamic networks that might affect the evolution of different behaviours (Rand, Arbesman & Christakis, 2011). Therefore, it is likely that a general theory of association of contexts should take into account at least role frequency and the dynamics of social networks. Future theoretical work might encapsulate these ideas in mathematical models.

Regarding the query of whether participants genuinely bring everyday heuristics into experimental setups, the existing body of work is equally limited, largely due to the complexities in discerning the associations participants make. An early study by Engel and Rand (2014) offers some insights. They discovered that participants play a neutrally-framed prisoner's dilemma similarly to its cooperatively-framed counterpart. By contrast, cooperative behaviour diminished in a competitively framed variant. This suggests that participants might perceive the neutrally framed dilemma as inherently cooperative. The authors interpreted this as a piece of evidence that experimental subjects bring cooperative strategies to the laboratory. While this is an important starting point, further investigations in this area are needed.

**The need for uniformizing measures and manipulations**

This field of research operates on the underlying assumption that all manipulations of cognitive process have similar effects on each social decision. Several meta-analyses conducted so far implicitly include this assumption by analyzing all cognitive manipulations together (Fromell et al., 2020; Köbis et al., 2019; Rand, 2016; Rand et al., 2016). However, an important question arises: Is this assumption valid? Is it true that all cognitive manipulations have similar effects? While there is evidence that time pressure, ego depletion, conceptual primes of intuition, and cognitive load have comparable effects on cooperation (Kvarven et al., 2020; Rand, 2019), these meta-analyses indicate that explicit conceptual primes of emotion have a more significant impact. Moreover, these meta-analyses are limited to cooperative behaviour, leaving the more general question of whether all cognitive manipulations have similar effects on all social decisions largely unanswered.

The meta-analyses discussed in this article frequently show significant heterogeneity across cognitive manipulations. However, the heterogeneity primarily pertains to the size of the effect, rather than its direction, indicating that cognitive manipulations may differ in their "quantitative strength" rather than their "qualitative effect". Isler and Yilmaz (2022) support this view, as they found that several cognitive manipulations, including time pressure, time delay, cognitive load, and conceptual primes of emotion, have the anticipated effects on intuitive or deliberative decision-making, albeit to varying degrees. However, it is crucial to approach these results with caution, given the limited statistical power resulting from the relatively small number of treatments for each cognitive manipulation.

Furthermore, as of now, dual-process theories lack a universally accepted threshold to differentiate between intuition and deliberation (e.g., longer than x seconds or less than x amount of cognitive load; De Neys, 2021, 2022). Consequently, assertions are confined to pragmatic, operational boundaries and any inconclusive or null result could potentially be attributed to the imposed constraints not being stringent enough, thereby permitting some degree of deliberation.

To add complexity to the topic, Newton, Feeney and Pennycook (2023) presented experimental evidence that challenges the assumption that people vary along a single dimension of thinking style. Their results suggest that there are four distinct thinking styles: actively open-minded thinking, close-minded thinking, preference for intuitive thinking, and preference for effortful thinking. This classification opens up new avenues for research, such as investigating the specific effects of cognitive manipulations on each of these styles and examining how they relate to social decision-making.

Additionally, there is a broader methodological issue *within* cognitive manipulations. For instance, when it comes to time pressure studies, some scholars use a time pressure of five seconds, while others use ten, fifteen, or even twenty seconds. Even if one accepts the idea that time pressure promotes intuitive processing, comparing studies that use different time constraints might be problematic. More profound inconsistencies arise in other cognitive manipulation techniques. Subjects might be cognitively loaded with different parallel tasks, depleted using different tasks, and so on. For instance, it is challenging to compare the effect of the *e*-hunting task with the effect of the Stroop task. It is as if different researchers were measuring the length of the same stick, but one is using meters, another is using miles, another is using yards, and another is using… kilograms! Since uniformizing the units of measurement is the first step to building a truly scientific theory, this is a fundamental direction for future research.

**Extending the dual-process approach to other social decisions**

While the decisions discussed in this article cover a broad range of social decisions, they are not exhaustive. In fact, there are other types of social decisions that warrant further investigation.

One such example is the equity-efficiency trade-off, which refers to the conflict between creating equity in the distribution of resources and maintaining efficiency in the process. This trade-off is often considered the "fundamental problem in distributive justice" (Hsu et al. 2008) or the "big trade-off" (Okun, 2015). Those responsible for resource distribution frequently face this conflict, as exemplified by Okun's "leaky bucket" argument, where taxation increases equity but also imposes costs on the institution implementing it. Understanding the cognitive basis of this trade-off is therefore of primary importance, yet little is known about it. Capraro et al. (2017) found that time pressure promotes equity while time delay favors efficiency. Similarly, Kessler et al. (2019) found that time delay promotes altruism when the cost of the altruistic action is smaller than its benefit. These works suggest that intuitive processing promotes equity, while deliberative processing favours efficiency. However, Krawczyk and Sylwestrzak (2018) argued that time pressure does not promote equity in general, but only envy, which is the desire to decrease the other's payoff when it is greater than one's own. Other studies using dictator games with varying cost-to-benefit ratios failed to detect significant effects of cognitive process manipulations (Balafoutas, Kerschbamer & Oexl, 2018; Friehe and Schildberg-Hörisch, 2017; Merkel & Lohse, 2019), or found that time pressure simply increased prosocial behaviour in prosocial individuals and self-regarding behaviour in individualists (Chen & Krajbich, 2018).

Another behavioural domain worth exploring is ingroup favouritism. Ingroup favouritism is considered one of the most pervasive biases among humans. Even when groups are formed randomly, people tend to favour those in their own group (Tajfel, 1970). Ingroup favouritism has been observed in numerous economic games, including public goods, dictator, ultimatum, and prisoner's dilemma games (see Everett et al., 2015, for a review). Despite the significance of this behaviour, however, only a handful of studies have investigated its cognitive underpinnings. Rand, Newman, and Wurzbacher (2015) found that time pressure increases cooperation with both ingroup and outgroup members. Similarly, Everett et al. (2017) found that time delay decreases cooperation with both ingroup and outgroup members. De Dreu, Dussel, and Ten Velden (2015) examined situations in which cooperating with ingroup members automatically entails competing with outgroup members. They found that cognitive load increases ingroup favouritism.

One more example is aggression. Aggression can manifest in different forms, depending on the underlying motivations and triggers. Reactive aggression, for instance, is often impulsive and defensive, while goal-oriented aggression tends to be more deliberative (Nelson & Trainor, 2007). Everett, Ingbretsen, Cushman, and Cikara (2018) developed a novel economic game, called the "pre-emptive strike game", to study whether defensive aggression is intuitive or deliberative. They found that time pressure increases aggression, especially against outgroup members.

In reality, there are situations in which cooperation is not costly, but risky, meaning that mutual cooperation is an equilibrium, but there are also "safer", less risky equilibria. The risk comes from the fact that the cooperator receives the benefit of cooperation only if the other partner cooperates (Skyrms, 2004). What are the cognitive basis of risky cooperation? To date, only one study addressed this question. Belloc, Bilancini, Boncinelli and D'Alessandro (2019) found that a 10-second time pressure increases cooperation in a one-shot stag-hunt game, suggesting that time constraints may promote cooperative behaviour in situations where cooperation is risky.

In recent years, behavioural scientists have begun studying the factors that lead people to share fake news on social media, taking a dual-process approach. In a study conducted by Bago, Pennycook and Rand (2020), participants were asked to rate the accuracy of news headlines using a two-response paradigm. The authors found that, when given the opportunity to reconsider their initial choice, participants were more likely to correct intuitive mistakes and thus better able to distinguish real news from fake news. Other studies have also shown that emphasizing the importance of accuracy can reduce the sharing of fake news (Pennycook & Rand, 2022), likely because it directs people's limited attention towards accuracy (Pennycook et al., 2021). These findings suggest that people may share fake news on social media partly because they tend to rely on their intuition and need to consciously deliberate in order to overcome this tendency.

**Small effect sizes and the need for well-powered replications**

It is worth emphasizing that, even when achieving statistical significance, the meta-analytic effect sizes were modest in magnitude. See Table 10 for a summary of the effect sizes. This observation highlights the necessity for future studies to be appropriately powered. For example, the preceding reviews catalogued numerous instances of studies reporting a null effect. In light of the meta-analytic results, it is plausible to hypothesize that a considerable portion of these studies may have suffered from inadequate statistical power.

## Limits on generality

Virtually all the studies discussed in this review have been conducted within WEIRD (Western Educated Industrialized Rich and Democratic; Henrich, Heine & Norenzayan, 2010) societies. Given that the internalization of strategies optimal for daily life varies based on social context, there is a pressing need for future research to explore cross-cultural differences. Moreover, apart from two exceptions that studied the cognitive underpinnings of cooperative behaviour among adolescents (Nava et al. 2023) and children (Corbit, Dockrill, Hartlin, & Moore, 2023), this research has mainly targeted adults. An essential future research avenue would be to trace the development of heuristics across different life stages. The previous subsections highlighted that variations in methodologies might yield different results. Furthermore, past research has predominantly centered on social decisions such as cooperation, altruism, truth-telling, reciprocity, and act utilitarianism. Yet, important social behaviours like aggression, risky cooperation, equity-efficiency trade-off remain underexplored. Therefore, it is important to exercise caution and not to generalize these results beyond the social decisions, cognitive manipulations, and populations described. Future research should expand the scope to ensure broader applicability of the results.

## Conclusion

This article reviewed the existing literature on the cognitive basis of cooperation, altruism, honesty, positive and negative reciprocity, expectations of reciprocity, and the "do no harm" deontological proscription. In cases where meta-analyses were lacking, a meta-analysis was conducted in order to be able to draw quantitative conclusions. A summary of the results is provided in Table 10.

*Table 10. Summary of the meta-analytic results presented in the previous sections. For each social decision, the second column reports the meta-analytic effect, where the sign*

*represents the effect of promoting intuition or inhibiting deliberation on the corresponding social decision.*

| **Social decision** | **Meta-analytic effect of promoting intuition or inhibiting deliberation** |
|---|---|
| Cooperation | Rand (2016): 6.1 percentage points, 95% CI = [3.4, 8.9]<br>Kvarven et al. (2020): 2.19 percentage points, CI not reported, but $p = 0.005$ |
| Altruism | Fromell et al. (2020): Cohen's d = - 0.015, 95% CI = [-0.070, 0.041]) |
| Truth-telling when dishonesty harms abstract others | Köbis et al. (2019): log OR = 0.38, 95% CrI = [−0.86, 1.63], p = 0.0004 |
| Truth telling when dishonesty harms concrete others | Köbis et al. (2019): log OR = −0.04, 95% CrI = [−1.32, 1.24], p = 0.861 |
| Negative reciprocity | d = 0.215, 95% CI = [0.107, 0.324] |
| Expectations of negative reciprocity | d = 0.195, 95% CI = [0.019, 0.371] |
| Positive reciprocity | d = -0.049, 95% CI = [-0.220, 0.121] |
| Expectations of positive reciprocity | d = -0.108, 95% CI = [-0.352, 0.135] |
| Utilitarianism in sacrificial dilemmas | d = -0.234, 95% CI = [-0.300, -0.168] |
| Utilitarianism in dilemmas where harm is a side-effect | d = -0.141, 95% CI = [-0.255, -0.028] |

These meta-analytic results were organized into a theoretical framework inspired by Rand and colleagues' Social Heuristics Hypothesis, which contends that intuitive processes promote the use of heuristics that have been internalized in everyday situations that resemble the situation at hand. It has been argued that three heuristics that many people are likely to have internalized are: don't get harmed; don't harm others, when they can harm you back; don't receive fewer resources than others. The framework is promising, as it makes predictions that are generally in line with the empirical evidence. However, a number of questions remain unanswered. Future work should aim to shed light on the areas in which empirical evidence is inconclusive. The hope is that this review can contribute to bring further attention to an area of research that has emerged over the last decades as one of the most vibrant and exciting in social science.

## Author note

Data and material, including the code to replicate the meta-analyses, are available at this link: https://osf.io/2v9uj/?view_only=ce8390f627104efeb737db61f32d61d3. This study was not pre-registered.

# References

Aarøe, L., & Petersen, M. B. (2013). Hunger games: Fluctuations in blood glucose levels influence support for social welfare. *Psychological Science*, *24*, 2550-2556.

Achtziger, A., Alós-Ferrer, C., & Wagner, A. K. (2015). Money, depletion, and prosociality in the Dictator Game. *Journal of Neuroscience, Psychology, and Economics, 8*, 1-14.

Achtziger, A., Alós-Ferrer, C., & Wagner, A. K. (2016). The impact of self-control depletion on social preferences in the ultimatum game. *Journal of Economic Psychology, 53*, 1-16.

Achtziger, A., Alós-Ferrer, C., & Wagner, A. K. (2018). Social preferences and self-control. *Journal of Behavioral and Experimental Economics, 74*, 161-166.

Aguiar, F., Brañas-Garza, P., Cobo-Reyes, R., Jimenez, N., & Miller, L. M. (2009). Are women expected to be more generous? *Experimental Economics, 12*, 93-98.

Aina, C., Battigalli, P., & Gamba, A. (2020). Frustration and anger in the Ultimatum Game: An experiment. *Games and Economic Behavior*, *122*, 150-167.

Ainsworth, S. E., Baumeister, R. F., Ariely, D., & Vohs, K. D. (2014). Ego depletion decreases trust in economic decision making. *Journal of Experimental Social Psychology, 54*, 40-49.

Alós-Ferrer, C., & Garagnani, M. (2020). The cognitive foundations of cooperation. *Journal of Economic Behavior & Organization*, *175*, 71-85.

Andersen, S., Gneezy, U., Kajackaite, A., & Marx, J. (2018). Allowing for reflection time does not change behavior in dictator and cheating games. *Journal of Economic Behavior and Organization, 145*, 24-33.

Anderson, C., & Dickinson, D. L. (2010). Bargaining and trust: The effect of 36hr sleep deprivation on socially interactive decisions. *Journal of Sleep Research, 19*, 54-63.

Andrighetto, G., Capraro, V., Guido, A., & Szekely, A. (2020). Cooperation, response time, and social value orientation: A meta-analysis. *Proceedings of the Cognitive Science Society.*

Artavia-Mora, L., Bedi, A. S., & Rieger, M. (2017). Intuitive help and punishment in the field. *European Economic Review, 92*, 133-145.

Artavia-Mora, L., Bedi, A. S., & Rieger, M. (2018). Help, prejudice and headscarvers. *Working Paper available at http://ftp.iza.org/dp11460.pdf*.

Axelrod, R., & Hamilton, W. D. (1981). The evolution of cooperation. *Science*, *211*, 1390-1396.

Bago, B., Bonnefon, J. F., & De Neys, W. (2021). Intuition rather than deliberation determines selfish and prosocial choices. *Journal of Experimental Psychology: General*, *150*, 1081-1094.

Bago, B., & De Neys, W. (2019). The intuitive greater good: Testing the corrective dual process model of moral cognition. *Journal of Experimental Psychology: General, 148*, 1782-1801.

Bago, B., Rand, D. G., & Pennycook, G. (2020). Fake news, fast and slow: Deliberation reduces belief in false (but not true) news headlines. *Journal of Experimental Psychology: General*, *149*, 1608-1613.

Balafoutas, L., Kerschbamer, R., & Oexl, R. (2018). Distributional preferences and ego depletion. *Journal of Neuroscience, Psychology, and Economics, 11*, 147-165.

Balafoutas, L., & Tarek, J. - L. (2018). Impunity under pressure: On the role of emotions as a commitment device. *Economics Letters, 168*, 112-114.

Ball, L., Thomson, V., & Stupple, E. (2017). Conflict and dual process theory: The case of belief bias. In W. De Neys (Ed.), *Dual Process Theory 2.0*. Oxon, UK: Routledge.

Banker, S., Ainsworth, S. E., Baumeister, R. F., Ariely, D., & Vohs, K. D. (2017). The sticky anchor hypothesis: Ego depletion increases susceptibility to situational cues. *Journal of Behavioral Decision Making, 30*, 1027-1040.

Banks, A. P., & Hope, C. (2014). Heuristic and analytic processes in reasoning: An event-related potential study of belief bias. *Psychophysiology, 51*, 290-297.

Baumeister, R. F., Heatherton, T., & Tice, D. M. (1994). *Losing control: How and why people fail at self-regulation.* San Diego, CA: Academic Press.

Baumeister, R. F., & Tierney, J. (2011). *Willpower: Rediscovering the greatest human strength.* New York, NY: Penguin Press.

Bear, A., & Rand, D. G. (2016). Intuition, deliberation, and the evolution of cooperation. *Proceedings of the National Academy of Sciences*, *113*, 936-941.

Becker, E. (1973). *The denial of death*. New York: Simon and Schuster.

Belloc, M., Bilancini, E., Boncinelli, L., & D'Alessandro, S. (2019). Intuition and deliberation in the Stag Hunt game. *Scientific Reports,9*, 14833.

Benjamin, D. J., Brown, S. A., & Shapiro, J. M. (2013). Who is "behavioral"? Cognitive ability and anomalous preferences. *Journal of the European Economic Association, 11*, 1231-1255.

Bentham, J. (1983). *The collected works of Jeremy Bentham: Deontology, together with a table of the springs of action; and the article on utilitarianism.* Oxford, England: Oxford University Press. (Original work published 1789).

Bereby-Meyer, Y., Hayakawa, S., Shalvi, S., Corey, J. D., Costa, A., & Keysar, B. (2020) Honesty speaks a second language. *Topics in Cognitive Science, 12*, 632-643.

Białek, M., & De Neys, W. (2017). Dual processes and moral conflict: Evidence for deontological reasoners' intuitive utilitarian sensitivity. *Judgment and Decision Making, 12*, 148-167.

Białek, M., Muda, R., Stewart, K., Niszczota, P., & Pieńkosz, D. (2020). Thinking in a foreign language distorts allocation of cognitive effort: Evidence from reasoning. *Cognition*, *205*, 104420.

Bieleke, M., Gollwitzer, P. M., Oettingen, G., & Fischbacher, U. (2017). Social value orientation moderates the effects of intuition versus reflection on responses to unfair ultimatum offers. *Journal of Behavioral Decision Making*, *30*, 569-581.

Bilancini, E., Boncinelli, L., Guarnieri, P., & Spadoni, L. (2021). Delaying and Motivating Decisions in the (Bully) Dictator Game. *Available at SSRN 4163676*.

Bilancini, E., Boncinelli, L., & Celadin, T. (2022). Social value orientation and conditional cooperation in the online one-shot public goods game. *Journal of Economic Behavior & Organization*, *200*, 243-272.

Bird, B. M., Geniole, S. N., Procyshyn, T. L., Ortiz, T. L., Carré, J. M., & Watson, N. V. (2019). Effect of exogenous testosterone on cooperation depends on personality and time pressure. *Neuropsychopharmacology*, *44*, 538.

Biziou van Pol, L., Haenen, J., Novaro, A., Occhipinti-Liberman, A., & Capraro, V. (2015). Does telling

white lies signal pro-social preferences? *Judgment and Decision Making, 10*, 538-548.

Bogliacino, F., & Gómez, C. (in preparation). The effect of negative emotions and cognitive load on trust and cooperation.

Bouwmeester, S., Verkoeijen, P. J. L., Acze, B., Barbosa, F., Bègue, L., et al. (2017). Registered replication report: Rand, Greene & Nowak (2012). *Perspectives on Psychological Science, 12,* 527-542.

Bowles, S., & Gintis, H. (2011). *A cooperative species: Human reciprocity and its evolution.* Princeton University Press.

Boyd, R., & Richerson, P. J. (2005). *The Origin and Evolution of Cultures.* Oxford, UK: Oxford University Press.

Brañas-Garza, P., Capraro, V., & Rascón-Ramírez, E. (2018). Gender differences in altruism on Mechanical Turk: Expectations and actual behaviour. *Economics Letters, 170*, 19-23.

Briers, B., Pandelaere, M., Dewitte, S., & Warlop, L. (2006). Hungry for money: The desire for caloric resources increases the desire for financial resources and vice versa. *Psychological Science*, *17*, 939-943.

Brosnan, S. F., & De Waal, F. B. (2003). Monkeys reject unequal pay. *Nature*, *425*, 297-299.

Brouwer, S. (2019). The auditory foreign-language effect of moral decision making in highly proficient bilinguals. *Journal of Multilingual and Multicultural Development*, *40*, 865-878.

Brouwer, S. (2021). The interplay between emotion and modality in the Foreign-Language effect on moral decision making. *Bilingualism: Language and Cognition*, *24*, 223-230.

Cabrales, A., Espín, A. M., Kujal, P., & Rassenti, S. (2022). Trustors' disregard for trustees deciding quickly or slowly in three experiments with time constraints. *Scientific Reports*, *12*, 12120.

Cacioppo, J. T., & Petty, R. E. (1982). The need for cognition. *Journal of Personality and Social Psychology*, *42*, 116-131.

Camerer, C. F., et al. (2018). Evaluating the replicability of social science experiments in Nature and Science between 2010 and 2015. *Nature Human Behaviour, 2*, 637.

Cantarero, K., & Van Tilburg, W. A. (2014). Too tired to taint the truth: Ego-depletion reduces other-benefiting dishonesty. *European Journal of Social Psychology, 44*, 743-747.

Cappelen, A. W., Sørensen, E. Ø., & Tungodden, B. (2013). When do we lie? *Journal of Economic Behavior and Organization, 93,* 258-265.

Cappelletti, D., Güth, W., & Ploner, M. (2011) Being of Two Minds: Ultimatum offers under cognitive constraints. *Journal of Economic Psychology, 32*, 940-950.

Capraro, V. (2013). A model of human cooperation in social dilemmas. *PLoS ONE, 8*, e72427.

Capraro, V. (2017). Does the truth come naturally? Time pressure increases honesty in one-shot deception games. *Economics Letters, 158,* 54-57.

Capraro, V., & Cococcioni, G. (2015). Social setting, intuition and experience in laboratory experiments interact to shape cooperative decision-making. *Proceedings of the Royal Society: Biological Sciences, 282*, 20150237.

Capraro, V., & Cococcioni, G. (2016). Rethinking spontaneous giving: Extreme time pressure and ego-depletion favor self-regarding reactions. *Scientific Reports, 6,* 27219.

Capraro, V., Corgnet, B., Espín, A. M., & Hernán-González, R. (2017). Deliberation favours social efficiency by making people disregard their relative shares: Evidence from USA and India. *Royal Society Open Science, 4*, 160605.

Capraro, V., Everett, J. A., & Earp, B. D. (2019b). Priming intuition disfavors instrumental harm but not impartial beneficence. *Journal of Experimental Social Psychology*, *83*, 142-149.

Capraro, V., & Rand, D. G. (2018). Do the right thing: Experimental evidence that preferences for moral behavior, rather than equity and efficiency per se, drive human prosociality. *Judgment and Decision Making, 13*, 99-111.

Capraro, V., Schulz, J., & Rand, D. G. (2019a). Time pressure and honesty in a deception game. *Journal of Behavioral and Experimental Economics, 79*, 93-99.

Carlson, R. W., Aknin, L. B., & Liotti, M. (2016). When is giving an impulse? An ERP investigation of intuitive prosocial behavior. *Social Cognitive and Affective Neuroscience, 11*, 1121-1129.

Carter, E. C., Kofler, L. M., Forster, D. E., & McCullough, M. E. (2015). A series of meta-analytic tests of the depletion effect: Self-control dfoes not seem to rely on a limited resource. *Journal of Experimental Psychology: General, 144,* 796-815.

Carter., E. C., & McCullough, M. E. (2014). Publication bias and the limited strength model of self-control: Has the evidence for ego-depletion been overestimated. *Frontiers in Psychology, 5,* 823.

Cellini, N., Lotto, L., Pletti, C., & Sarlo, M. (2017). Daytime REM sleep affects emotional experience but not decision choices in moral dilemmas. *Scientific Reports*, *7*, 1-11.

Chaiken, S. (1980). Heuristic versus systematic information processing and the use of source versus message cues in persuasion. *Journal of Personality and Social Psychology, 39*, 752-766.

Chen, F., & Krajbich, I. (2018). Biased sequential sampling underlies the effects of time pressure and delay in social decision making. *Nature Communications, 9*, 3557.

Chen, S., & Chaiken, S. (1999). The heuristic-systematic model in its broader context. In S. Chaiken & Y. Trope (Eds.), *Dual-process theories in social psychology* (pp. 73-96). New York, NY, US: Guilford Press.

Chiou, W. – B., Wu, W. – H., & Cheng, W. (2017). Self-control and honesty depend on exposure to pictures of the opposite sex in men but not women. *Evolution and Human Behavior, 38*, 616-625.

Chuan, A., Kessler, J. B., Milkman, K. L. (2018). Field study of charitable giving reveals that reciprocity decays over time. *Proceedings of the National Academy of Sciences, 115*, 1766-1771.

Cipolletti, H., McFarlane, S., & Weissglass, C. (2016). The moral foreign-language effect. *Philosophical Psychology, 29*, 23–40.

Cone, J., & Rand, D. G. (2014). Time pressure increases cooperation in competitively framed social dilemmas. *Plos One, 9*, e115756.

Conway, P., & Gawronski, B. (2013). Deontological and utilitarian inclinations in moral decision making: A process dissociation approach. *Journal of Personality and Social Psychology, 104*, 216-235.

Conway, P., Goldstein-Greenwood, J., Polacek, D., & Greene, J. D. (2018). Sacrificial utilitarian judgments do reflect concern for the greater good: Clarification via process dissociation and the judgments of philosophers. *Cognition, 179*, 241-265.

Corbit, J., Dockrill, M., Hartlin, S., & Moore, C. (2023). Intuitive cooperators: Time pressure increases children's cooperative decisions in a modified public goods game. *Developmental Science*, *26*, e13344.

Corey, J. D., Hayakawa, S., Foucart, A., Aparici, M., Botella, J., Costa, A., & Keysar, B. (2017). Our moral choices are foreign to us. *Journal of Experimental Psychology: Learning, Memory, and Cognition, 43*, 1109-1128.

Corgnet, B., Espín, A. M., Hernán-González, R. (2015). The cognitive basis of social behavior: Cognitive reflection overrides antisocial but not always prosocial motives. *Frontiers in Behavioral Neuroscience, 9*, 287.

Cornelissen, G., Dewitte, S., & Warlop. L. (2011). Are social value orientations expressed automatically? Decision making in the Dictator Game. *Personality and Social Psychology Bulletin, 37*, 1080-1090.

Costa, A., Foucart, A., Arnon, I., Aparici, M., & Apesteguia, J. (2014a). “Piensa” twice: On the foreign language effect in decision making. *Cognition, 130*, 236–254.

Costa, A., Foucart, A., Hayakawa, S., Aparici, M., Apesteguia, J., Heafner, J., & Keysar, B.(2014b). Your morals depend on language. *PLoS ONE*, 9, e94842.

Crockett, M. J., Clark, L., Tabibnia, G., Lieberman, M. D., & Robbins, T. W. (2008). Serotonin modulates behavioral reactions to unfairness. *Science, 320*, 1739-1739.

Crockett, M. J., Clark, L., Lieberman, M. D., Tabibnia, G., & Robbins, T. W. (2010a). Impulsive choice and altruistic punishment are correlated and increase in tandem with serotonin depletion. *Emotion, 10*, 855-862

Crockett, M. J., Clark, L., Hauser, M. D., & Robbins, T. W. (2010b). Serotonin selectively influences moral judgment and behavior through effects on harm aversion. *Proceedings of the National Academy of Sciences, 107*, 17433-17438.

Cummins, D. D., & Cummins, R. C. (2012). Emotion and deliberative reasoning in moral judgment. *Frontiers in Psychology*, *3*, 328.

Curtin, J., & Fairchild, B. A. (2003). Alcohol and cognitive control: Implications for regulation of behavior during response conflict. *Journal of Abnormal Psychology, 112*, 424-436.

Danziger, S., Levav, J., Avnaim-Pesso, L. (2011). Extraneous factors in judicial decisions. *Proceedings of the National Academy of Sciences, 108*, 6889-6892.

Darling, M. (2010). *The effect of cognitive load on third party punishment* (Doctoral dissertation, Tufts University). Available at: https://www.proquest.com/docview/756891970

Darwin, C. (1871). *The descent of man, and selection in relation to sex* (Vol. 1). Murray.

Dawes, R. (1980). Social dilemmas. *Annual Review of Psychology, 31*, 169-193.

Debey, E., Verschuere, B., & Crombez, G. (2012). Lying and executive control: An experimental investigation using ego depletion and goal neglect. *Acta Psychologica, 140*, 133-141.

De Dreu, C. K. W., Dussel, B. D., & Ten Velden, F. S. (2015) In intergroup conflict, self-sacrifice is stronger among pro-social indviduals, and parochial altruism emerges especially among cognitively taxed individuals. *Frontiers in Psychology, 6*, 572.

De Neys, W. (2012). Bias and conflict: A case for logical intuitions. *Perspectives on Psychological Science, 7*, 28-38.

De Neys, W., & Pennycook, G. (2019). Logic, fast and slow: Advances in dual-process theorizing. *Current Directions in Psychological Science.* Online first: https://doi.org/10.1177/0963721419855658.

De Neys, W. (2021). On dual-and single-process models of thinking. *Perspectives on Psychological Science*, *16*, 1412-1427.

De Neys, W. (2022). The cognitive unconscious and dual process theories of reasoning. *The Cognitive Unconscious: The First Half Century*, 223.

De Neys, W. (2023) Advancing theorizing about fast-and-slow thinking. *Behavioral and Brain Sciences, 46*, e111: 1–71. doi:10.1017/S0140525X2200142

Del Maschio, N., Del Mauro, G., Bellini, C., Abutalebi, J., & Sulpizio, S. (2022). Foreign to whom? Constraining the moral foreign language effect on bilinguals' language experience. *Language and*

*Cognition*, *14*, 511-533.

Dickinson, D. L., & Masclet, D. (2021). Unethical Decision Making and Sleep Restriction: Experimental Evidence. *Available at: https://papers.ssrn.com/sol3/papers.cfm?abstract_id=3892565*

Dickinson, D. L., & McElroy, T. (2017) Sleep restriction and circadian effects on social decisions. *European Economic Review, 97*, 57-71.

Døssing, F., Piovesan, M., & Wengström, E. (2017). Cognitive load and cooperation. *Review of Behavioral Economics, 4*, 69-81.

Driver, M. Y. (2022). Switching codes and shifting morals: How code-switching and emotion affect moral judgment. *International Journal of Bilingual Education and Bilingualism*, *25*, 905-921.

Duval, S., & Tweedie, R. (2000). A nonparametric "trim and fill" method of accounting for publication bias in meta-analysis. *Journal of the American Statistical Association*, *95*, 89-98.

Dylman, A. S., & Champoux-Larsson, M. F. (2020). It's (not) all Greek to me: Boundaries of the foreign language effect. *Cognition*, *196*, 104148.

Eagly, A. H. (1987). *Sex differences in social behavior: A social-role interpretation*. Mahwah, New Jersey: L. Erlbaum Associates.

Engel, C. (2011). Dictator games: A meta study. *Experimental Economics*, *14*, 583-610.

Epstein, S. (1994). Integration of the cognitive and psychodynamic unconscious. *American Psychologist, 49*, 709-724.

Epstein, S., & Pacini, R. (1999). Some basic issues regarding dual-process theories from the perspective of cognitive–experiential self-theory. In S. Chaiken & Y. Trope (Eds.), *Dual-process theories in social psychology* (pp. 462-482). New York, NY, US: Guilford Press.

Epstein, S., Pacini, R., Denes-Raj, V., & Heier, H. (1996). Individual differences in intuitive-experiential and analytical-rational thinking styles. *Journal of Personality and Social Psychology*, 71, 390-405.

Evans, A. M., Dillon, K. D., & Goldin, G., & Krueger, J. I. (2011). Trust and self-control: The moderating role of the default. *Judgment and Decision Making, 6*, 697-705.

Evans, A. M., Dillon, K. D., & Rand, D. G. (2015). Fast but not intuitive, slow but not deliberative:

Decision conflict drives reaction times in social dilemmas. *Journal of Experimental Psychology: General, 144*, 951-966.

Evans, A. M., & Rand, D. G. (2018). Cooperation and decision time. *Current Opinion in Psychology, 26*, 67-71.

Evans, J. St. B. T. (1989). *Bias in Human Reasoning: Causes and Consequences*. Brighton, UK: Erlbaum.

Evans, J. St. B. T. (2006). The heuristic-analytic theory of reasoning: extension and evaluation. *Psychonomic Bulletin and Review, 13*, 378-395.

Evans, J. St. B. T. (2008). Dual-processing accounts of reasoning, judgment, and social cognition. *Annual Review of Psychology, 29*, 255-278.

Evans, J. St. B. T., & Over, D. E. (1996). *Rationality and Reasoning*. Hove, UK: Psychol. Press.

Evans, J. St. B. T., & Stanovich, K. E. (2013). Dual-process theories of higher cognition: Advancing the debate. *Perspectives on Psychological Science, 8*, 223-241.

Everett, J. A. C., Faber, N. S., & Crockett, M. (2015). Preferences and beliefs in ingroup favoritism. *Frontiers in Behavioral Neuroscience, 9*, 15.

Everett, J. A. C., Ingbretsen, Z., Cushman, F., & Cikara, M. (2017). Deliberation erodes cooperative behavior – Even towards competitive outgroups, even when using control condition, and even when eliminating selection bias. *Journal of Experimental Social Psychology, 73*, 76-81.

Everett, J. A. C., Ingbretsen, Z., Cushman, F., & Cikara, M. (2018). Aggression, fast and slow: Intuition favors defensive aggression. *Available at* https://psyarxiv.com/4v39b/

Fabio, R. A., Marcuzzo, A., & Calabrese, C. (2023). Cognitive Load and Peace Attitude Influence Cooperative Choices in the Iterated Prisoner's Dilemma Game. *Psychological Reports*, 00332941231166604.

Fehr, E., & Fischbacher, U. (2003). The nature of human altruism. *Nature, 425*, 785-791.

Fehr, E., & Fischbacher, U. (2004). Social norms and human cooperation. *Trends in Cognitive Sciences, 8*, 185-190.

Fehr, E., & Gächter, S. (2002). Altruistic punishment in humans. *Nature, 415*, 137-140.

Fehr, E., & Schmidt, K. M. (1999). A theory of fairness, competition, and cooperation. *The Quarterly Journal of Economics, 114*, 817-868.

Ferrara, M., Bottasso, A., Tempesta, M., Carrieri, M., De Gennaro, L., & Ponti, G. (2015). Gender differences in sleep deprivation effects on risk and inequality aversion: Evidence from an economic experiment. *PLoS ONE, 10*, e0120029.

Fillmore, M., Carscadden, J., Vogel-Sprott, M. (1998). Alcohol, cognitive impairment, and expectancies. *Journal of Studies on Alcohol, 59*, 174-179.

Fischbacher, U., & Föllmi-Heusi, F. (2013). Lies in disguise – An experimental study on cheating. *Journal of the European Economic Association, 11*, 525-547.

Fodor, J. (1983). *The Modularity of Mind*. Scranton, PA: Crowell.

Fodor, J. (2001). *The Mind Doesn't Work That Way*. Cambridge, MA: MIT Press.

Foerster, A., Pfister, R., Schmidts, C., Dignath, D., & Kunde, W. (2013). Honesty saves time (and justifications). *Frontiers in Psychology, 4*, 473.

Francis, K. B., Gummerum, M., Ganis, G., Howard, I. S., & Terbeck, S. (2019). Alcohol, empathy, and morality: acute effects of alcohol consumption on affective empathy and moral decision-making. *Psychopharmacology*, *236*, 3477-3496.

Fraser, S., & Nettle, D. (2020). Hunger affects social decisions in a multi-round Public Goods Game but not a single-shot Ultimatum Game. *Adaptive Human Behavior and Physiology*, *6*, 334-355.

Frey, V., De Mulder, H. N., Ter Bekke, M., Struiksma, M. E., Van Berkum, J. J., & Buskens, V. (2022). Do self-talk phrases affect behavior in ultimatum games?. *Mind & Society*, *21*, 89-119.

Friese, M., Loschelder, D. D., Gieseler, K., Frankenbach, J., & Inzlicht, M. (2019). Is ego depletion real? An analysis of arguments. *Personality and Social Psychology Review, 23*, 107-131.

Friehe, T., & Schildberg-Hörisch, H. (2017). Self-control and crime revisited: Disentangling the effect of self-control on risk taking and antisocial behavior. *International Review of Law and Economics, 49*, 23-32.

Fromell, H., Nosenzo, D., & Owens, T. (2020). Altruism, fast and slow? Evidence from a meta-analysis

and a new experiment. *Experimental Economics*, *23*, 979-1001.

Gai, P. J., & Puntoni, S. (2021). Language and consumer dishonesty: a self-diagnosticity theory. *Journal of Consumer Research*, *48*, 333-351.

Gao, Q., Jia, X., Liu, H., Wang, X., & Liu, Y. (2020). Attachment style predicts cooperation in intuitive but not deliberative response in one-shot public goods game. *International Journal of Psychology*, *55*, 478-486.

Gargalianou, V., Urbig, D., & van Witteloostuijn, A. (2017). Cooperating or competing in three languages: Cultural accommodation or alienation? *Cross Cultural & Strategic Management, 24*, 167-191.

Gärtner, M. (2018). The prosociality of intuitive decisions depends on the status quo. *Journal of Behavioral and Experimental Economics, 74*, 127-138.

Gärtner, M., Andersson, D., Västfjäll, D., & Tinghög, G. (2022). Affect and prosocial behavior: The role of decision mode and individual processing style. *Judgment and Decision making*, *17*, 1-13.

Geipel, J., Hadjichristidis, C., & Surian, L.(2015a). The foreign language effect on moral judgment: The role of emotions and norms. *PLoS ONE, 59*, 8–17.

Geipel, J., Hadjichristidis, C., & Surian, L.(2015b). How foreign language shapes moral judgment. *Journal of Experimental Social Psychology, 10*, e0131529.

Gevins, A. S., & Cutillo, B. C. (1993). Neuroelectric evidence for distributed processing in human working memory. *Electroencephalography and Clinical Neurophysiology, 87,* 128-143.

Gilbert, D. T., & Hixon, J. G. (1991). The trouble of thinking: Activation and application of stereotypic beliefs. *Journal of Personality and Social Psychology, 60,* 509-517.

Gino, F., Schweitzer, M. E., Mead, N. L., & Ariely, D. (2011). Unable to resist temptation: How self-control depletion promotes unethical behavior. *Organizational Behavior and Human Decision Processes, 115,* 191-203.

Gintis, H., Bowles, S., Boyd, R., & Fehr, E. (2003). Explaining altruistic behavior in humans. *Evolution and Human Behavior, 24*, 153-172.

Glöckner, A., & Betsch, T. (2008). Modeling option and strategy choices with connectionist networks: Towards an integrative model of automatic and deliberate decision making. *Judgment and Decision Making, 3*, 215-228.

Glöckner, A., & Witteman, C. L. M. (2010). Beyond dual-process models: A categorization of processes underlying intuitive judgment and decision making. *Thinking & Reasoning, 16*, 1-25.

Goeschl, T., & Lohse, J. (2018). Cooperation in public good games. Calculated or confused? *European Economic Review, 107*, 185-203.

Gotlib, T., & Converse, P. (2010). Dishonest Behavior: The impact of prior self-regulatory exertion and personality. *Journal of Applied Social Psychology*, *40*, 3169–3191.

Gneezy, U. (2005). Deception: The role of consequences. *The American Economic Review, 95*, 384-394.

Gneezy, U., Rockenbach, B., & Serra-Garcia, M. (2013). Measuring lying aversion. *Journal of Economic Behavior & Organization, 93*, 293-300.

Gravelle, J. G. (2009). Tax havens: Internation tax avoidance and evasion. *National Tax Journal, 68*, 727-753.

Greenberg, J., Pyszczynski, T., & Solomon, S. (1986). The causes and consequences of a need for self-esteem: A terror management theory. In *Public self and private self* (pp. 189-212). Springer, New York, NY.

Greene, J. D., Morelli, S. A., Lowenberg. K., Nystrom, L. E., & Cohen, J. D. (2008). Cognitive load selectively interferes with utilitarian moral judgment. *Cognition, 107,* 1144-1154.

Greene, J. D., Sommerville, R. B., Nystrom, L. E., Darley, J. M., & Cohen, J. D. (2001). An fMRI investigation of emotional engagement in moral judgment. *Science, 293*, 2105-2108.

Grimm, V., & Mengel, F. (2011). Let me sleep on it: Delay reduces rejection rates in ultimatum games. *Economics Letters, 111*, 113-115.

Grolleau, G., Sutan, A., El Harbi, S., Jedidi, M. (2018). Do we need more time to give less? Experimental evidence from Tunisia. *Bulletin of Economic Research, 70*, 400-409.

Grossman, Z., & Van der Weele, J. J. (2017). Dual-process reasoning in charitable giving: Learning from non-results. *Games, 8*, 36.

Gunia, B. C., Wang, L., Huang, L., Wang, J. and Murnighan, J. K. (2012). Contemplation and conversation: Subtle influences on moral decision making. *Academy of Management Journal*, 55, 13–33.

Gürcay, B., & Baron, J. (2017). Challenges for the sequential two-system model of moral judgement. *Thinking & Reasoning*, *23*, 49-80.

de Haan, T., & van Veldhuizen, R. (2015). Willpower depletion and framing effects. *Journal of Economic behavior and Organization, 117*, 47-61.

Hagger, M. S., Chatzisarantis, N. L. D., Alberts, H., Anggono, C. O., Batailler, C., et al. (2016). A multilab preregistered replication of the ego-depletion effect. *Perspectives on Psychological Science, 11,* 546-573.

Halali, E., Bereby-Meyer, Y., Meiran, N. (2014). Between self-interest and reciprocity: The social bright side of self-control failure. *Journal of Experimental Psychology: General, 143*, 745-754.

Halali, E., Bereby-Meyer, Y., Ockenfels, A. (2013). Is it all about the self? The effect of self-control depletion on ultimatum game proposers. *Frontiers in Human Neuroscience, 7*, 240.

Hallsson, B. G., Siebner, H. R., & Hulme, O. J. (2018). Fairness, fast and slow: A review of dual process models of fairness. *Neuroscience & Biobehavioral Reviews, 89*, 49-60.

Ham, J., & van den Bos, K. (2010). On Unconscious Morality: The Effects of Unconscious Thinking on Moral Decision Making. *Social Cognition, 28*, 74–83.

Hammond, K. R. (1996). *Human Judgment and Social Policy*. New York: Oxford Univ. Press.

Handley, S. J., Newstead, S. E., & Trippas, D. (2011). Logic, beliefs, and instruction: A test of default interventionist account of belief bias. *Journal of Experimental Psychology: Learning, Memory, and Cognition, 37*, 28-43.

Hardin, G. (1968). The tragedy of the commons. *Science, 162*, 1243-1248.

Harel, I., & Kogut, T. (2015). Visceral needs and donation decisions: Do people identify with suffering or with relief?. *Journal of Experimental Social Psychology*, *56*, 24-29.

Harris, A., Young, A., Hughson, L., Green, D., Doan, S. N., Hughson, E., & Reed, C. L. (2020). Perceived relative social status and cognitive load influence acceptance of unfair offers in the Ultimatum Game. *Plos one*, *15*, e0227717.

Hauge, K. E., Brekke, K. A., Johansson, L. – O., Johansson-Stenman, O., & Svedsäter, H. (2016). Keeping others in our mind or in our heart? Distribution games under cognitive load. *Experimental Economics, 19*, 562-576.

Häusser, J. A., Stahlecker, C., Mojzisch, A., Leder, J., Van Lange, P. A., & Faber, N. S. (2019). Acute hunger does not always undermine prosociality. *Nature Communications*, *10*, 4733.

Hayakawa, S., Tannenbaum, D., Costa, A., Corey, J. D., & Keysar, B. (2017). Thinking more or feeling less? Explaining the foreign language effect on moral judgment. *Psychological Science, 28*, 1387–1397.

Heilman, M. E., & Okimoto, T. G. (2007). Why are women penalized for success at male tasks?: The implied communality deficit. *Journal of Applied Psychology, 92*, 81-92.

Henrich, J., Boyd, R., Bowles, S., Camerer, C., Fehr, E., Gintis, H., & McElreath, R. (2001). In search of homo economicus: behavioral experiments in 15 small-scale societies. *American Economic Review*, *91*, 73-78.

Henrich, J., Heine, S. J., & Norenzayan, A. (2010). Most people are not WEIRD. *Nature*, *466*, 29-29.

Herrmann, B., Thoni, C., & Gachter, S. (2008). Antisocial punishment across societies. *Science*, *319*, 1362-1367.

Hinson, J. M., Jameson, T. L., & Whitney, P. (2003). Impulsive decision making and working memory. *Journal of Experimental Psychology: Learning, Memory, and Cognition, 29,* 298-306.

Hochman, G., Ayal, S., & Ariely, D. (2015). Fairness requires deliberation: The primacy of economic over social considerations. *Frontiers in Psychology, 6*, 747.

Holbein, J. B., Schafer, J. P., & Dickinson, D. L. (2019). Insufficient sleep reduces voting and other prosocial behaviours. *Nature Human Behaviour, 3*, 492-500.

Houston, C. J. (2010). Short-Term Retention, Time Pressure, and Accessibility Tasks

Do Not Interfere with Utilitarian Moral Judgment. Master thesis, Harvard University. *Available at: https://nrs.harvard.edu/URN-3:HUL.INSTREPOS:37367557*

Hsu, M., Anen, C., & Quartz, S. R. (2008). The right and the good: Distributive justice and neural encoding of equity and efficiency. *Science, 320*, 1092-1095.

Inzlicht, M., Schmeichel, B. J. (2012). What is ego depletion? Toward a mechanistic revision of the resource model of self-control. *Perspectives on Psychological Science, 7,* 450-463.

Inzlicht, M., Schmeichel, B. J., & Macrae, C. N. (2014). Why self-control seems (but may not be) limited. *Trends in Cognitive Sciences, 18,* 127-133.

Isler, O., Maule, J., & Starmer, C. (2018). Is intuition really cooperative? Improved tests support the social heuristics hypothesis. *PloS One*, *13*, e0190560.

Isler, O., Gächter, S., Maule, A. J., & Starmer, C. (2021a). Contextualised strong reciprocity explains selfless cooperation despite selfish intuitions and weak social heuristics. *Scientific Reports*, *11*, 13868.

Isler, O., & Yilmaz, O. (2019). Intuition and deliberation in morality and cooperation: An overview of the literature. In Liebowitz, J. (Ed.), *Developing informed intuition for decision-making* (pp. 101-115). Boca Raton, FL, US: CRC Press.

Isler, O., Yilmaz, O., & John Maule, A. (2021b). Religion, parochialism and intuitive cooperation. *Nature Human Behaviour*, *5*, 512-521.

Isler, O., & Yilmaz, O. (2022). How to activate intuitive and reflective thinking in behavior research? A comprehensive examination of experimental techniques. *Behavior Research Methods*, 1-20.

Itzchakov, G., Uziel, L., & Wood, W. (2018). When attitudes and habits don't correspond: Self-control depletion increases persuasion but not behavior. *Journal of Experimental Social Psychology, 75*, 1-10.

Jacoby, L. L. (1991). A process dissociation framework: Separating auto- matic from intentional uses of memory. *Journal of Memory and Language, 30,* 513–541.

Jaeger, B., Evans, A. M., Stel, M., & van Beest, I. (2019). Explaining the persistent influence of facial

cues in social decision-making. *Journal of Experimental Psychology: General, 148*, 1008-1021.

Jarke-Neuert, J., & Lohse, J. (2016). I'm in a hurry, I don't want to know! The effects of time pressure and transparency on self-serving behavior. *Available at SSRN: https://ssrn.com/abstract=2823678*.

Jenni, K., & Loewenstein, G. (1997). Explaining the identifiable victim effect. *Journal of Risk and Uncertainty*, *14*, 235-257.

Haidt, J., & Joseph, C. (2004). Intuitive ethics: How innately prepared intuitions generate culturally variable virtues. *Daedalus*, *133*, 55-66.

Kahane, G., Everett, J. A. C., Earp, B. D., Caviola, L., Faber, N. S., Crockett, M. J., & Savulescu, J. (2018). Beyond sacrificial harm: A two-dimensional model of utilitarian psychology. *Psychological Review*, 125, 131-164.

Kahneman, D. (2011). *Thinking, fast and slow.* New York, NY: Farrar, Straus and Giroux.

Kahneman, D., & Frederick, S. (2005). A Model of Heuristic Judgment. In K. J. Holyoak & R. G. Morrison (Eds.), *The Cambridge handbook of thinking and reasoning* (pp. 267-293). New York, NY, US: Cambridge University Press.

Kamphuis, J., Meerlo, P., Koolhaas, J. M., & Lancel, M. (2012). Poor sleep as a potential causal factor in aggression and violence. *Sleep Medicine, 13*, 327-334.

Kant, I. (2002). *Groundwork for the metaphysics of morals.* New Haven, CT: Yale University Press. (Original work published 1797).

Kaplan, H., Hill, J., Lancaster, J., & Hurtado, A. M. (2000). A theory of human life history evolution: Diet, intelligence, and longevity. *Evolutionary Anthropology, 9,* 156-185.

Katzir, M., Cohen, S., & Halali, E. (2021). Is it all about appearance? Limited cognitive control and information advantage reveal self-serving reciprocity. *Journal of Experimental Social Psychology*, *97*, 104192.

Kessler, J. B., Kivimaki, H., & Niederle, M. (2017). Thinking fast and slow: Generosity over time. *Available at*

*https://users.nber.org/~kesslerj/papers/KesslerKivimakiNiederle_GenerosityOverTime.pdf*

Kessler, J. B., & Meier, S. (2014). Learning from (failed) replications: Cognitive load manipulation and charitable giving. *Journal of Economic Behavior & Organization, 102*, 10-13.

Keysar, B., Hayakawa, S. L., & An, S. G. (2012). The foreign-language effect: Thinking in a foreign tongue reduces decision biases. *Psychological Science, 23*, 661-668.

Killgore, W. D., Killgore, D. B., Day, L. M., Li, C., Kamimori, G. H., & Balkin, T. J. (2007). The effects of 53 hours of sleep deprivation on moral judgment. *Sleep*, *30*, 345-352.

Kinnunen, S. P., & Windmann, S. (2013). Dual-processing altruism. *Frontiers in Psychology, 4*, 193.

Kirk, D., Gollwitzer, P. M., & Carnevale, P. J. (2011). Self-regulation in ultimatum bargaining: Goals and plans help accepting unfair but profitable offers. *Social Cognition*, *29*, 528-546.

Klenk, M. (2022). The Influence of Situational Factors in Sacrificial Dilemmas on Utilitarian Moral Judgments: A Systematic Review and Meta-Analysis. *Review of Philosophy and Psychology*, *13*, 593-625.

Köbis, N., Verschuere, B., Bereby-Meyer, Y., Rand, D. G., & Shalvi, S. (2019). Intuitive honesty versus dishonesty: Meta-analytic. *Perspectives on Psychological Science, 14*, 778-796.

Kohlber, L. (1963). The development of children's orientation towards moral order. *Human Development, 6*, 11-33.

Kohlberg, L. (1981). *The philosophy of moral development: Moral stages and the idea of justice* (Vol. 1). San Francisco: Harper & Row.

Kollock, P. (1988). Social dilemmas: The anatomy of cooperation. *Annual Review of Sociology, 24*, 183-214.

Körner, A., & Volk, S. (2014). Concrete and abstract ways to deontology: Cognitive capacity moderates construal level effects on moral judgments. *Journal of Experimental Social Psychology*, *55*, 139-145.

Krajbich, I., Bartling, B., Hare, T., & Fehr, E. (2015). Rethinking fast and slow based on a critique of reaction-time reverse inference. *Nature Communications, 6*, 7455.

Krawczyk, M., & Sylwestrzak, M. (2018). Exploring the role of deliberation time in non-selfish behavior:

The double response method. *Journal of Behavioral and Experimental Economics, 72*, 121-134.

Kroneisen, M., & Steghaus, S. (2021). The influence of decision time on sensitivity for consequences, moral norms, and preferences for inaction: Time, moral judgments, and the CNI model. *Journal of Behavioral Decision Making*, *34*, 140-153.

Kruglanski, A. W. (2013). Only one? The default interventionist perspective as a unimodel—Commentary on Evans & Stanovich (2013). *Perspectives on Psychological Science*, *8*, 242-247.

Kruglanski, A. W., & Gigerenzer, G. (2018). Intuitive and deliberate judgments are based on common principles. In *the Motivated Mind* (pp. 104-128). Routledge.

Kvaran, T., Nichols, S., & Sanfey, A. (2013). The effect of analytic and experiential modes of thought on moral judgment. *Progress in Brain Research, 202,* 187-196.

Kvarven, A., Strømland, E., Wollbrant, C., Andersson, D., Johannesson, M., Tinghög, G., ... & Myrseth, K. O. R. (2020). The intuitive cooperation hypothesis revisited: a meta-analytic examination of effect size and between-study heterogeneity. *Journal of the Economic Science Association*, *6*, 26-42.

Lee, J. J., & Gino, F. (2015). Poker-faced morality: Concealing emotions leads to utilitarian decision making. *Organizational Behavior and Human Decision Processes*, *126*, 49-64.

Levine, E., Barasch, A., Rand, D. G., Berman, J. Z., & Small, D. A. (2018). Signaling emotion and reason in cooperation. *Journal of Experimental Psychology: General*, 5, 702-719.

Li, Z., Gao, L., Zhao, X., & Li, B. (2021). Deconfounding the effects of acute stress on abstract moral dilemma judgment. *Current Psychology*, *40*, 5005-5018.

Li, Z., Xia, S., Wu, X., & Chen, Z. (2018). Analytical thinking style leads to more utilitarian moral judgments: An exploration with a process-dissociation approach. *Personality and Individual Differences*, *131*, 180–184.

Lieberman, M. D. (2003). Reflective and reflexive judgment processes: A social cognitive neuroscience approach. In J. P. Forgas, K. R. Williams, & W. von Hippel (Eds.). *Social Judgments: Implicit and Explicit Processes* (pp. 44-67). New York: Cambridge Univ. Press.

Liu, S., & Hao, F. (2011). An application of a dual-process approach to decision making in social dilemmas. *American Journal of Psychology, 124,* 203-212.

Liu, Y., He, N., & Dou, K. (2015). Ego-depletion promotes altruistic punishment. *Open Journal of Social Science, 3*, 62-69.

Liu, C., & Liao, J. (2021). Stand up to action: The postural effect of moral dilemma decision-making and the moderating role of dual processes. *PsyCh Journal*, *10*, 587-597.

Lohse, J. (2016). Smart or selfish – When smart guys finish nice. *Journal of Behavioral and Experimental Economics, 64*, 28-40.

Lohse, T., Simon, S. A., & Konrad, K. A. (2018). Deception under time pressure: Conscious decision or a problem of awareness? *Journal of Economic Behavior & Organization, 146*, 31-42.

Lotz, S. (2015). Spontaneous giving under structural inequality: Intuition promotes cooperation in asymmetric social dilemmas. *PLoS ONE, 10*, e0131562.

Lyrintzis, E. A. (2017). *Calibrating the Moral Compass: The Effect of Tailored Communications on Non-Profit Advertising*. The Claremont Graduate University.

Martin, F. A. (2020). *Mortality salience, group membership, and reciprocity* (Doctoral dissertation, Loyola University Chicago). Available at https://www.proquest.com/docview/2427517613

Mazar, N., Amir, O., & Ariely, D. (2008). The dishonesty of honest people: A theory of self-concept maintenance. *Journal of Marketing Research, 45*, 633-644.

McEvoy, D. M., Bruner, D. M., Dickinson, D. L., & Drummond, S. P. (2021). Sleep restriction and strategy choice in cooperation and coordination games. *Economics Letters*, *208*, 110049.

Mead, N. L., Baumeister, R. F., Gino, F., Schweitzer, M. E., & Ariely, D. (2009). Too Tired to Tell the Truth: Self-Control Resource Depletion and Dishonesty. *Journal of Experimental Social Psychology*, *45*, 594–597.

Merkel, A., & Lohse, J. (2019). Is fairness intuitive? An experiment accounting for subjective utility differences under time pressure. *Experimental Economics, 22*, 24-50.

Milczarski, W., Borkowska, A., Paruzel-Czachura, M., & Białek, M. (2022). Using a foreign language does not make you think more: Null effects of using a foreign language on cognitive reflection and numeracy. *Available on PsyArxiv.*

Milinski, M., Semman, D., & Krambeck, H. – J. (2002). Reputation helps solve the "tragedy of the commons". *Nature, 415*, 424-426.

Mill, J. S. (1863). *Utilitarianism.* London, England: parker, Son, and Bourne.

Mischkowski, D., & Glöckner, A. (2016). Spontaneous cooperation for prosocials, but not proselfs: Social value orientation moderates spontaneous cooperative behavior. *Scientific Reports, 6*, 21555.

Mitkidis, P., Lindeløv, J. K., Elbaek, C. T., Porubanova, M., Grzymala-Moszczynska, J., & Ariely, D. (2022). Morality in the time of cognitive famine: The effects of memory load on cooperation and honesty. *Acta Psychologica*, *228*, 103664.

Miyake, A., & Shah, P. (Eds.). (1999). *Models of working memory: Mechanisms of active maintenance and executive control*. Cambridge University Press.

Moller, A. C., Deci, E. L., & Ryan, R. M. (2006). Choice and ego-depletion: The moderating role of autonomy. *Personality and Social Psychology Bulletin, 32,* 1024-1036.

Moore, C., Tenbrunsel, A. E. (2014). "Just think about it"? Cognitive complexity and moral choice. *Organizational Behavior and Human Decision Processes, 123,* 138-149.

Moskowitz, H., Burns, M. M., Williams, A. F. (1985). Skills performance at low blood alcohol levels. *Journal of Studies on Alcohol, 46*, 482-485.

Motro, D., Ordonez, L. D., Pittarello, A., & Welsh, D. T. (2018). Investigating the Effects of Anger and Guilt on Unethical Behavior: A Dual-Process Approach. *Journal of Business Ethics, 152*, 133-148.

Mrkva, K. (2017). Giving, fast and slow: Reflection increases costly (but not uncostly) charitable giving. *Journal of Behavioral Decision Making, 30*, 1052-1065.

Muda, R., Niszczota, P., Białek, M., & Conway, P. (2018). Reading dilemmas in a foreign language reduces both deontological and utilitarian response tendencies. *Journal of Experimental Psychology, Learning, Memory, & Cognition, 44*, 321-326.

Muraven, M., Pogarsky, G., & Shmueli, D. (2006). Self-control depletion and the general theory of crime. *Journal of Quantitative Criminology*, *22*, 263–277.

Nava, F., Margoni, F., Herath, N., & Nava, E. (2023). Age-dependent Changes in Intuitive and Deliberative Cooperation. *Scientific Reports, 13*, 4457.

Nelson, R. J., & Trainor, B. C. (2007). Neural mechanisms of aggression. *Nature Reviews Neuroscience*, *8*, 536-546.

Neo, W. S., Yu, M., Weber, R., & Gonzalez, C. (2013). The effects of time delay in reciprocity games. *Journal of Economic Psychology, 34,* 20-35.

Newton, C., Feeney, J., & Pennycook, G. (2023). On the disposition to think analytically: Four distinct intuitive-analytic thinking styles. *Personality and Social Psychology Bulletin*. Online first: https://doi.org/10.1177/01461672231154886

Nisbett, R., Peng, K., Choi, I., & Norenzayan, A. (2001). Culture and systems of thought: holistic vs. analytic cognition. *Psychological Review, 108*, 291–310.

Nockur, L., & Pfattheicher, S. (2021). The beautiful complexity of human prosociality: on the interplay of honesty-humility, intuition, and a reward system. *Social Psychological and Personality Science*, *12*, 877-886.

Nowak, M. A. (2006). Five rules for the evolution of cooperation. *Science, 314*, 1560-1563.

Nowak, M. A., & Sigmund, K. (2005). Evolution of indirect reciprocity. *Nature, 437,* 1291-1298.

Oechssler, J., Roider, A., & Schmitz, P. W. (2015). Cooling off in negotiations: Does it work? *Journal of Institutional and Theoretical Economics, 171*, 565-588.

Okun, A. M. (2015). *Equality and efficiency: The big tradeoff*. Brookings Institution Press.

Olschewski, S., Rieskamp, J., & Scheibehenne, B. (2018). Taxing cognitive capacities reduces choice consistency rather than preference: A model-based test. *Journal of Experimental Psychology: General, 147*, 462-484.

Olson, M. (1965). *The logic of collective action: Public goods and the theory of groups*. Cambridge, MA: Harvard University Press.

Osgood, M. J., & Muraven, M. (2015). Self-control depletion does not diminish attitudes about being prosocial but diminish prosocial behaviors. *Basic Applied Social Psychology, 37*, 68-80.

Ostrom, E. (2000). Collective action and the evoltion of social norms. *Journal of Economic Perspectives, 14*, 137-158.

Paruzel-Czachura, M., Pypno, K., Everett, J. A., Białek, M., & Gawronski, B. (2023). The Drunk Utilitarian Revisited: Does Alcohol Really Increase Utilitarianism in Moral Judgment? *Personality and Social Psychology Bulletin*, *49*, 20-31.

Patil, I., Zucchelli, M. M., Kool, W., Campbell, S., Fornasier, F., Calò, M., ... & Cushman, F. (2021). Reasoning supports utilitarian resolutions to moral dilemmas across diverse measures. *Journal of Personality and Social Psychology*, *120*, 443-460.

Paxton, J. M., Bruni, T., & Greene, J. D. (2014). Are "counter-intuitive" deontological judgments really counter-intuitive? An empirical reply to Kahane et al. (2012). *Social Cognitive and Affective Neuroscience, 9*, 1368–1371.

Paxton, J. M., Ungar, L., & Greene, J. D. (2012). Reflection and reasoning in moral judgment. *Cognitive Science, 36*, 163-17.

Pennycook, G. (2023). A framework for understanding reasoning errors: From fake news to climate change and beyond. *Advances in Experimental Social Psychology, 67*, 131-208.

Pennycook, G., Epstein, Z., Mosleh, M., Arechar, A. A., Eckles, D., & Rand, D. G. (2021). Shifting attention to accuracy can reduce misinformation online. *Nature*, *592*, 590-595.

Pennycook, G., Fugelsang, J. A., & Koehler, D. J. (2015). What makes us think? A three-stage dual-process model of analytic engagement. *Cognitive Psychology, 80*, 34-72.

Pennycook, G., & Rand, D. G. (2022). Accuracy prompts are a replicable and generalizable approach for reducing the spread of misinformation. *Nature Communications*, *13*, 2333.

Perc, M., Jordan, J. J., Rand, D. G., Wang, Z., Boccaletti, S., & Szolnoki, A. (2017). Statistical physics of human cooperation. *Physics Reports, 687*, 1-51.

Pfattheicher, S., Keller, J., Knezevic, G. (2017). Sadism, the intuitive system, and antisocial punishment in the public goods game. *Personality and Social Psychology Bulletin, 43*, 337-346.

Pitesa, M., Thau, S., & Pillutla, M. M. (2013). Cognitive control and socially desirable behavior: The role of interpersonal impact. *Organizational Behavior and Human Decision Processes, 122*, 232–243.

Ramsden, M. (2015). Think About It: Cognitive Load and the Trolley Scenario as an Analogue of Gun Violence. *Psychological Research*, 136-142.

Rand, D. G. (2016). Cooperation, fast and slow: Meta-analytic evidence for a theory of social heuristics and self-interested deliberation. *Psychological Science, 27,* 1192-1206.

Rand, D. G. (2019). Intuition, deliberation, and cooperarion: Further meta-analytic evidence from 91 experiments on pure cooperation. *Available at: https://papers.ssrn.com/sol3/papers.cfm?abstract_id=3390018*

Rand, D. G., Arbesman, S., & Christakis, N. A. (2011). Dynamic social networks promote cooperation in experiments with humans. *Proceedings of the National Academy of Sciences*, *108*, 19193-19198.

Rand, D. G., Brescoll, V. L., Everett, J. A. C., Capraro, V., & Barcelo, H. (2016). Social heuristics and social roles: Intuition favors altruism for women but not for men. *Journal of Experimental Psychology: General, 145*, 389-396.

Rand, D. G., Greene, J. D., & Nowak, M. A. (2012). Spontaneous giving and calculated greed. *Nature, 489,* 427-430.

Rand, D. G., & Kraft-Todd, G. T. (2014). Reflection does not undermine self-interested prosociality. *Frontiers in Behavioral Neuroscience, 8*, 300.

Rand, D. G., Newman, G. E., & Wurzbacher, O. M. (2015). Social context and the dynamic of cooperative choices. *Journal of Behavioral Decision Making, 28,* 159-166.

Rand, D. G., & Nowak, M. A. (2013). Human cooperation. *Trends in Cognitive Sciences, 17*, 413-425.

Rand, D. G., Peysakhovich, A., Kraft-Todd, G., Newman, G. E., Wurzbacher, O., Nowak, M. A., & Greene, J. D. (2014). Social heuristics shape intuitive cooperation. *Nature Communications, 5*, 3677.

Rantapuska, E., Freese, R., Jääskeläinen, I. P., & Hytönen, K. (2017) Does short-term hunger increase trust and trustworthiness in a high trust society? *Frontiers in Psychology, 8*, 1944.

Reber, A. S. (1993). *Implicit Learning and Tacit Knowledge*. Oxford, UK: Oxford Univ. Press.

Rehren, P. The Effect of Cognitive-Processing Manipulations on Moral Judgment: A Systematic Review and Meta-Analysis. *Available at SSRN 4149203*.

Reis, M., Pfister, R., & Foerster, A. (2022). Cognitive load promotes honesty. *Psychological Research*, 1-19.

Roch, S. G., Lane, J. A. S., Samuelson, C. D., Allison, S. T., Dent, J. L. (2000). Cognitive load and the equality heuristic: A two-stage model of resource overconsumption in small groups. *Organizational Behavior and Human Decision Processes, 83*, 185-212.

Rosas, A., & Aguilar-Pardo, D. (2020). Extreme time-pressure reveals utilitarian intuitions in sacrificial dilemmas. *Thinking & Reasoning*, *26*, 534-551.

Rosas, A., Bermúdez, J. P., & Aguilar-Pardo, D. (2019). Decision conflict drives reaction times and utilitarian responses in sacrificial dilemmas. *Judgment and Decision Making*, *14*, 555-564.

Rubinstein, A. (2007). Istinctive and cognitive reasoning: A study of response times. *The Economic Journal, 117*, 1243-1259.

Sætrevik, B., & Sjåstad, H. (2022). Mortality salience effects fail to replicate in traditional and novel measures. *Meta-Psychology*, *6*.

Samuelson, W., & Zeckhauser, R. (1988). Status quo bias in decision making. *Journal of Risk and Uncertainty, 1*, 7-59.

Santa, J. C., Exadaktylos, F., & Soto-Faraco, S. (2018). Beliefs about others' intentions determine whether cooperation is the faster choice. *Scientific Reports*, *8*, 7509.

Scherlippens, E. (2019). *Look into my eyes.* Doctoral dissertation, Ghent University. Available at https://libstore.ugent.be/fulltxt/RUG01/002/791/129/RUG01-002791129_2019_0001_AC.pdf

Schneider, W., & Shiffrin, R. M. (1977). Controlled and automatic human information processing I: detection, search and attention. *Psychological Review, 84*, 1-66.

Schulz, J. F., Fischbacher, U., Thöni, C., & Utikal, V. (2014). Affect and fairness: Dictator games under cognitive load. *Journal of Economic Psychology, 41*, 77-87.

Schwitzgebel, E., & Cushman, F. (2015). Philosophers' biased judgments persist

despite training, expertise and reflection. *Cognition, 141*, 127–137.

Shalvi, S., Eldar, O., & Bereby-Meyer, Y. (2012). Honesty requires time (and lack of justifications). *Psychological Science*, *23*, 1264–1270.

Shi, R., Qi, W. G., Ding, Y., Liu, C., & Shen, W. (2020). Under what circumstances is helping an impulse? Emergency and prosocial traits affect intuitive prosocial behavior. *Personality and Individual Differences*, *159*, 109828.

Shin, H. I., & Kim, J. (2017). Foreign language effect and psychological distance. *Journal of Psycholinguistic Research*, *46*, 1339-1352.

Shiv, B., & Fedorikhin, A. (1999). Heart and mind in conflict: The interplay of affect and cognition in consumer decision making. *Journal of Consumer Research, 26,* 278-292.

Singer, P. (1979). *Practical Ethics*. Cambridge University Press, Cambridge, UK.

Skyrms, B. (2004). *The Stag Hunt and the Evolution of Social Structure.* Cambridge University Press, Cambridge.

Sloman, S. A. (1996). The empirical case for two systems of reasoning. *Psychological Bulletin, 119*, 3-22.

Soon, C. S., Brass, M., Heinze, H. J., & Haynes, J. D. (2008). Unconscious determinants of free decisions in the human brain. *Nature Neuroscience,11,* 543-545.

Small, D. A., Loewenstein, G., & Slovic, P. (2007). Sympathy and callousness: The impact of deliberative thought on donations to identifiable and statistical victims. *Organizational Behavior and Human Decision Processes*, *102*, 143-153.

Smith, E. R., & DeCoster, J. (2000). Dual-process models in social and cognitive psychology: conceptual integration and links to underlying memory systems. *Personality and Social Psychology Review, 4,* 108-131.

Spears, D., Fernández-Linsenbarth, I., Okan, Y., Ruz, M., & González, F. (2018). Disfluent fonts lead to more utilitarian decisions in moral dilemmas. *Psicológica Journal*, *39*, 41-63.

Speer, S. P., Smidts, A., & Boksem, M. A. (2020). Cognitive control increases honesty in cheaters but cheating in those who are honest. *Proceedings of the National Academy of Sciences*, *117*, 19080-

19091.

Speer, S. P., Smidts, A., & Boksem, M. A. (2022). Cognitive control and dishonesty. *Trends in Cognitive Sciences, 26*, 796-808.

Spence, S. A., Farrow, T. F. D., Herford, A. E., Wilkinson, I. D., Xheng, Y., & Woodruff, P. W. R. (2001). Behavioural and functional anatomical correlates of deception in humans. *Neuroreport, 12*, 2849-2853.

Stanovich, K. E. (1999). *Who is Rational? Studies of Individual Differences in Reasoning*. Mahwah, NJ: Elrbaum.

Stanovich, K. E. (2004). *The Robot's Rebellion: Finding Meaning in the Age of Darwin*. Chicago: Chicago Univ. Press.

Stanovich, K. E. (2011). *Rationality and the Reflective Mind*. New York, NY: Oxford University Press.

Starcke, K., Ludwig, A. C., & Brand, M. (2012). Anticipatory stress interferes with utilitarian moral judgment. *Judgment and Decision making*, *7*, 61-68.

Strack, F., & Deutsch, R. (2004). Reflective and impulsive determinants of social behavior. *Personality and Social Psychology Review, 8,* 220-247.

Strohminger, N., Lewis, R. L., & Meyer, D. E. (2011). Divergent effects of different positive emotions on moral judgment. *Cognition, 119*, 295-300.

Strømland, E., Tjotta, S., & Torsvik, G. (2016). Cooperating, fast and slow: Testing the social heuristics hypothesis. *Available at SSRN:* https://ssrn.com/abstract=2780877

Strømland, E., & Torsvik, G. (2019). Intuitive prosociality: Heterogeneous treatment effects or false positive? *Available at* https://osf.io/hrx2y/

Stroop, J. R. (1935). Studies on interferences in serial verbal reactions. *Journal of Experimental Psychology, 18,* 643-662.

Suter, R. S., & Hertwig, R. (2011). Time and moral judgment. *Cognition, 119,* 454-458.

Sutter, M. (2009). Deception through telling the truth?! Experimental evidence from individuals and teams. *The Economic Journal, 119*, 47-60.

Swann, W. B., Hixon, J. G., Stein-Seroussi, A., & Gilbert, D. T. (1990). The fleeting gleam of praise: Cognitive processes underlying behavioral reactions to self-relevant feedback. *Journal of Personality and Social Psychology, 59,* 17-26.

Tajfel, H. (1970). Experiments in intergroup discrimination. *Scientific American, 223*, 96-103.

Thielmann, I., Spadaro, G., & Balliet, D. (2020). Personality and prosocial behavior: A theoretical framework and meta-analysis. *Psychological Bulletin*, *146*, 30-90.

Thompson, V. A., Turner, J. A. P., & Pennycook. (2011). Intuition, reason, and metacognition. *Cognitive Psychology, 63*, 107-140.

Timmons, S., & Byrne, R. M. (2019). Moral fatigue: The effects of cognitive fatigue on moral reasoning. *Quarterly Journal of Experimental Psychology*, 72, 943-954.

Tinghög, G., Andersson, D., Bonn, C., Böttiger, H., Josephson, C., Lundgren, G., Västfjäll, D., Kirchler, M., & Johannesson, M. (2013). Intuition and cooperation reconsidered. *Nature, 498,* E1.

Tinghög, G., Andersson, D., Bonn, C., Johannesson, M., Kirchler, M., Koppel, L., & Västfjäll, D. (2016). Intuition and moral decision-making – The effect of time pressure and cognitive load on moral judgment and altruistic behavior. *PLoS ONE, 11,* e0164012.

Toates, F. (2006). A model of the hierarchy of behaviour, cognition and consciousness. *Consciousness and Cognition, 15,* 75–118.

Tomasello, M., Carpenter, M., Call, J, Behne, T., & Moll, H. (2005). In search of the uniquely human. *Behavioral and Brain Sciences, 28,* 721-727.

Trémolière, B., & Bonnefon, J.-F. (2014). Efficient kill-save ratios ease up the cognitive demands on counterintuitive moral utilitarianism. *Personality and Social Psychology Bulletin, 40,* 923-930.

Trémolière, B., De Neys, W., & Bonnefon, J.-F. (2012). Mortality salience and morality: Thinking about death makes people less utilitarian. *Cognition*, 124, 379-384.

Trivers, R. (1971). The evolution of reciprocal altruism. *The Quarterly Journal of Biology, 46,* 35-57.

Tversky, A., & Kahneman, D. (1974). Judgment under Uncertainty: Heuristics and Biases. *Science*, *185*, 1124-1131.

Ugazio, G., Lamm, C., & Singer, T. (2012). The role of emotions for moral judgments depends on the type of emotion and moral scenario. *Emotion, 12*, 579-590.

Urbig, D., Terjesen, S., Procher, V., Muehlfeld, K., & van Witteloostuijn, A. (2016). Come on and take a free ride: Contributing to public goods in native and foreign language settings. *Academy of Management Learning and Education, 15*, 268-286.

Valdesolo, P., & DeSteno, D. (2006). Manipulations of emotional context shape moral judgment. *Psychological Science, 17*, 476-477.

Van der Cruyssen, I., D'hondt, J., Meijer, E., & Verschuere, B. (2020). Does honesty require time? Two preregistered direct replications of experiment 2 of Shalvi, Eldar, and Bereby-Meyer (2012). *Psychological Science*, *31*, 460-467.

van't Veer, A. E., Stel, M., & Van Beest, I. (2014). Limited capacity to lie: Cognitive load interferes with being dishonest. *Judgment and Decision Making*, *9*, 199–206.

Verkoeijen, P. P. J. L., & Bouwmeester, S. (2014). Does intuition cause cooperation? *PLoS ONE, 9*, e96654.

Verschuere, B., & Shalvi, S. (2014). The truth comes naturally! Does it? *Journal of Language and Social Psychology, 33*, 417-423.

Vives, M. – L-, Aparici, M., & Costa, A. (2018). The limits of the foreign language effect on decision-making: The case of outcome bias and the representativeness heuristic. *PLoS ONE, 13*, e0203528.

Vohs, K. D., Glass, B. D., Maddox, W. T., & Markman, A. B. (2011). Ego depletion is not just fatigue: Evidence from a total sleep deprivation experiment. *Social Psychological and Personality Science, 2*, 166-173.

Volk, S., Köhler, T., & Pudelko, M. (2014). Brain drain: The cognitive neuroscience of foreign language processing in multinational corporations. *Journal of International Business Studies, 45*, 862-885.

Walczyk, J. J., Mahoney, K. T., Doverspike, D., Griffith-Ross, D. A. (2009). Cognitive lie detection: Response time and consistency of answers as cues to deception. *Journal of Business Psychology, 24*, 33-49.

Wang, C. S., Sivanathan, N., Narayanan, J., Ganegoda, D. B., Bauer, M., Bodenhausen, G. V., & Murnighan, K. (2011). Retribution and emotional regulation: The effects of time delay in angry economic interactions. *Organizational Behavior and Human Decision Processes, 116*, 46-54.

Wang, Y., Zhang, X., Li, J., & Xie, X. (2019). Light in darkness: Low self-control promotes altruism in crises. *Basic and Applied Social Psychology, 41*, 201-213.

Weippert, M., Rickler, M., Kluck, S., Behrens, K., Bastian, M., Mau-Moeller, A., ... & Lischke, A. (2018). It's harder to push, when I have to push hard—Physical exertion and fatigue changes reasoning and decision-making on hypothetical moral dilemmas in males. *Frontiers in Behavioral Neuroscience*, *12*, 268.

Welsh, D. T., Ellis, A. P. J., Christian, M. S., & Mai, K. M. (2014). Building a self- regulatory model of sleep deprivation and deception: The role of caffeine and social influence. *The Journal of Applied Psychology*, *99*(6), 1268–1277.

Wilson, T. D. (2002). *Strangers to Ourselves*. Cambridge, MA: Belknap.

Xu, H., Bègue, L., Sauve, L., & Bushman, B. J. (2014). Sweetened blood sweetens behavior. Ego depletion, glucose, guilt, and prosocial behavior. *Appetite*, *81*, 8-11.

Yam, K. C. (2018). The effects of thought suppression on ethical decision making: Mental rebound versus ego depletion. *Journal of Business Ethics, 147*, 65-79.

Yam, K. C., Chen, X. – P., & Reynolds, S. J. (2014). Ego depletion and its paradoxical effects on ethical decision making. *Organizational Behavior and Human Decision Processes, 124*, 204-214.

Yamagishi, T., Matsumoto, Y., Kiyonari, T., Takagishi, H., Li, Y., Kanai, R., & Sakagami, M. (2017). Response time in economic games reflects different types of decision conflict for prosocial and proself individuals. *Proceedings of the National Academy of Sciences, 114*, 6394-6399.

Yudkin, D. A., Rothmund, T., Twardawski, M., Thalla, N., & Van Bavel, J. J. (2016). Reflexive intergroup bias in third-party punishment. *Journal of Experimental Psychology: General, 145*, 1448-1459.

Zhong, C.-B. (2011). The ethical dangers of deliberative decision making. *Administrative Science*

*Quarterly*, *56*, 1–25.

# Footnotes

[1] Acknowledgements hidden for double-blind review.

[2] This review will focus only on experimental manipulations, which allows to draw a causal link between the independent and the dependent variables, as opposed to correlational studies that can be driven by unobserved variables. For example, since intuitive processes are typically fast and deliberative processes are typically slow, one might think that looking at response times in the control condition, rather than experimentally manipulating response time, would be enough to draw conclusions regarding which choice is typically driven by intuition and which is typically driven by deliberation. However, it has been noticed that response time predicts decision conflict and choice discriminability, rather than deliberation, at least in some contexts (Evans, Dillon & Rand, 2015; Evans & Rand, 2018; Krajbich, Bartling, Hare & Fehr, 2015). Therefore, this article will not include studies that have looked at the correlation between response times (or other typical correlates of the cognitive processes) and social behaviour.

[3] There are also studies exploring how inducing a particular emotion affects a subsequent decision (Motro, Ordóñez, Pittarello & Welsh, 2018; Strohminger, Lewis & Meyer, 2011; Ugazio, Lamm & Singer, 2012). These works have shown that different emotions affect social decisions differently. For example, Strohminger et al. (2011) and Ugazio et al. (2012) found that anger tends to increase consequentialist judgments while disgust tends to increase deontological judgments. Motro et al. (2018) let participants write about a time in their life in which they felt angry vs guilty; they found that inducing anger made people more likely to lie, while inducing guilt made them more honest. This emotion-dependent effect of priming suggests that inducing a particular emotion does not simply promote intuitive processes. For this reason, this review will include only studies inducing people to “rely on emotion”, *in general*, and not those inducing a particular emotion.

[4] Some authors argued that sleep deprivation and ego depletion are different (Vohs, Glass, Maddox & Markman, 2011). However, a review found a positive effect of sleep deprivation on

aggression, a behaviour typically driven by lack of self-control, suggesting that sleep deprivation successfully impairs self-control (Kamphuis, Meerlo, Koolhaas & Lancel, 2012).

[5] For example, Danziger, Levav and Avnaim-Pesso (2011) found that judges' favourable rulings drop from nearly 65% to nearly 0% from the morning to lunch time, and then again from after lunch time to dinner time.

[6] Alcohol intoxication is known to diminish the executive cognitive functions necessary for effortful and controlled processing (Moskowitz, Burns & Williams, 1985; Fillmore, Carscadden & Vogel-Sprott, 1998; Curtin & Fairchild, 2003).

[7] Mortality salience is a cognitive manipulation based on Terror Management Theory (Becker, 1973; Greenberg, Pyszczynski & Solomon, 1986). According to this theory, people rely on their cognitive resources to avoid the anxiety provoked by death-related thoughts. Therefore, reminding people about their mortality has the effect of depleting their cognitive resources. Several studies supported this view (Greenberg, Pyszczynski & Solomon, 1986). However, mortality salience also has been criticized, as a pre-registered replication of the original study failed to find an effect (Sætrevik & Sjåstad, 2022).

[8] The Need for Cognition is a scale introduced by Cacioppo and Petty (1982) to measure the extent to which a person is inclined towards effortful cognitive activities.

[9] The Faith in Intuition is a scale introduce by Epstein, Pacini, Denes-Raj and Heier (1996) to measure the extent to which a person is inclined towards effortless, intuitive, activities.

[10] There is some work looking at the correlation between response time and cooperation in the traveller's dilemma (Rubinstein, 2007; Santa, Exadaktylos, & Soto-Faraco, 2018), but, to the best of current knowledge, no studies exploring the causal link.

[11] There are also some correlational studies that are broadly in line with this view. Mischkowski and Glöckner (2016) found that response time is positively correlated with cooperation among prosocials, but not among pro-selves, for whom there is no correlation. Yamagishi et al. (2017) found the same result for prosocials; however, among pro-selves, they found response time to be negatively correlated with cooperation. However, a meta-analysis found that the relationship between social value orientation and

response time is moderated by extremity of choice, suggesting that it might be driven by decision conflict (Andrighetto, Capraro, Guido & Szekely, 2020).

[12] Fromell et al. (2020) included in their meta-analysis also studies using the dictator game with varying cost to benefit ratio. However, in a private communication, one of the authors (Daniele Nosenzo) confirmed that their results remain qualitatively the same when restricting the analysis to the standard dictator games.

[13] Some authors have also considered situations in which participants are not incentivized to lie. Several works have provided evidence that, in this case, honesty tends to be intuitive (Debey, Verschuere & Crombez, 2012; Spence et al. 2001; Walczyk, Mahoney, Doverspike & Griffith-Ross, 2009).

[14] Note that this is not inconsistent with the result mentioned in Footnote 13, that honesty is intuitive when participants are not incentivized to lie. As observed by Verschuere and Shalvi (2014), "truth telling may be the natural response absent clear motivations to lie […] and that lying may prevail as the automatic reaction when it brings about important self-profit".

[15] The purpose of this article is not to take a stance on the best way to define utilitarianism in philosophical terms or how to measure it. However, it is important to note that there is an ongoing debate on this topic. For instance, Conway, Goldstein-Greenwood, Polacek & Greene (2018) used a process dissociation approach to argue that sacrificial utilitarian judgments do reflect genuine moral concern. According to their work, "antisociality predicts reduced deontological inclinations, not increased utilitarian inclinations".

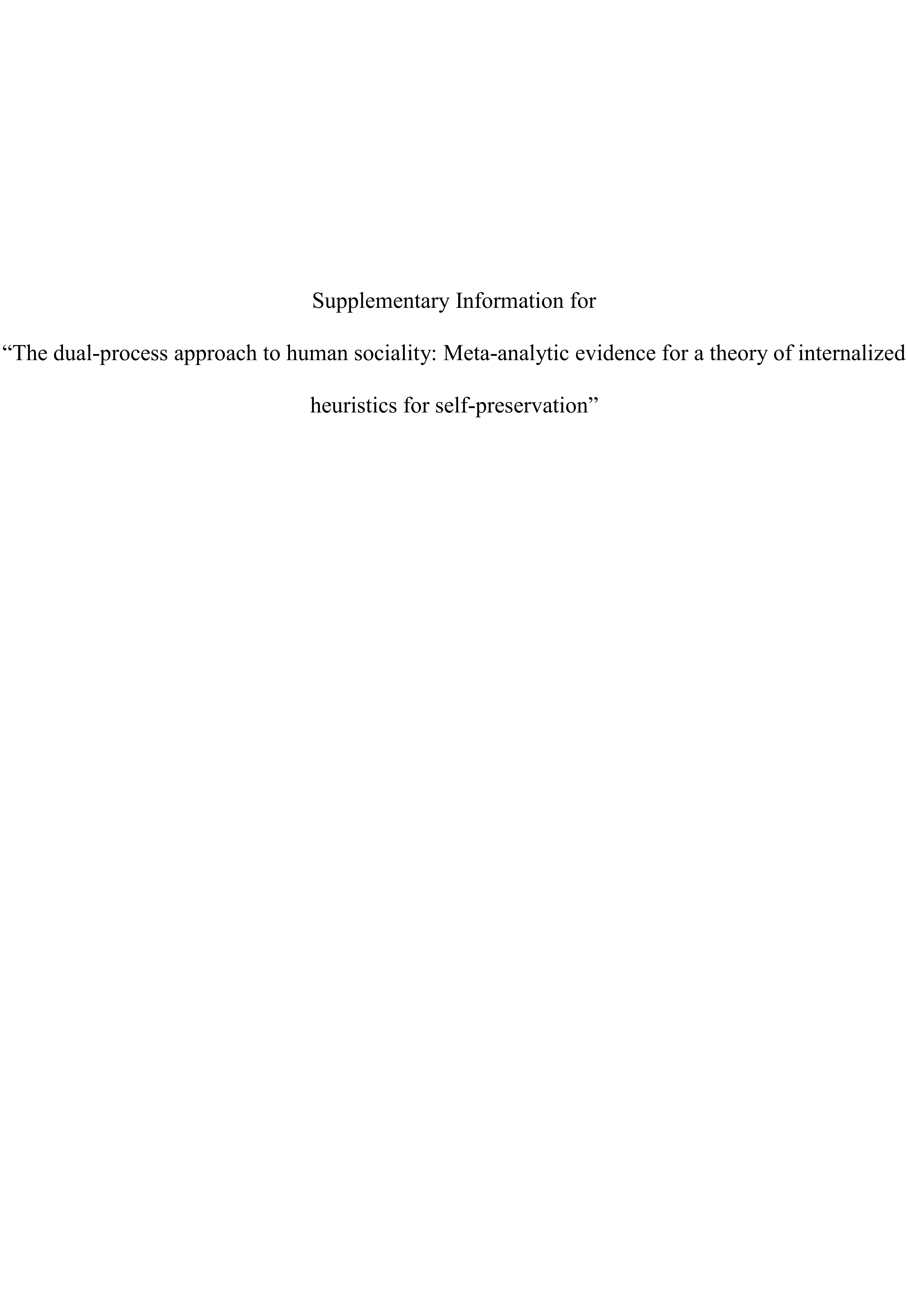

Supplementary Information for

"The dual-process approach to human sociality: Meta-analytic evidence for a theory of internalized heuristics for self-preservation"

**Systematic review and meta-analysis of the cognitive basis of negative reciprocity**

The systematic review was conducted as follows:

- On April 24, 2023, a search on Google Scholar was performed using the following query: ("ultimatum game" OR "second-party punishment" OR "third-party punishment") AND ("time pressure" OR "time delay" OR "cognitive load" OR "ego depletion" OR "two response" OR "intuition priming" OR "intuition recall" OR "emotion induction") AND "one shot". This search yielded a total of 825 items.
- Going through the Google Scholar list, the titles, abstracts, and, in some cases, the methods of each article were reviewed, with the following a stopping rule: after encountering 50 consecutive papers that did not meet the eligibility criteria for the meta-analysis, the review was stopped. This happened at page 14 in Google Scholar. The eligibility criteria for the review were: between-subject cognitive manipulation, one-shot game or iterated with random re-matching, and inclusion of a measure of negative reciprocity or expectations of negative reciprocity.
- The remaining papers were carefully screened, leading to the exclusion of those that did not meet the inclusion criteria. See Figure S1 for the flow diagram.
- To conduct the meta-analysis, each paper was divided into various treatments based on the cognitive manipulations and the dependent measure. Specifically, every possible pair of experimental conditions where one condition was relatively more Type 1 (e.g., time pressure) and the other condition was relatively more Type 2 (e.g., time delay) was considered. Additionally, conditions were separated based on whether the outcome measure was negative reciprocity or expectations of negative reciprocity. For each pair of conditions, the standardized mean difference and its corresponding 95% confidence interval were calculated.

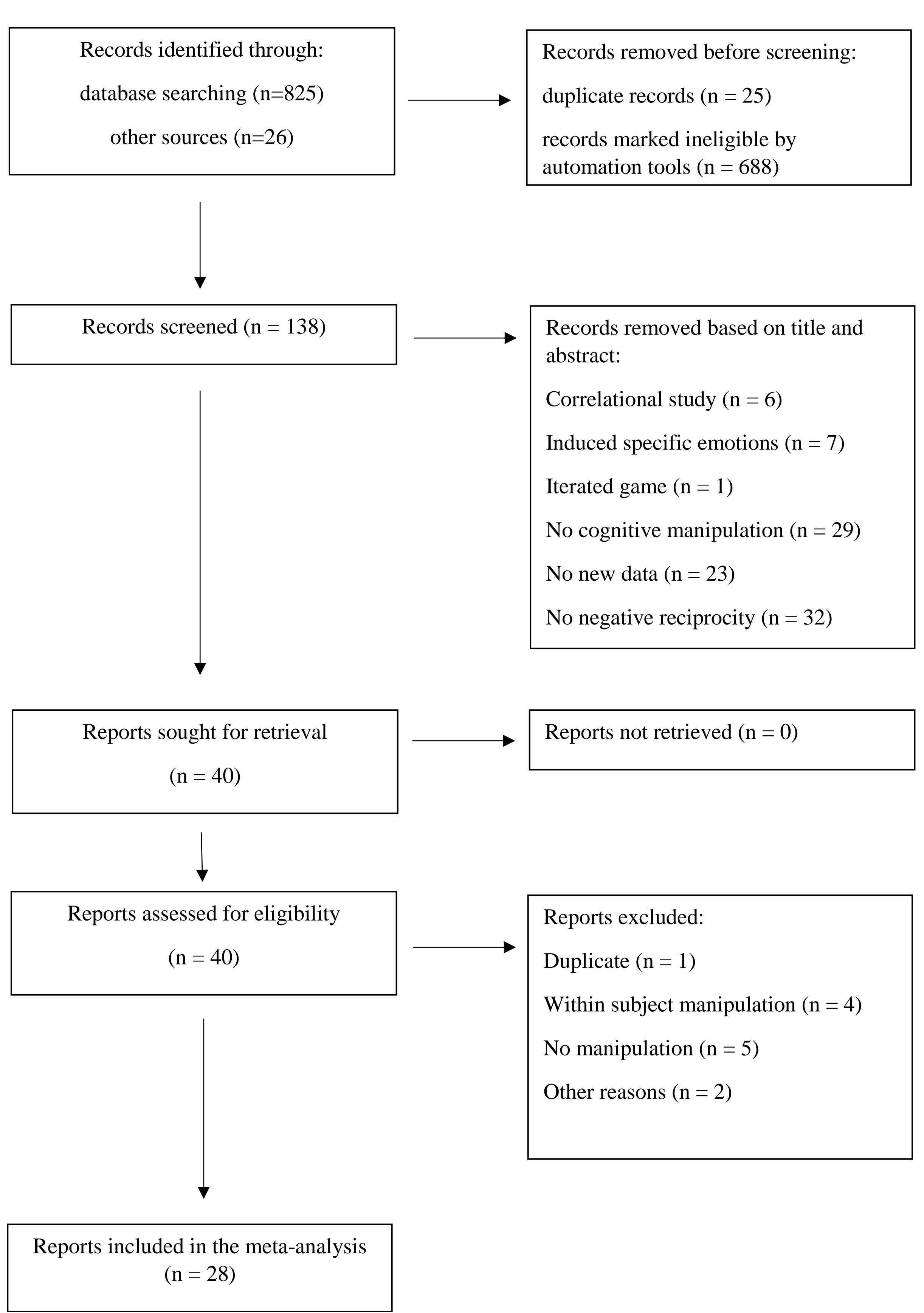

**Figure S1.** Flow diagram of the systematic review and meta-analysis of experimental studies testing the cognitive basis of negative reciprocity.

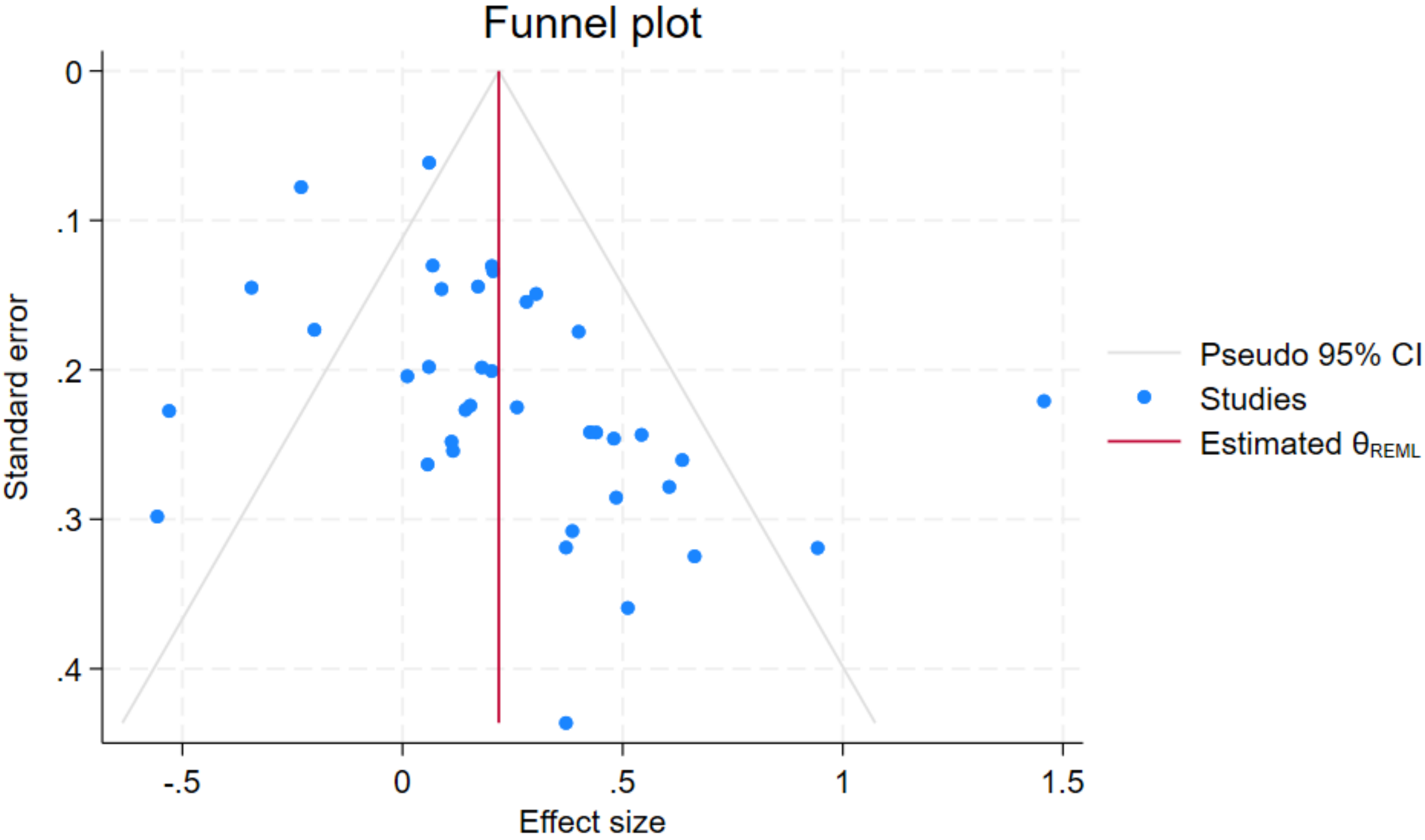

**Figure S2**. Funnel plot (after trim-and-fill) of the studies testing the effect of cognitive manipulations on negative reciprocity.

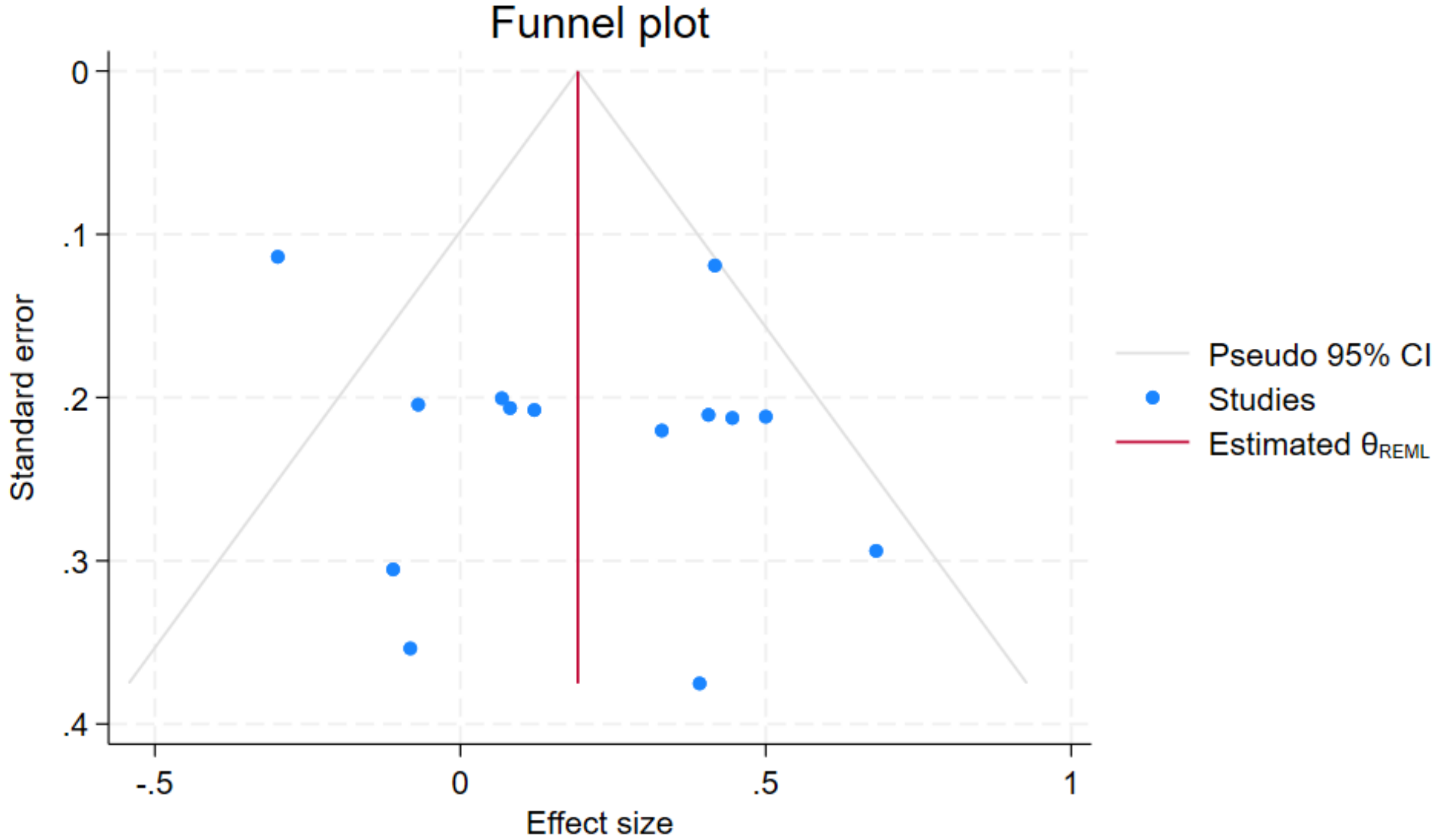

**Figure S3.** Funnel plot (after trim-and-fill) of the studies testing the effect of cognitive manipulations on expectations of negative reciprocity

**Systematic review and meta-analysis of the cognitive basis of positive reciprocity**

The systematic review was conducted as follows:

- On April 25, 2023, a search on Google Scholar was performed using the following query: ("ultimatum game" OR "second-party punishment" OR "third-party punishment") AND ("time pressure" OR "time delay" OR "cognitive load" OR "ego depletion" OR "two response" OR "intuition priming" OR "intuition recall" OR "emotion induction") AND "one shot". This search yielded a total of 552 items.
- Going through the Google Scholar list, the titles, abstracts, and, in some cases, the methods of each article were reviewed, with the following a stopping rule: after encountering 50 consecutive papers that did not meet the eligibility criteria for the meta-analysis, the review was stopped. This happened at page 20 in Google Scholar. The eligibility criteria for the review were: between-subject cognitive manipulation, one-shot game or iterated with random re-matching, and inclusion of a measure of positive reciprocity or expectations of positive reciprocity.
- The remaining papers were carefully screened, leading to the exclusion of those that did not meet the inclusion criteria. See Figure S4 for the flow diagram.
- To conduct the meta-analysis, each paper was divided into various treatments based on the cognitive manipulations and the dependent measure. Specifically, every possible pair of experimental conditions where one condition was relatively more Type 1 (e.g., time pressure) and the other condition was relatively more Type 2 (e.g., time delay) was considered. Additionally, conditions were separated based on whether the outcome measure was positive reciprocity or expectations of positive reciprocity. For each pair of conditions, standardized mean difference and its corresponding 95% confidence interval were calculated.

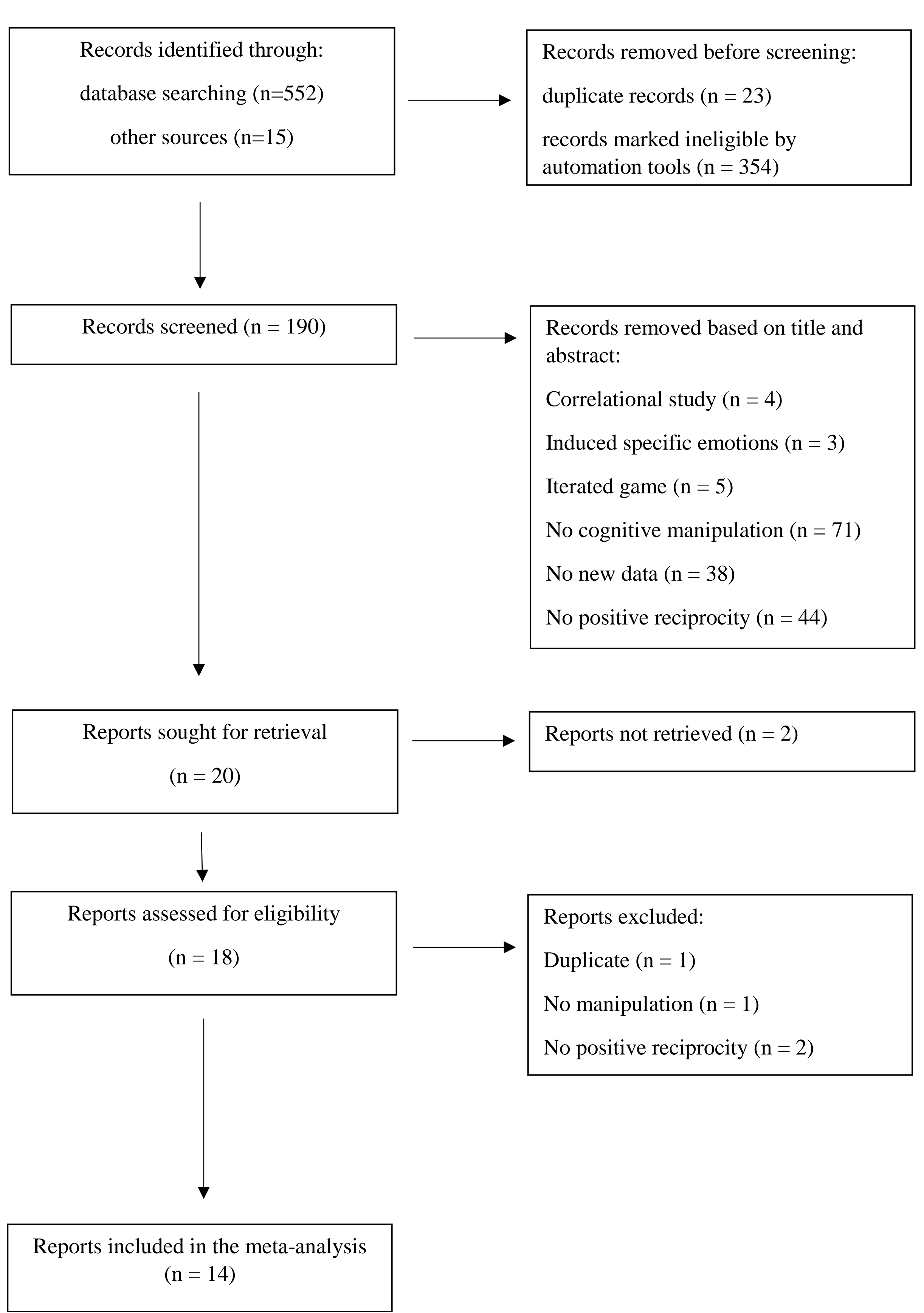

**Figure S4.** Flow diagram of the systematic review and meta-analysis of experimental studies testing the cognitive basis of positive reciprocity.

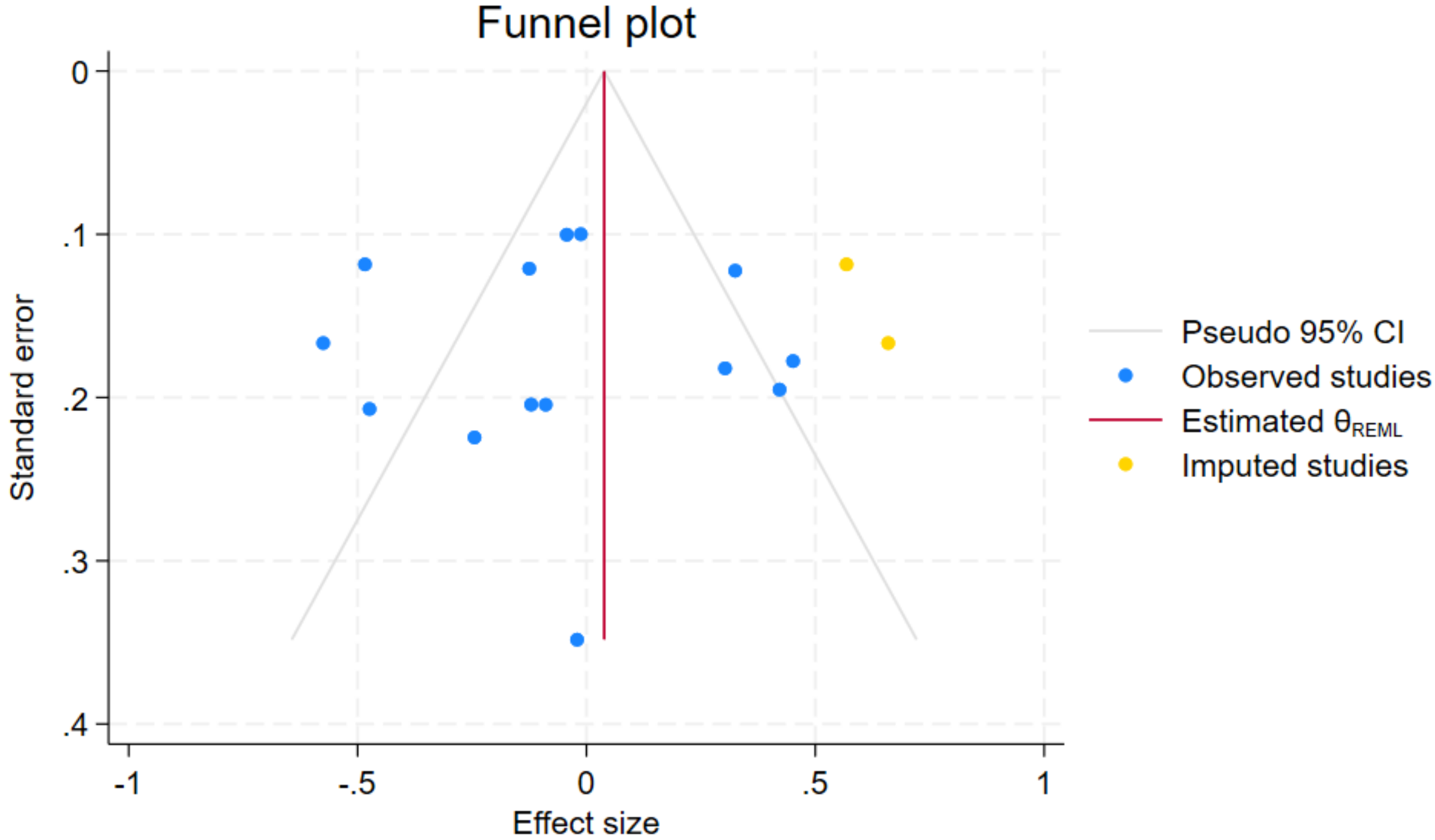

**Figure S5.** Funnel plot of the studies testing the effect of cognitive manipulations on positive reciprocity.

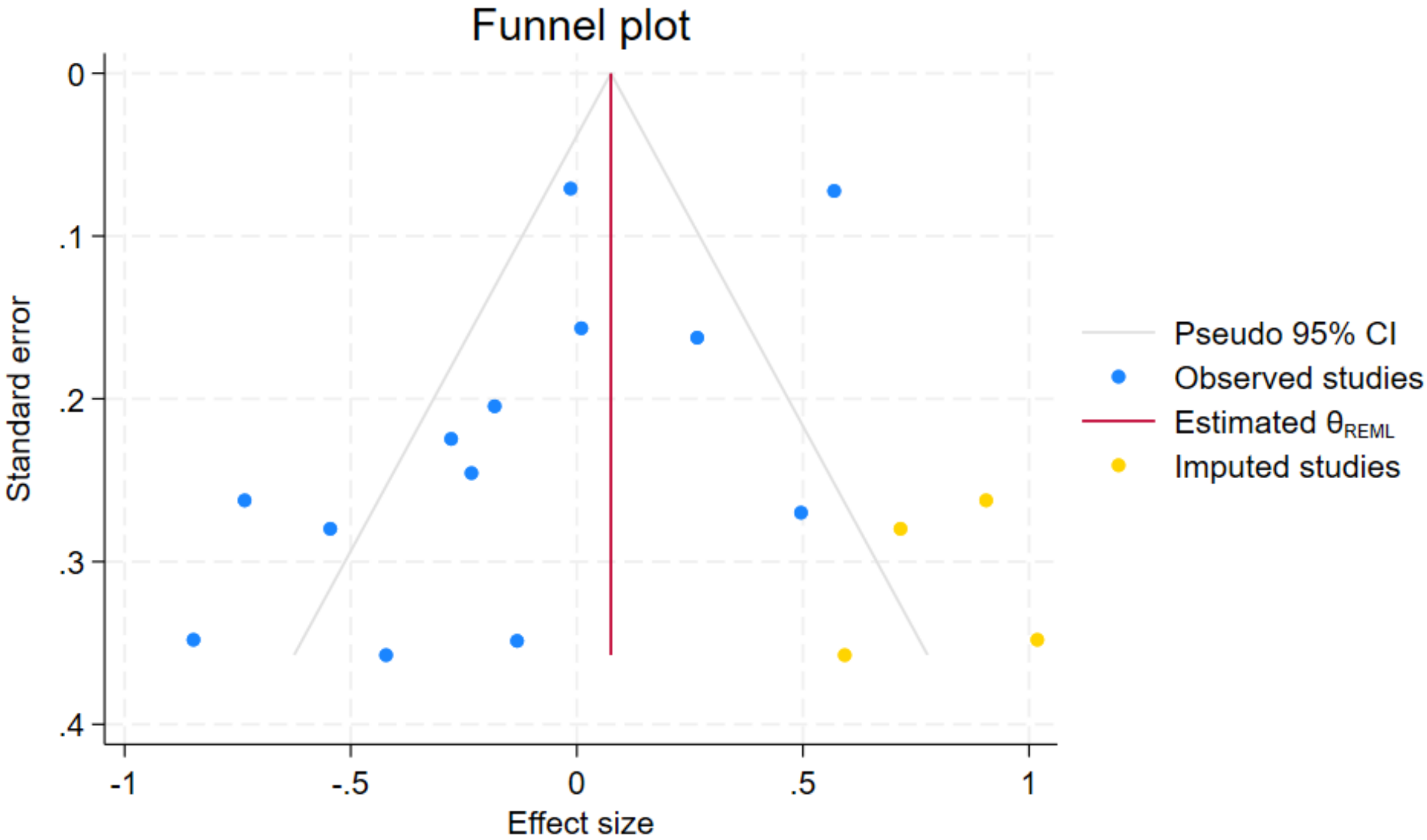

**Figure S6.** Funnel plot of the studies on the cognitive basis of expectations of positive reciprocity, including also the missing studies estimated by the trim-and-fill method (reported in orange).

**Systematic review and meta-analysis of the cognitive basis of act utilitarianism**

The systematic review was conducted as follows:

- On April 25, 2023, a search on Google Scholar was performed using the following query: ("personal moral dilemmas" OR "sacrificial dilemmas") AND ("time pressure" OR "time delay" OR "cognitive load" OR "ego depletion" OR "two response" OR "intuition priming" OR "intuition recall" OR "emotion induction"). This search yielded a total of 599 items.
- Going through the Google Scholar list, the titles, abstracts, and, in some cases, the methods of each article were reviewed, with the following a stopping rule: after encountering 50 consecutive papers that did not meet the eligibility criteria for the meta-analysis, the review was stopped. This happened at page 47 in Google Scholar. The eligibility criteria for the review were: between-subject cognitive manipulation, inclusion of a moral dilemmas involving directly harming an innocent person to save a greater number of people. Dilemmas in which harm was a side-effect as well as dilemmas were the decision maker is one of the people that would be saved from the utilitarian action were excluded.
- The remaining papers were carefully screened, leading to the exclusion of those that did not meet the inclusion criteria. See Figure S7 for the flow diagram.
- To conduct the meta-analysis, each paper was divided into various treatments based on the cognitive manipulations. Specifically, every possible pair of experimental conditions where one condition was relatively more Type 1 (e.g., time pressure) and the other condition was relatively more Type 2 (e.g., time delay) was considered. For each pair of conditions, the standardized mean difference and its corresponding 95% confidence interval were calculated.

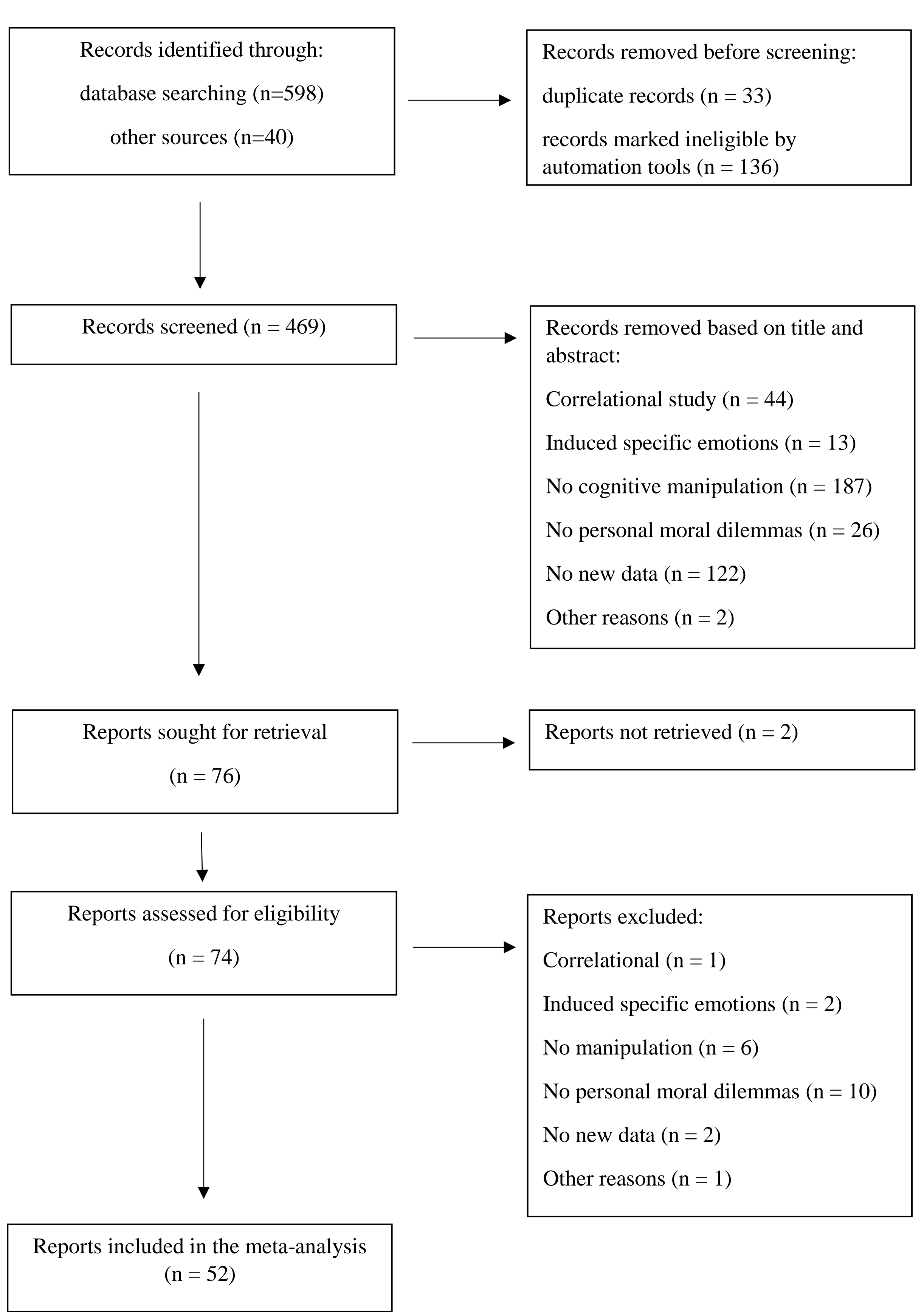

**Figure S7.** Flow diagram of the systematic review and meta-analysis of experimental studies testing the cognitive basis of act utilitarianism.

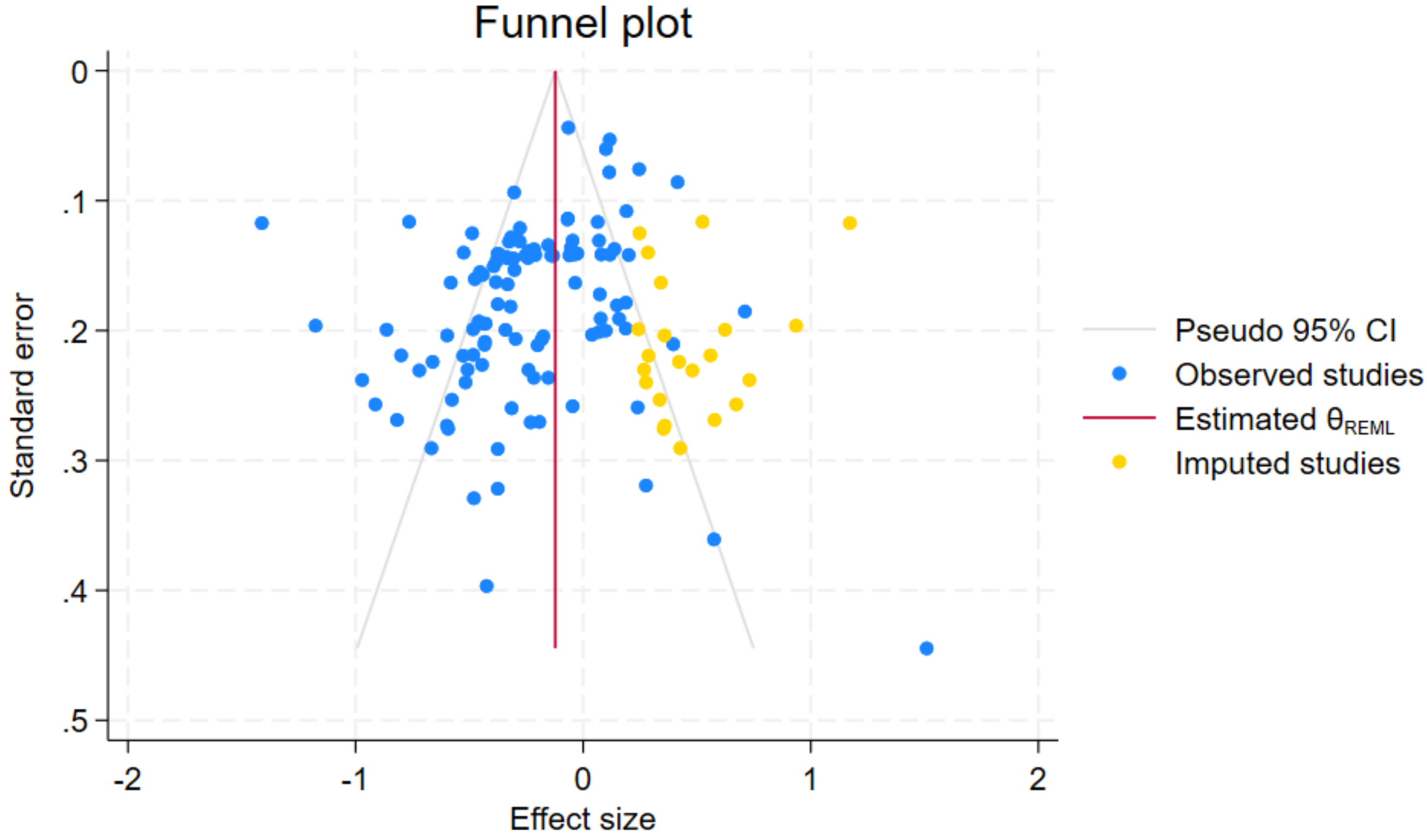

**Figure S8.** Funnel plot of the studies on the cognitive basis of utilitarianism using sacrificial dilemmas; in orange the missing studies estimated by the trim-and-fill method.

**Systematic review and meta-analysis of the cognitive basis of harm as a side effect**

The systematic review was conducted as follows:

- On April 29, 2023, a search on Google Scholar using the following query: ("impersonal moral dilemmas") AND ("time pressure" OR "time delay" OR "cognitive load" OR "ego depletion" OR "two response" OR "intuition priming" OR "intuition recall" OR "emotion induction") was performed. This search yielded a total of 227 items.
- Going through the Google Scholar list, the titles, abstracts, and, in some cases, the methods of each article were reviewed, with the following a stopping rule: after encountering 50 consecutive papers that did not meet the eligibility criteria for the meta-analysis, the review was stopped. The first 50 items were already included, as they were found in the previous systematic review. The eligibility criteria for the review were: between-subject cognitive manipulation, inclusion of a moral dilemmas involving indirectly harming an innocent person (i.e., as a side effect) to save a greater number of people. Dilemmas in which harm was direct as well as dilemmas were the decision maker is one of the pople that would be saved from the utilitarian action were excluded.

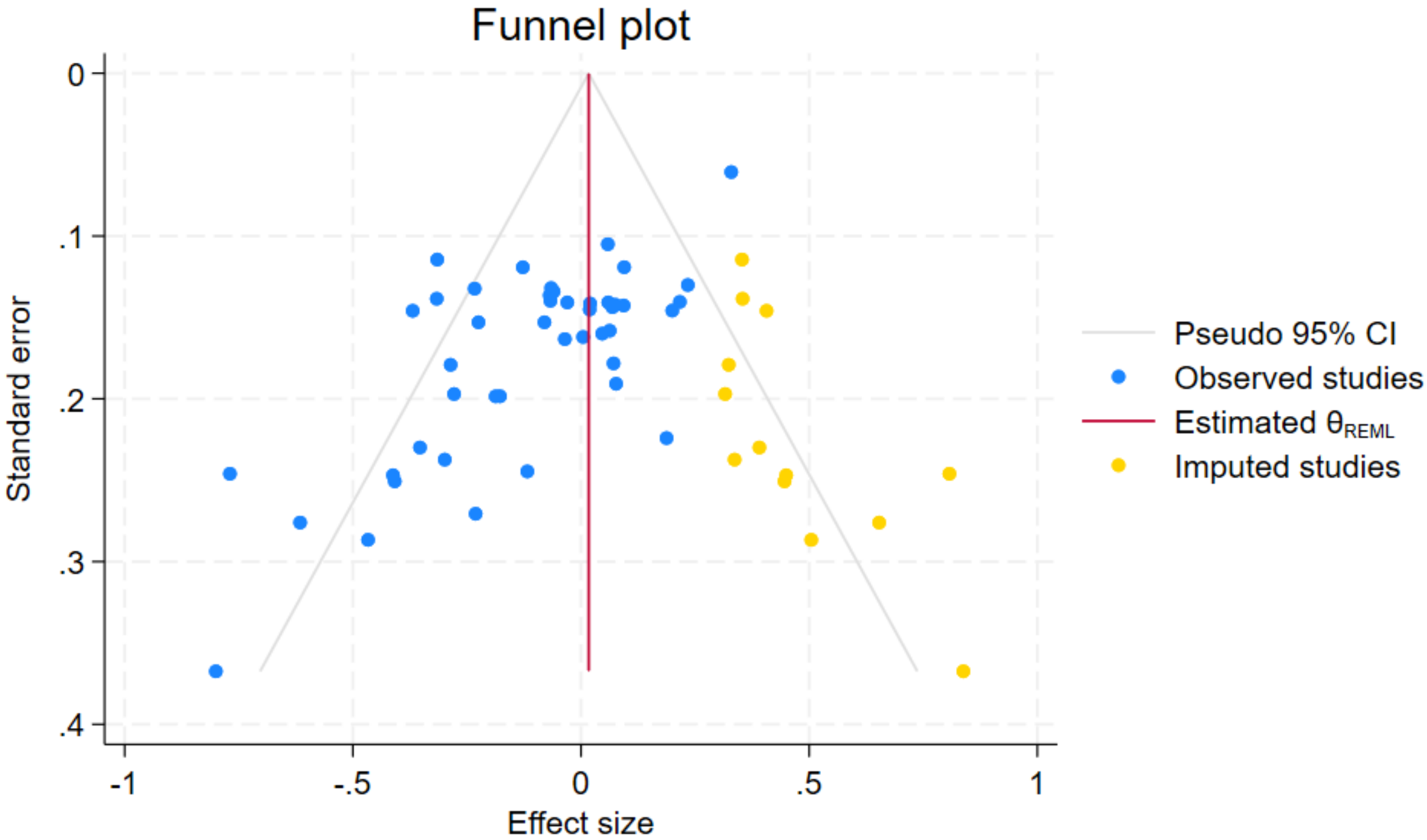

**Figure S9.** Funnel plot of the studies testing the cognitive basis of utilitarianism using side-effect dilemmas.